\documentclass[prx,twocolumn,superscriptaddress]{revtex4-1}
\usepackage{graphicx}
\usepackage{subfigure}
\usepackage{latexsym}
\usepackage{amsmath,amssymb}
\usepackage{bm}
\usepackage{color}
\usepackage{stackrel}
\usepackage{epsfig}
\usepackage{slashed}

\definecolor{myblue}{rgb}{.93, .93, 1}

\newcommand{\bsub}{\begin{subequations}}
\newcommand{\esub}{\end{subequations}}

\newcommand{\vex}[1]{\bm{\mathrm{#1}}}

\renewcommand{\v}[1]{\textbf{#1}}

\begin{document}

\title{Magic-Angle Semimetals with Chiral Symmetry}

\author{Yang-Zhi~Chou}\email{yzchou@umd.edu}
\affiliation{Condensed Matter Theory Center and the Joint Quantum Institute, Department of Physics,
	University of Maryland, College Park, MD 20742 USA}
\affiliation{Department of Physics and Center for Theory of Quantum Matter, University of Colorado Boulder, Boulder, CO 80309 USA}
\author{Yixing Fu}
\affiliation{Department of Physics and Astronomy, Center for Materials Theory, Rutgers University, Piscataway, NJ 08854 USA}
\author{Justin H. Wilson}
\affiliation{Department of Physics and Astronomy, Center for Materials Theory, Rutgers University, Piscataway, NJ 08854 USA}
\author{E. J. K\"onig}
\affiliation{Department of Physics and Astronomy, Center for Materials Theory, Rutgers University, Piscataway, NJ 08854 USA}
\author{J. H. Pixley}
\affiliation{Department of Physics and Astronomy, Center for Materials Theory, Rutgers University, Piscataway, NJ 08854 USA}

\date{\today}

\begin{abstract}
We construct and solve a two-dimensional, chirally symmetric model of Dirac cones subjected to a quasiperiodic modulation. In real space, this is realized with a quasiperiodic hopping term.
This hopping model, as we show, at the Dirac node energy has a rich phase diagram with a semimetal-to-metal phase transition at intermediate amplitude of the quasiperiodic modulation, and a transition to a phase with a diverging density of states and sub-diffusive transport when the quasiperiodic hopping is strongest.
We further demonstrate that the semimetal-to-metal phase transition can be characterized by the multifractal structure of eigenstates in momentum space and can be considered as a unique ``unfreezing'' transition.  This unfreezing transition in momentum space generates flat bands with a dramatically renormalized bandwidth in the metallic phase similar to the phenomena of the band structure of twisted bilayer graphene at the magic angle. We characterize the nature of this transition numerically as well as analytically in terms of the formation of a band of topological zero modes. 
For pure quasiperiodic hopping,
we provide strong numerical evidence that the low-energy density of states develops a
divergence and the eigenstates exhibit Chalker (quantum-critical) scaling despite the model not being random. At particular commensurate limits the model realizes higher-order topological insulating phases. We discuss how these systems can be realized  in experiments on ultracold atoms and metamaterials.
\end{abstract}

\maketitle

\section{Introduction}
\label{sec:introduction}

Quantum phase transitions are ubiquitous in condensed matter systems~\cite{Sachdev-book}. For conventional symmetry breaking quantum transitions, that are described within a Landau-Ginzburg-Wilson paradigm~\cite{Goldenfeld-book}, macroscopic thermodynamic observables show singularities associated with a fundamental change of the ground state with critical exponents dictated by the universality class. There are also quantum phase transitions that take place in the energy spectrum, not necessarily in the ground state, that do not have to have any effect on thermodynamic observables but can affect transport or thermalization properties, such as Anderson \cite{Anderson1958,Abrahams-1979,Lee-1985,Evers2008_RMP} or many-body localization \cite{Basko2006,Gornyi2005,Nandkishore2015,Abanin-2019}, respectively. Interestingly, these transitions fall outside the conventional Landau-Ginzburg paradigm for symmetry breaking thermodynamic phase transitions. Moreover, since they are associated with a fundamental change in the wavefunctions, it is more apt to call them eigenstate phase transitions (EPTs).
These are inherently dynamical phase transitions
and can be
driven by either randomness or deterministic quasiperiodicity.

The majority of the known examples of eigenstate phase transitions involve localization. These transitions do not necessarily have an effect on the density of states (DOS), therefore they do not need to coincide with a thermodynamic phase transition~\cite{Evers2008_RMP,Nandkishore2015}. However, in some cases EPTs can affect both eigenstates and  thermodynamics by fundamentally changing the low-energy DOS.
There are various examples where an EPT gives rise to a ``pile up'' of states near zero energy which creates a diverging low-energy DOS, with the chiral symmetry classes (purely ``off-diagonal'' matrices) of both one- and two-dimensional disordered conductors being prominent examples~\cite{Dyson1953,Gade1991,Gade1993}.
The form of the divergence can depend sensitively on the model under consideration and the presence of rare region effects \cite{Motrunich2002,Mudry2003,HafnerEvers2014,Ostrovsky-2014,FerreiraMucciolo2015,WeikEvers2016,Sanyal2016}.

There is naturally, a completely separate question of EPTs that \emph{generates} a non-zero DOS, namely where an EPT leads to the DOS going from zero to a non-zero value.
This question is particularly poignant to the case of semimetals that have a power-law vanishing DOS
at the nodal energy.
For instance, both two- and three-dimensional Dirac semimetal lattice models are unstable to disorder as indicated by a finite DOS~\cite{Aleiner-2006,Altland-2006,Pixley-2016,PixleyBR-2016,PixleyB-2017,Wilson-2017}.
In the case of quasiperiodicity, however, it has recently been shown numerically~\cite{Pixley2018,FuPixley2018} and rigorously proven mathematically~\cite{Mastropietro2020} that an infinitesimal potential strength is not sufficient to generate finite DOS. Instead, the semimetallic phase survives, albeit with a perturbatively reduced velocity, over an extended regime
where the quasiperiodicity is sufficiently weak.
At the phase boundary, which at fixed potential strength will be called the (first) ``magic-angle'', in analogy with twisted bilayer graphene~\cite{li_observation_2010,BistritzerMacDonald2011, DosSantosNeto2012},
the semimetal undergoes a quantum phase transition into a metallic phase with a finite DOS. This transition is sharp, with the DOS developing true non-analytic behavior, a feature that is rounded out in the presence of randomness~\cite{PixleyBR-2016}.
The semimetal is ballistic and composed of a subextensive number of plane wave states, which corresponds to localized wavefunctions in momentum space. The development of a finite DOS coincides with a delocalization transition in momentum space \cite{Pixley2018,FuPixley2018} and is sufficient to generate diffusive dynamics and random matrix theory level statistics in three-dimensions~\cite{Pixley2018}. As we have shown recently in Ref.~\onlinecite{FuPixley2018}, this kind of EPT
is 
related to
the single-particle physics of ``magic-angle'' twisted bilayer graphene~\cite{li_observation_2010,BistritzerMacDonald2011,DosSantosNeto2012}. 
We use the term ``magic-angle semimetals'' to describe this transition as it occurs along the line of vanishing Dirac cone velocity due to a moir\'e structure in generic Dirac semimetals.
Moreover, two-dimensional Dirac points are straightforward to generate using ultra-cold atom setups~\cite{Tarruell-2012,Aidelsburger-2015,Flaschner-2016,Weinberg-2016,Gonzalez2019}, which make this setting an ideal platform to study to these phenomena in experiments.
While randomness becomes stronger as the dimension decreases, quasiperiodicity evades this and can achieve transitions forbidden in the random problem.

\begin{figure}[t!]
	\centering
	\includegraphics[width=0.45\textwidth]{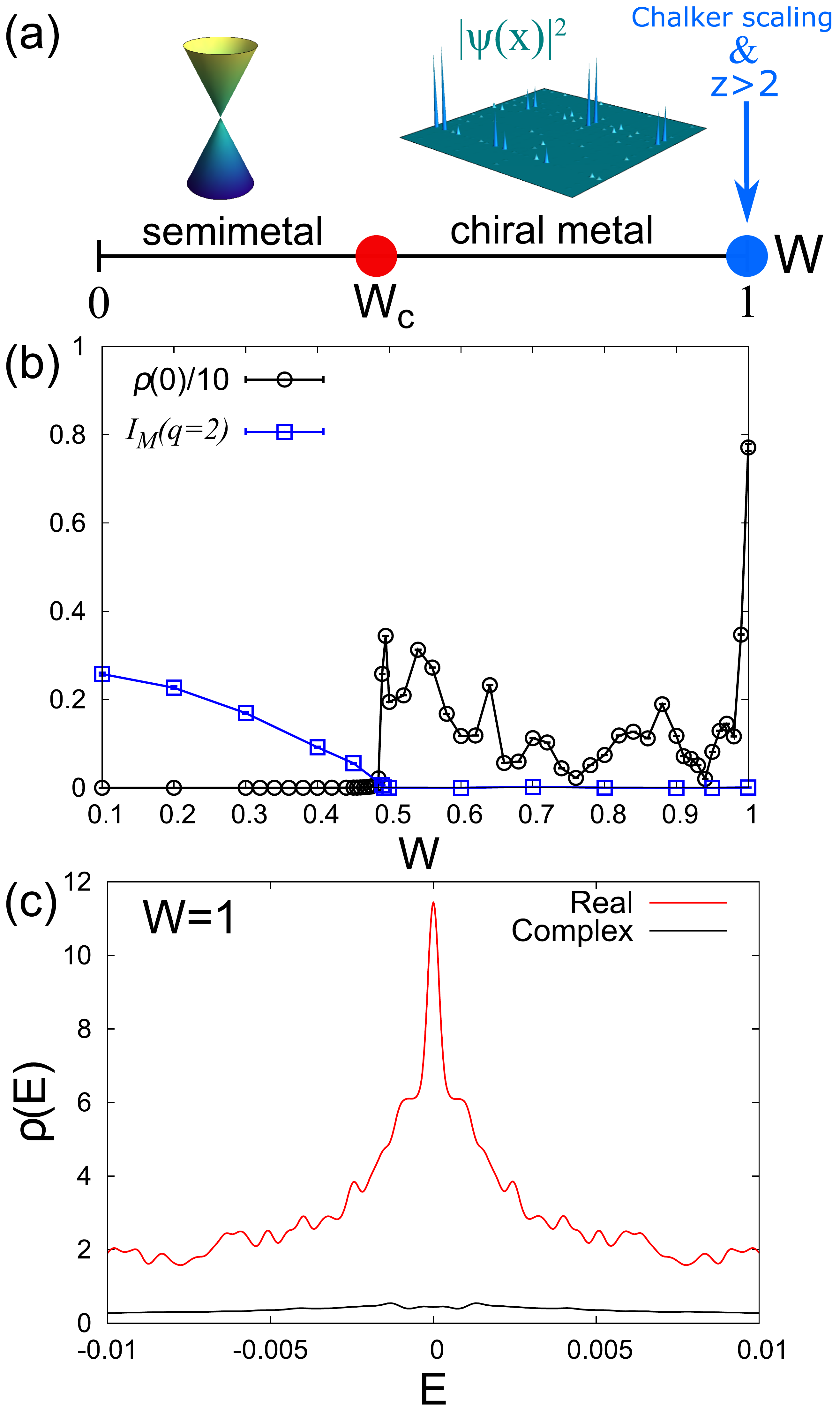}
	\caption{(a) Schematic phase diagram at the band center ($E=0$) extracted from our work. In the semimetal phase the linearly dispersing Dirac cone is stable in the low-energy regime. In the chiral metal phase a band of hybridized zero modes qualitatively explain the sparse (yet still delocalized) structure of the wave functions at the band center. The point $W=1$ is critical, with a diverging low-energy density of states, a dynamic exponent $z>2$, and multifractal eigenstates that obey Chalker scaling.
		(b) The zero energy DOS $\rho(0)$ for a linear system size $L=233$ and KPM expansion order \cite{Weisse-2006} $N_C=2^{14}$ and the momentum space inverse participation ratio $I_M(q=2)$, as defined in Eq.~(\ref{Eq:Pm_tauq}), at $E=0$ with $Q=2\pi F_{n-2}/L$ and $L=144$  versus the hopping strength $W$ on a  linear scale. The zero energy DOS becomes non-zero when the momentum-space IPR vanishes. (c) The low-energy DOS $\rho(E)$ as a function of energy $E$ for pure QP hopping ($W=1$) for the case of real and complex hopping amplitudes for system sizes $L=987$ and $L=233$ respectively. For the real QP hopping amplitudes we find the zero energy density of states diverges, which is cut off by the finite KPM expansion order $N_C$, here we take $N_C=2^{16}$.
	}
	\label{Fig1}
\end{figure}

With this in mind, a formidable task is to classify various universality classes within the family of ``magic-angle'' transitions. As a first step,
an interesting open question
is how does such an EPT depend
on the symmetries of the model. Indeed, it is now well understood that symmetries \cite{AltlandZirnbauer1997,Evers2008_RMP} dictate the universality class of conventional Anderson localization transitions with disorder.
Moreover, as we mentioned previously, symmetries can give rise to dramatic effects in random systems, for example the diverging DOS at low energies in the chiral symmetry classes \cite{Dyson1953,Gade1991,Gade1993}.
However, it is currently unclear what role symmetry plays in the quasiperiodic (QP) semimetal to metal transition and even at quasiperiodicity driven localization transitions in general \cite{Devakul2017}.
Does the diverging DOS also exist in the QP models with chiral symmetry?
The semimetallic model that we investigate in this paper is an ideal setting to numerically investigate this question because in the homogeneous limit the DOS is zero.
This allows for any potential divergence to show up clearly and not be hidden or obscured by the band structure. Finally, a comprehensive understanding of EPTs requires disentangling the effects of strong randomness, such as rare regions, and the effects of symmetry, which is usually a non-trivial task.
The comparison of QP and random systems with the same symmetries is a natural way to study this: Quasiperiodic systems do not possess rare regions due to the lack of large scale statistical fluctuations.

In this work, we show two-dimensional semimetals with Dirac points and QP hopping have (1) a ``magic-angle'' semimetal to metal phase transition \cite{FuPixley2018} at weak quasiperiodicity and (2) a diverging DOS similar to the random case induced by chiral symmetry at the strongest quasiperiodic strength.
This model is constructed and solved using a combination of numerical and analytic techniques. We show that at the semimetal to metal phase transition the DOS jumps, discontinuously to within our resolution, and develops a sharp non-analyticity. This generalizes the transition driven by a QP potential  to the chiral symmetry class.
This transition is accompanied by even flatter bands that what was seen in Ref. \cite{FuPixley2018}, a phenomenon even seen in the so-called chiral model of twisted bilayer graphene \cite{Tarnopolsky-2019}. Therefore, this model provides a route towards significantly increased correlations due to the quenching of kinetic energy, and we will see that in the strong QP limit, other mechanisms could additionally lead to strong correlations. A schematic phase diagram of the model is shown in Fig.~\ref{Fig1}(a), demonstrating the existence of a semimetal phase with Dirac cones in the band structure, a ``chiral metal'' phase with non-trivial real space structure in the wavefunctions, as well as the pure QP hopping limit, which is critical exhibiting sub-diffusive dynamics.

In addition to the phases described above, we generalize the multifractal analysis of real space wavefunctions to momentum space and demonstrate that the semimetal to metal phase transition can be described by a unique kind of an ``unfreezing'' transition \cite{FuPixley2018}. Due to the chiral symmetry in the model we are able to qualitatively describe the metallic phase as the formation of a band of topological zero modes.
As a complementary analysis, we use wavepacket dynamics across the phase diagram to determine (non-energy resolved) transport properties and find crossovers from ballistic, to super-diffusive, and lastly to sub-diffusive dynamics. Importantly, even for the strongest possible QP hopping strength we find that some delocalized states remain.
In the limit of pure QP hopping we show that  the low-energy DOS diverges in a power law fashion with corresponding eigenstates that exhibit Chalker scaling \cite{Chalker1988,Chalker1990}. This potentially implies that interactions are a relevant perturbation to the pure QP hopping model~\cite{Feigelman2007,Foster2012,Foster2014,BurmistrovMirlin2012}.

The rest of the paper is organized as follows. In Sec.~\ref{sec:model} we introduce the QP hopping model. In Sec.~\ref{sec:obs}, we define the main observables of interest, while in Sec.~\ref{sec:results} we present, in detail, our numerical and analytical results. We discuss
the experimental aspects of realizing the theory in Sec.~\ref{sec:expt} and summarize
our results and the remaining open questions in the conclusion, Sec.~\ref{sec:discussion}. In Appendix~\ref{sec:cqph} we analyze a complex QP hopping model, in Appendix~\ref{app:LowEnergy} we provide detailed derivations of our analytic results, and in Appendix~\ref{App:alpha0} we describe the numerical method we use to extract the multifractal exponent. Lastly, in Appendix~\ref{app:QTI} we analyze commensurate limits of the model that can be described as a higher-order topological insulator.

\section{Model}
\label{sec:model}
The general form of the Hamiltonian that we focus on can be written as
\begin{equation}
H = H_0 + H_{\mathrm{QP}},
\label{eqn:H}
\end{equation}
where $H_0$ denotes a bare, translationally invariant hopping model and $H_{QP}$ is the non-trival part of the model that has the QP structure. The model we consider is on the square lattice and the bare hopping model is given by
\begin{equation}
H_0 = \sum_{\vex{r},\mu=x,y}iJ_{0}\psi^{\dagger}_{\vex{r}+\hat\mu}\sigma_{\mu}\psi_{\vex{r}}+\text{H.c.},
\label{eqn:H0}
\end{equation}
where $J_{0}$ is the bare hopping amplitude between site $\vex{r}$ and $\vex{r}+\hat\mu$,
$\sigma_{x,y}$ are the Pauli matrices, and $\psi_{\vex{r}}$ is a two-component spinor of annihilation operators. The dispersion relation for $H_0$ is $E_0(\vex{k}) = \pm 2 J_0 \sqrt{\sin k_x^2 + \sin k_y^2}$, which contains four Dirac points at $(0,0), (0,\pi), (\pi,0),$ and $(\pi,\pi)$, and a low-energy DOS $\rho(E) \sim |E|$. Thus, this spinful model on the square lattice describes a two dimensional semimetal with linearly dispersing excitations.
This model naturally captures the universal low-energy physics of two-dimensional semimetals and is convenient for performing both analytical as well as numerical calculations.
It is important to realize that, on the single-particle level,
the model in
Eq.~\eqref{eqn:H0}  describes the direct sum of two $\pi$-flux models~\cite{FuPixley2018} which are readily implemented using shaken optical lattices~\cite{Aidelsburger-2015}. And indeed, much of our analysis and conclusions apply equally well for a single copy of pi-flux.

\subsection{Quasiperiodic perturbation}

The QP  part of the Hamiltonian on the square lattice is given by
\begin{align}
H_{QP}=\sum_{\vex{r},\mu=x,y}iJ_{\mu}(\vex{r})\psi^{\dagger}_{\vex{r}+\hat\mu}\sigma_{\mu}\psi_{\vex{r}}+\text{H.c.},
\label{eqn:HQPH}
\end{align}
where $J_{\mu}(\vex{r})$ is the QP hopping amplitude between site $\vex{r}$ and $\vex{r}+\hat\mu$.
We construct the hopping matrix elements by considering a two-dimensional surface [e.g. $\cos(Q x) + \cos(Q y)$] with a quasiperiodic wavevector $Q$   (i.e. incommensurate with the underlying lattice) that we evaluate at the mid-point of each bond on the lattice, this yields
\begin{equation}
J_{\mu}(\vex{r}) =W\sum_{\nu=x,y}\cos \left[Q \left(r_{\nu}+\hat{\mu}\cdot\hat{\nu}/2\right)+\phi_{\nu}\right],
\label{eqn:hop}
\end{equation}
where $Q$ is an incommensurate wavevector, $\phi_x$ and $\phi_y$ are random phases sampled uniformly between $[0,2\pi]$ that are the same at each site, and we have set the lattice spacing to unity. We take the linear system size to be given by a Fibonacci number $L=F_n$ and take a rational approximate for the QP wave vector $Q=Q_{L}\equiv2\pi F_{n-2}/L$ (unless otherwise stated) such that as $n\rightarrow \infty$, $Q/2\pi \rightarrow 4/(\sqrt{5}+1)^2$.

In order to reach the pure QP hopping model with finite model parameters we find it convenient to parameterize the bare hopping to be given by
\begin{equation}
J_0 = \sqrt{1-W^2},
\label{eqn:J0}
\end{equation}
such that at $W=0$, $H\rightarrow H_0$ and for $W=1$, $H\rightarrow H_{QP}$. To test for the possibility of a divergence in the low-energy DOS it is ideal to start from a semimetal model where we know \textit{a priori} there is (strictly speaking) zero DOS in the bare model, thus any potential finite or divergent DOS we find is strictly due to the QP hopping.

\subsection{Commensurate limit and higher order topological insulator phases}

In a commensurate limit, the model in Eq.~\eqref{eqn:H} can realize a higher order topological phase. Higher order topological insulators have a gapped topological bulk as well as a gapped topological surface. This induces corner modes in two-dimensions and hinge modes in three-dimensions~\cite{benalcazar2017HOTI}. In particular, in the present model for $Q = \pi n/2$ for $n=1,3$ ($n=2$) the hopping is commensurate with a sixteen (four) site unit cell and  perfect nesting induces a gap at the Dirac nodes. As a result, the model  realizes a higher-order topological insulator phase for a sufficiently strong $W$, which we describe in more detail in Appendix~\ref{app:QTI}. We will sketch the results in this subsection.

As a concrete example, for $Q=\pi$ we analytically show that the model we consider is a quadrupole topological insulator~\cite{benalcazar2017HOTI} (QTI). The hopping for $Q=\pi$ induces a two-sublattice unit cell. The Bloch Hamiltonian is then $h(\vex{k}) = W(\cos(k_x)\tau_x\sigma_0 -\sin(k_x)\tau_y\sigma_x - \cos(k_y)\tau_y\sigma_y-\sin(k_y)\tau_y\sigma_x) +E_0(\vex{k})\tau_z\sigma_0$, where $\sigma, \tau$ are Pauli matrices parametrizing an effective 4-dimensional Hilbert space, see Appendix~\ref{app:QTI}.
Interestingly, this Bloch Hamiltonian is equivalent to the QTI model in Ref.~\cite{benalcazar2017HOTI} without intracell coupling for $W>0$, and as we demonstrate in Appendix~\ref{app:QTI} this phase has topological corner modes at zero energy that lie within the surface and bulk band gap.

In Sec.~\ref{subsec:dos} and Appendix~\ref{app:QTI} we show
similar HOTI behavior also show up when $Q=2\pi m/n$, where $n$ is an even factor of $L$, and $\gcd (m,n) = 1$. These can be interpreted by considering a unit cell of $n^2$ sites. For larger $n$, there are fewer unit cells in our finite size calculation, making the HOTI character more challenging to observe.
Interestingly, in a similar vein, recent work on twisted bilayer graphene predicts the existence of HOTI with large twist angles \cite{Park2019}.
It is interesting to note that the quasiperiodic model we investigate here can be regarded as tuning away from a higher-order topological phase via an incommensurate flux.

\section{Observables}
\label{sec:obs}
We solve the Hamiltonian in Eq.~\eqref{eqn:H} using a combination of numerically exact methods. To compute the DOS and wave packet dynamics we use the Chebyshev expansion techniques including the kernel polynomial method (KPM)~\cite{Weisse-2006,Fehske-2007}, which allows us to reach sufficiently large system sizes ($L=987$ is the largest system size considered here). In addition, we obtain wavefunctions via Lanczos or exact diagonalization.
In this section, we define various observables that are used in this work.

\subsection{The structure of eigenvalues}

To study the transition out of the semimetal phase and the effect of strong QP hopping, we compute the average density of states (DOS), which is defined as
\begin{equation}
\rho(E) = \frac{1}{L^2}\left [ \sum_i \delta(E-E_i) \right]
\label{eqn:dos}
\end{equation}
where $[ \dots ]$ denotes an average  over random phases and twists. The KPM expands the DOS in terms of Chebyshev polynomials up to an order $N_C$, and as a result any non-analytic behavior in the DOS will be rounded by the finite expansion order (in addition to the finite system size).
For the DOS calculations we use twisted boundary conditions, e.g. a phase $e^{i \theta_\mu}$ along the $\mu$ direction, which we incorporate by multiplying each hopping element $J_0 + J_\mu(\v r) \rightarrow e^{i\theta_\mu/L}[J_0 + J_\mu(\v r)]$ in Eqs.~\eqref{eqn:H0} and \eqref{eqn:HQPH}. We average over random twists and phases sampled uniformly between $[0,2\pi]$; for the KPM data we average over 500 samples.
In certain regimes of the model we  use the power law scaling of the low-energy DOS
\begin{equation}
\rho(E) \sim |E|^{d/z-1}
\label{eqn:zdos}
\end{equation}
to extract the dynamic exponent $z$.
The finite KPM expansion order leads to a broadening of the Dirac delta functions in the definition of the DOS [see Eq.~\eqref{eqn:dos}] into Gaussians with a width $\delta E = \pi D /N_C$ for a bandwidth $D$ (this holds for the Jackson kernel~\cite{Weisse-2006} that we are using for all of the calculations presented here). Thus, we also use the scaling of $\rho(E=0)$ with $N_C$, where Eq.~\eqref{eqn:zdos} implies that $\rho(E=0) \sim (N_C)^{1-d/z}$, to analyze the scaling of the low-energy density of states.

To study the real-space localization properties of the model we study the typical DOS, which is the geometric mean of the local DOS. This is defined as
\begin{eqnarray}
\rho_{\mathrm{typ}}(E) &=& \exp \left( \frac{1}{N_s}\left[ \sum_i^{N_s} \log \rho_i(E) \right] \right)
\label{eqn:tdos}
\end{eqnarray}
and the local DOS is given by
\begin{eqnarray}
\rho_i (E) &=& \sum_{n,\alpha} |\langle  n| i,\alpha  \rangle |^2\delta(E-E_n),
\end{eqnarray}
where $|n \rangle$ and $E_n$ denote exact eigenstates and eigenenergies, $\alpha$ denotes the two spin states due to the spinor structure of the Hamiltonian, and $N_s \ll L^2$ is a small number of randomly chosen sites that we average over to improve the statistics. In the thermodynamic limit, the typical density of states is non-zero in the extended phase and will go to zero in an Anderson insulating phase, which thus serves as a diagnostic for real-space localization.

\subsection{The structure of eigenstates}\label{Sec:eigenstates}

We connect the physical properties of the model to its low-energy eigenstates by studying their structure in both real and momentum space. The semimetal phase is characterized by stable plane-wave states that are localized in momentum space. As shown in Refs.~\cite{Pixley2018,FuPixley2018}, a unique feature of the ``magic-angle'' semimetal to metal transition is that it coincides with a delocalization of the momentum-space wavefunctions. This implies that the critical momentum-space wavefunctions are developing non-trivial structure that we should be able to describe using methods to treat localization transitions in real space.

The properties of the probability distribution of an eigenstate can be characterized by a multifractal analysis \cite{Huckestein1995,Evers2008_RMP}.
We first define a ``coarse grained'' real-space wavefunction ($\psi_b$) with its resolution controlled by a binning size $b\ge 1$.
The spatial region is divided into $(L/b)\times(L/b)$ boxes. We assign a position vector $\vex{X}_j$ to indicate the position of the $j$th box. The binned wavefunction is given by $\psi_b(\vex{X}_j)\equiv\sum_{\vex{x}}'\psi(\vex{x})$ where $\psi$ is the original normalized wavefunction, and $\sum_{\vex{x}}'$ runs over the positions inside the $j$th box.
Then,
we define the real-space (generalized) inverse participation ratio (IPR) and multifractal exponent via
\begin{align}\label{Eq:Pr_tauq}
\mathcal{I}_R(E;q,b,L)=\sum_{\vex{X}_j}|\psi_b(E,\vex{X}_j)|^{2q}\propto\left(\frac{b}{L}\right)^{\tau_R(q)},
\end{align}
where $\mathcal{I}_R(E,q,b,L)$ is the $q$th real-space IPR with a binning size $b$, $E$ is the energy of the wavefunction, and we use a subscript $R$ to denote real space.
Note that the sum in Eq.~(\ref{Eq:Pr_tauq}) is running over the positions of boxes ($\vex{X}_j's$) rather than the full lattice points. The quantity $\tau_R(q)$ is  the multifractal exponent associated with the $q$th IPR in real space, and
$b=1$ is the finest resolution in the IPR measure. The exponent $\tau_R(q)$ is extracted via varying values of $b$ for $b\ll L$.
To obtain $\tau_R(q)$ in the finite-size system, we vary the binning size $b$ for a given $L$.
The exponent $\tau_R(q)$ is known to be a self-averaging quantity in the studies of disordered free-fermion models \cite{Chamon1996}.
In addition, $\tau_R(q=0)=-d=-2$ (the trivial limit which corresponds to counting binning boxes) and $\tau_R(q=1)=0$ (normalization of the wavefunction) must hold for arbitrary wavefunctions.
Conventionally, one sets $b=1$ and $q=2$ for studying the second IPR as a proxy of spatial ergodicity/non-ergodicity in a wavefunction.

We now generalize the multifractal analysis to momentum-space wavefunctions and focus on the Dirac node energy $E=0$ and therefore drop the energy label. Similar to our work in Ref.~\cite{FuPixley2018}, we Fourier transform the zero energy wavefunction from real to momentum space $\phi(\vex{k}) = (1/L) \sum_{\vex{x}}e^{-i \vex{x}\cdot\vex{k}}\psi(E=0,\vex{x})$. Then, we set up momentum-space boxes of size $B$ and the binned wavefunction ($\phi_B$) in momentum space. We note that the box size $B$ in the momentum space determines the effective infrared scale while $b$ in real space is related to the effective ultraviolet scale.
The momentum-space IPR and multifractal exponent are given by
\begin{align}\label{Eq:Pm_tauq}
\mathcal{I}_M(q,B,N)=\sum_{\vex{K}_j}|\phi_{B}(\vex{K}_j)|^{2q}\propto\left(\frac{B}{N}\right)^{\tau_M(q)},
\end{align}
where $\mathcal{I}_M(q,B,N)$ is the $q$th momentum-space IPR with a momentum binning size $B$, a linear size of the momentum grid $N=L$, and we use a subscript $M$ to denote momentum space. Using this definition we can study localization transitions in momentum space by either fixing $q=2$ (Ref.~\cite{Pixley2018}) or in more detail by analyzing the behavior of the multifractal exponent $\tau_M(q)$ (Ref.~\cite{FuPixley2018}). $\tau_M(q)$ also obeys the conditions $\tau_M(q=0)=-d=-2$ and $\tau_M(q=1)=0$.

The multifractal exponents $\tau_{R}(q)$ and $\tau_M(q)$ provide systematic ways of characterizing the properties of the wavefunction probability distributions in the in the real- and momentum-space bases respectively. For a plane wave in real space, the spectrum is simply $\tau_R(q)=2(q-1)$, i.e. a straight line. The corresponding momentum-space wavefunction generically contains a few of sharp peaks (due to a linear combination of the degenerate eigenstates) and is characterized by $\tau_M(q)=0$ for $q\ge q_c$ where the termination value $q_c\ge 1$, indicates a ``frozen'' spectrum \cite{Evers2008_RMP}. In the limit of a single peak, the spectrum is reduced to a localization spectrum with $q_c\rightarrow 0$.
We will focus on an ``unfreezing'' transition in $\tau_M(q)$ which is related to the semimetal-metal transition.
In addition, we adopt a variant of the real-space multifractal exponent $\alpha_0$ (see Appendix~\ref{App:alpha0}) for characterizing the localization properties for finite-energy wavefunctions in the strong QP hopping limit.

Lastly, we test for Chalker scaling by defining a two-wavefunction correlation function as follows \cite{Chalker1988,Chalker1990,Cuevas2007,Chou2014}:
\begin{align}\label{Eq:C_E}
C(E)\equiv\sum_{\vex{x}}|\psi_{E_0}(\vex{x})|^2|\psi_{E}(\vex{x})|^2,
\end{align}
where $E_0$ is a reference energy and $\psi_{E}$ is the eigenstate with energy $E$. Note that the sum runs over all the positions and the internal degrees of freedom have been integrated over.
We are interested in energies near the Dirac node so we set $E_0=0$. The two-wavefunction correlation $C(E)$ characterizes the degree of overlapping probability  among two eigenstates separated by an energy $E$ in a fixed realization. In particular, $C(E)\sim 0$ for localized states with $0\le E \ll \delta_l$ ($\delta_l$ is the mean level spacing in a localization volume). For states near a mobility edge, $C(E)$ shows nontrivial scaling in the energy separation \cite{Chalker1988,Chalker1990,Fyodorov1997,Cuevas2007}.
States that obey a power law scaling
\begin{equation}
C(E)\sim |E|^{-\mu}
\end{equation}
with $\mu=[d-\tau_R(2)]/z>0$ exhibit Chalker scaling. (Note that the exponent $\mu$ here has been generalized to the system with a low-energy power law DOS \cite{Chou2014}.)
The existence of the power-law scaling potentially implies an enhancement of interactions \cite{Feigelman2007,Foster2012,BurmistrovMirlin2012,Foster2014}.
We adopt such a diagnostic to study the correlations among the low-energy states in the pure QP hopping limit.

\subsection{Dynamics}

We study transport properties of the model via wavepacket dynamics. We initialize a wave packet to be localized at a single site (${\bf r}_0=(0,0)$) in real space $\Psi_0({\bf r}) = \langle {\bf r} |\Psi_0 \rangle = \delta_{{\bf r}_0,{\bf r}}$ with zero initial velocity (in this case, a spin up/down state suffices), then time evolve that state $|\Psi(t) \rangle = e^{-iHt}|\Psi_0\rangle$, which we evaluate using a Chebyshev expansion~\cite{Fehske-2007}.  We compute the spread of the wavepacket
\begin{equation}
\langle \delta r(t)^2 \rangle \equiv \langle \Psi(t)| [ \hat {\bf r}-{\bf r}_0]^2| \Psi(t) \rangle
\label{eqn:r2}
\end{equation}
 where $ \hat {\bf r} = (\hat x,\hat y) = \sum_{{\bf r}}  ( x,y)| {\bf r} \rangle \langle {\bf r} |$ and ${\bf r} = (x,y)$.
The initialized wave packet has weight across the spectrum of eigenstates and is not energy resolved. Therefore it will not be particularly sensitive to the semimetal to metal phase transition at $E=0$. As a result any estimate we make will be averaged over all energy eigenstates. With this in mind, we use the scaling of wavepacket spreading at long times
\begin{equation}
\langle \delta r(t)^2 \rangle \sim t^{2/\tilde z}
\label{eqn:zwp}
\end{equation}
to extract an ``average'' estimate of of the dynamic exponent $\tilde z$ (and hence use a tilde) to distinguish this from our energy resolved DOS estimate of $z$ in Eq.~\eqref{eqn:zdos}.
We note here that the Chebyshev expansion order $N_C$ does not lead to a broadening of levels; it instead dictates the final time that can be reached accurately. Here we track this by requiring the norm of the wavefunction be preserved for all times. In all the results presented here we choose $N_C$ such that the wavepacket has enough time to spread out as far as possible ($= L/2$ in each direction due to periodic boundary conditions) so that the only finite-size effect in our data is due to system size and not $N_C$.

\section{Results}
\label{sec:results}
While we study all energies and quasiperiodicity strengths, our principle consideration is the Dirac node energy $(E=0)$.
At weak quasiperiodicity, we study the development of a non-zero DOS at the Dirac node, which coincides with a delocalization of the wavefunction in momentum space \cite{Pixley2018,FuPixley2018}.
At strong quasiperiodicity, we study the evolution of the low-energy eigenstates and wavepacket dynamics that contribute to a clear divergence in the low-energy DOS in the limit of pure QP hopping ($W=1$).

\subsection{Transition out of the semimetal phase}

\begin{figure}[t!]
	\centering
	\includegraphics[width=0.49\textwidth]{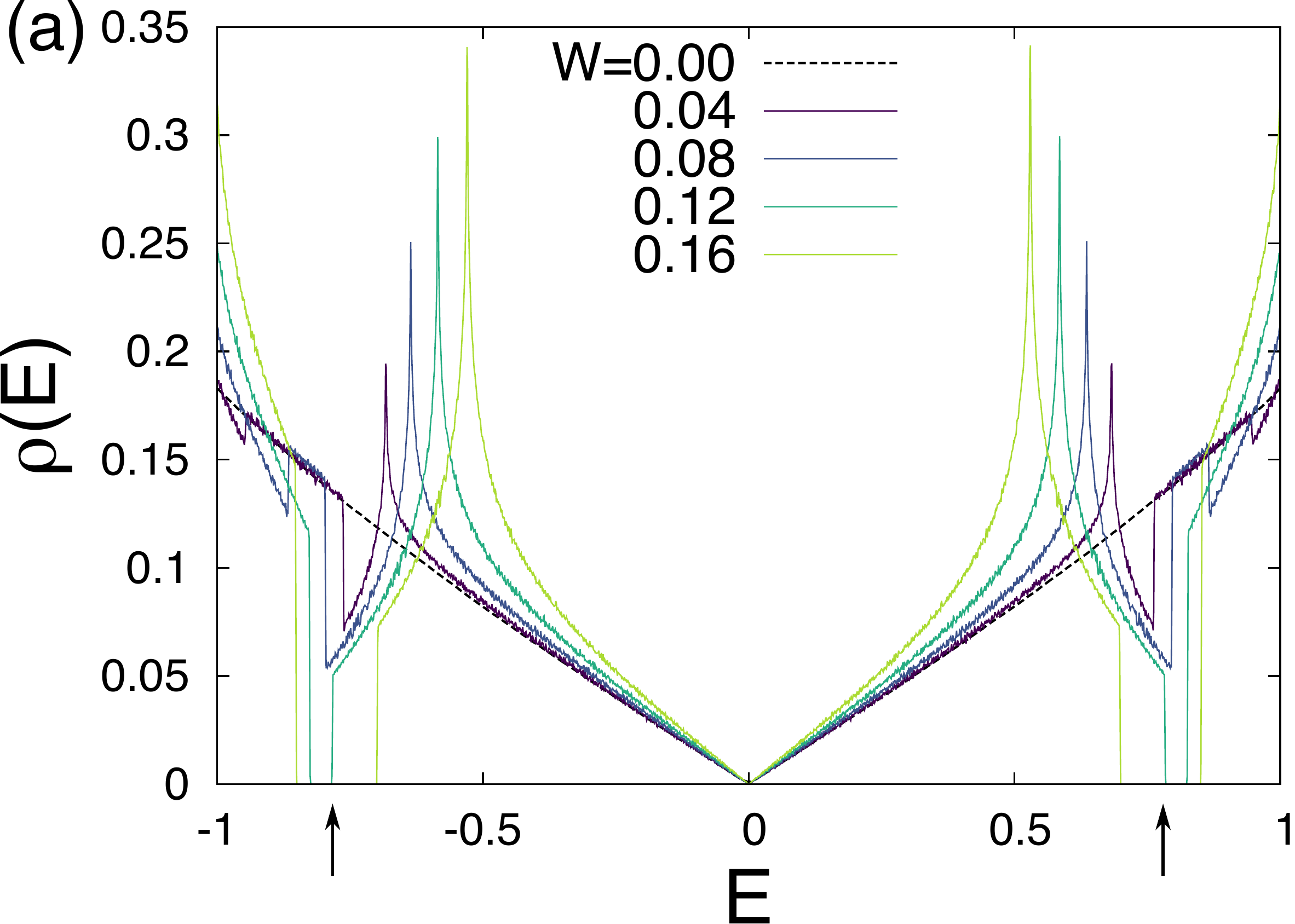}
	\includegraphics[width=0.49\textwidth]{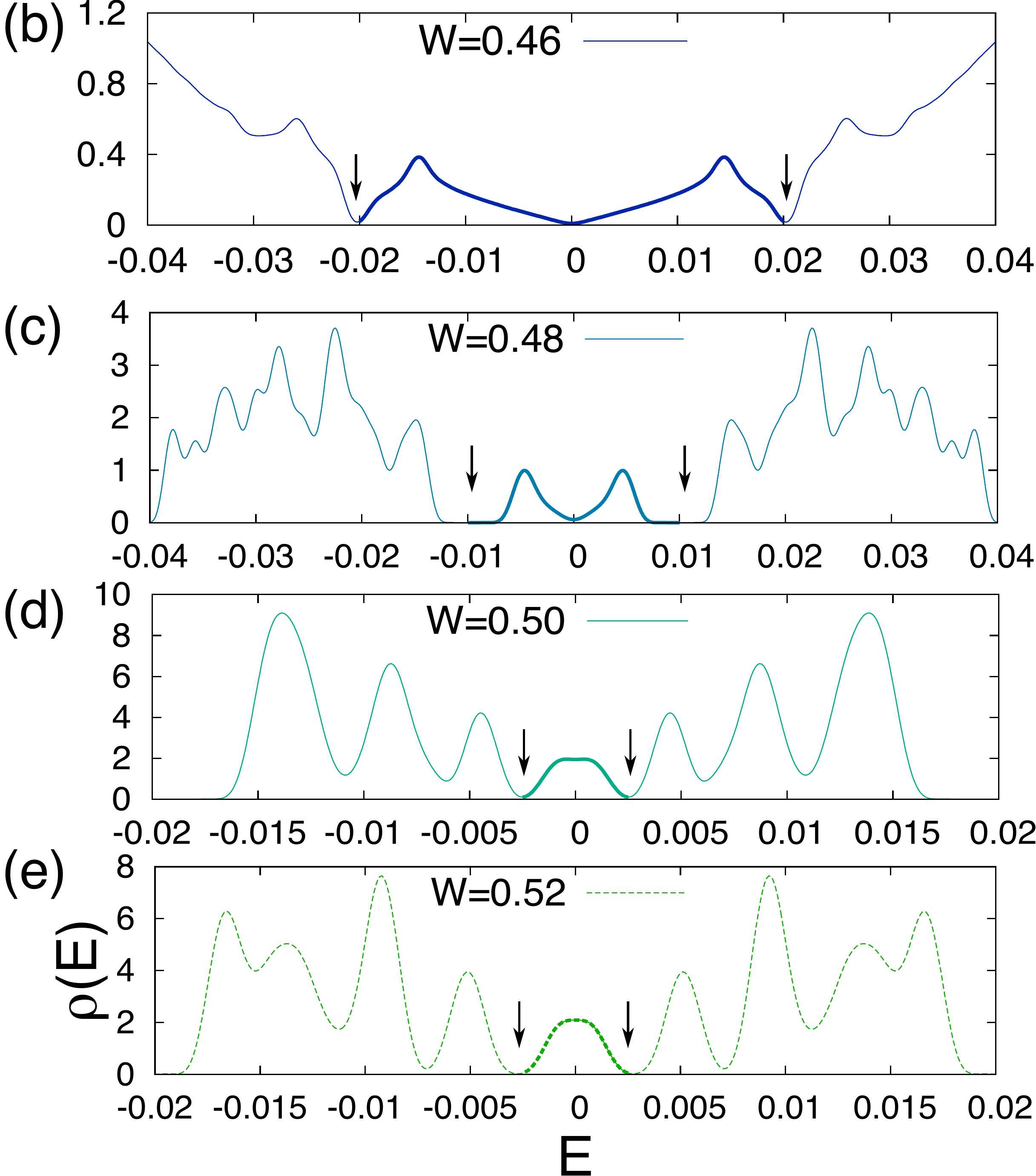}
	\caption{DOS versus $E$ for $L=233$ and $N_C=2^{14}$ with different QP hopping strengths $W$. (a)  Formation of the first miniband with increasing values of $W$ (vertical black arrows marking the gap that separates this miniband from the rest of the states). (b)-(e) Formation of the second miniband and semimetal to metal transition (vertical black arrows mark the location of the gap to the second miniband). The second miniband is displayed as a thicker line for clarity. Note that the full bandwidth  for $W=0$ is $4\sqrt{2}\approx 5.7$ and all of these results are obtained for $Q=2\pi\times 89/233$ with a critical value of $W$ for this $Q$ given by $W_c = 0.485 \pm 0.005$.
	}
	\label{Fig:DOS0}
\end{figure}

\subsubsection{Formation of the Miniband(s)}

Introducing
a weak QP hopping with $Q$ close to $\pi$,
creates dominant
internode scattering that transfers momentum $Q_L$ and mixes degenerate states of equivalent spin. This leads to the formation of hard gaps at finite energy that separates a semimetal miniband near $E=0$ described by a DOS $\rho(E) \approx \rho'(0) |E|$ with the rest of the spectrum. We note that this defines the slope $\rho'(0)$ and formally we only focus on $\rho'(0^+)$.
As $W$ increases, higher-order processes
gain importance, hybridize with lower-energy eigenstates, and, therefore, open additional smaller mini bands, see Fig.~\ref{Fig:DOS0}.  Similar to what was reported in Refs.~\cite{Pixley2018,FuPixley2018} for semimetals in a QP potential, these minibands can be described perturbatively in the QP strength, and the states in the miniband can be counted by considering the number of states near the Dirac cones that cannot be mixed via a momentum transfer that is restricted to a size $Q_L$ (or smaller for higher order perturbative processes). For $Q_L = 2\pi F_{n-2}/F_n$ we find that there $N_1 = 2 (F_{n - 3})^2$ states in the first miniband  and $N_4 =  2 (F_{n - 6})^2$ states in the second miniband, which are generated by  a momentum transfer of $Q_L$ (from first order in perturbation theory)  and $4Q_L - 3\pi$ (from fourth order in perturbation), respectively. This matches our numerical results,  which we compute using either exact diagonalization on small sizes or integrating the DOS over the energy window of the miniband.
The formation of the first and second miniband is shown in Fig.~\ref{Fig:DOS0} for a potential strength $W \approx 0.1$  and $W \approx 0.48$ respectively. 
The van Hove peaks in each miniband are conventional and we have checked that they diverge logarithmically in the thermodynamic limit (not shown).
Interestingly, this is a similar result to what was found in Refs.~\cite{Pixley2018,FuPixley2018}, thus the development of minibands at weak QP hopping is not distinct from those generated by a QP potential or from ``twisting'' two layers of graphene.

\begin{figure}
    \includegraphics[width=.9\columnwidth]{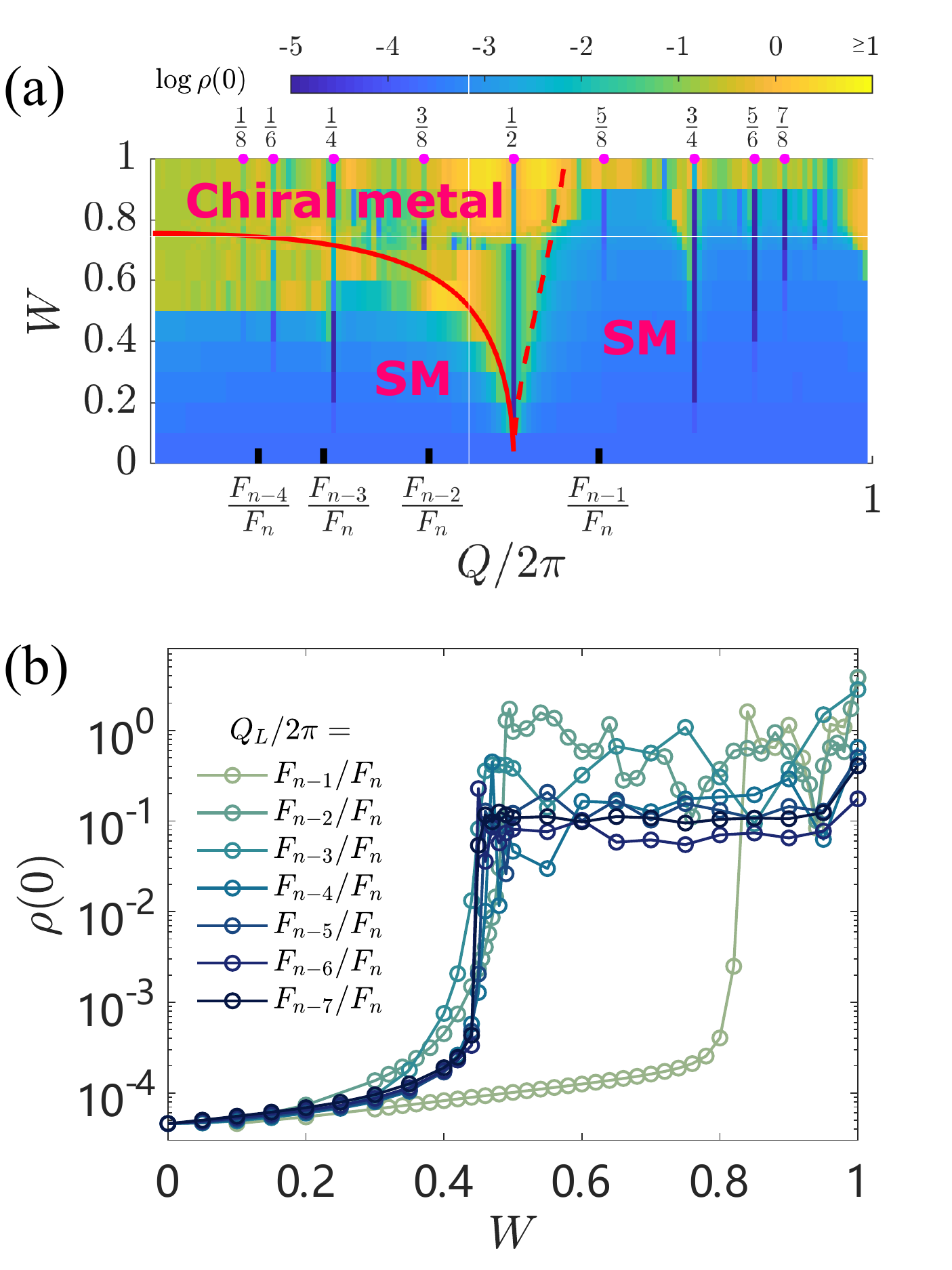}
\caption{
    The dependence of the DOS at zero energy on the choice of the wavevector $Q_L$. (a) A phase diagram in the space of $W$ and $Q$ specifying the semimetallic regime (SM), the gapped higher order topological insulating phases  (indicated by the sharp drops in DOS on vertical lines indicating rational Q labeled on top), and the chiral metal phase, where the color plot denotes the value of $\log \rho(0)$. Each data point is calculated for a  system size $L=144$ and KPM expansion order of $N_C=2^{12}$. For these finite sizes, $\rho(0)$ around $10^{-3}$ corresponds to the SM phase, while larger DOS signals the metallic phase.
    At $Q=n\pi/2$ and other highly commensurate ratios with even denominators, however,
    the model is gapped (as indicated by the sharp drop in DOS) and is a higher order topological insulator
    as shown in more detail in Appendix~\ref{app:QTI}.
    The solid red curve shows the result of perturbation theory for the critical $W_c$, given by $v=0$ in Eq.~\eqref{eqn:v}. For $Q > \pi$ the estimate of $W_c$ from Eq.~\eqref{eqn:v} becomes imaginary, we plot the magnitude of this as a dashed red curve.
    (b) The $Q_L/(2\pi)=F_{n-m}/F_n$ cuts (marked by the black ticks in top panel) with system sizes $L=144$, and $N_C=2^{14}$. We see the transition persists for very small $Q_L$. Note that the finite value of $\rho(0)$ in the semimetal regime is just a finite-size effect and the transition appears when this rises over several orders of magnitude, see Fig.~\ref{Fig:rho0real}.
}
\label{Fig:varQ}
\end{figure}

If we instead focus on a small $Q_L$ (relative to $\pi$) then internode scattering is no longer the dominant effect and intranode scattering also plays a prominent role in the low-energy description. In this case, the hard gaps can be softened into pseudogaps or smeared out altogether. Nonetheless, we still find a semimetal to metal phase transition persists at small $Q_L$.
For $Q_L \lesssim 2\pi F_{n-3}/F_{n}$ the location of semimetal-to-metal transition is roughly the same, as shown in Fig.~\ref{Fig:varQ}.

\subsubsection{Density of states and velocity renormalization}
\label{subsec:dos}

\begin{figure}[b]
	\centering
	\includegraphics[width=0.33\textwidth,angle=-90]{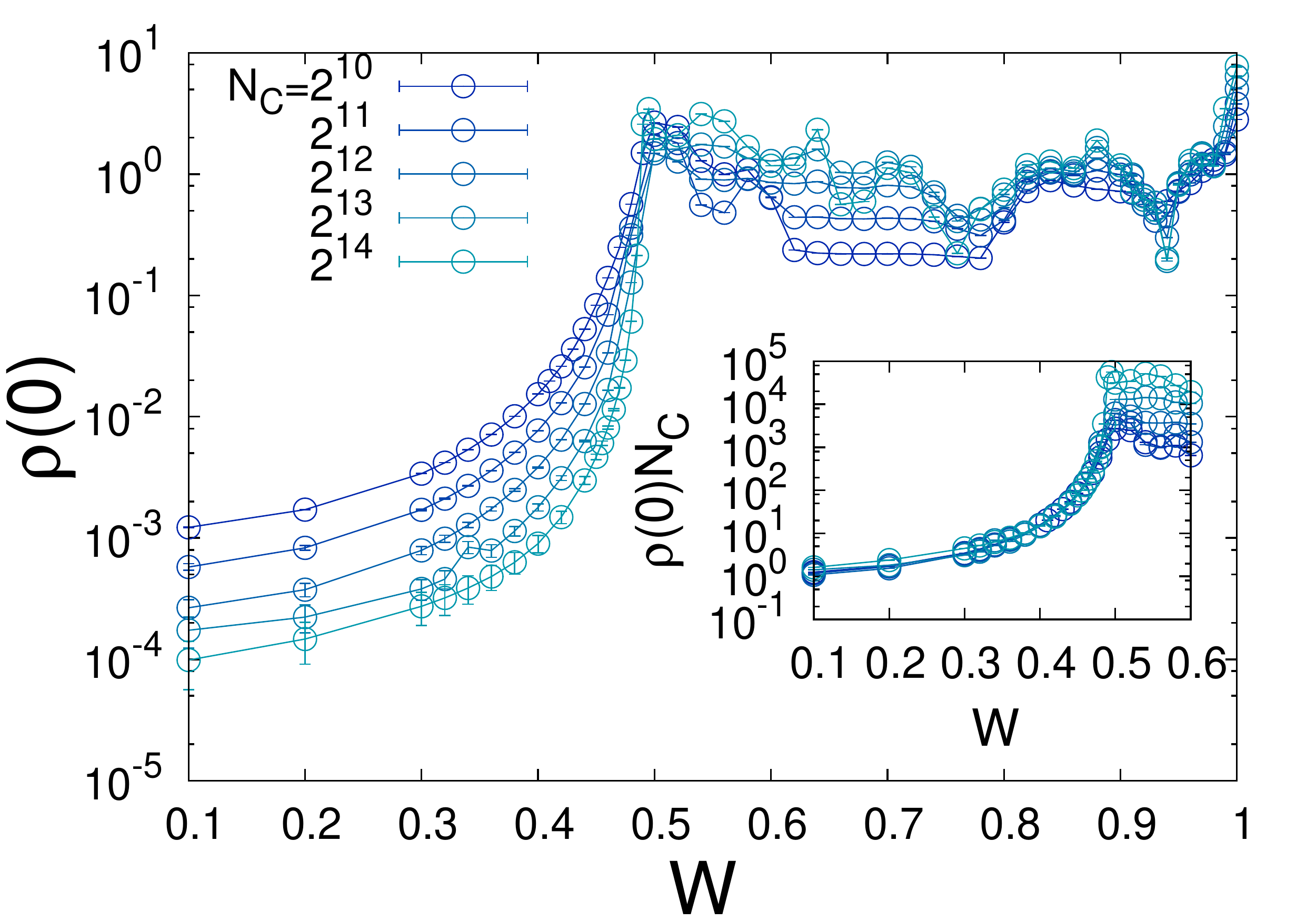}
	\caption{The zero-energy DOS $\rho(0)$ as a function of $W$ for various KPM expansion orders $N_C$ and a system size of $L=233$. In the semimetal regime $\rho(0)$ goes to zero for increasing $N_C$ like $\rho(0) \sim 1/N_C$, which allows us to identify a sharp semimetal to metal transition at $W_c = 0.485 \pm 0.005$. (Inset) The $N_C$ independence of $\rho(0) N_C$  allows us to identify the semimetal phase boundary and demonstrates the robustness of the semimetal phase to quasiperiodicity. This data for $N_C=2^{14}$ on a linear scale is shown in Fig.~\ref{Fig1}(b). }
	\label{Fig:rho0real}
\end{figure}

We first focus on the low-energy DOS at weak QP hopping strength. The semimetal is defined as having zero DOS at $E=0$, and we find this is stable over a finite range of $W$ (as shown in Figs.~\ref{Fig1}, \ref{Fig:varQ}, and \ref{Fig:rho0real}). This can be seen clearly from the scaling of the zero-energy DOS with the KPM expansion order; in the semimetal regime $\rho(E)\sim |E|$ implies that $\rho(E=0) \sim 1/N_C$ (see inset of Fig.~\ref{Fig:rho0real}) and we use this to locate the boundary of the semimetal phase. Note that this is completely different then the random model, where DOS is always non-zero due to the perturbative (marginal) relevance of disorder in two-dimensions~\cite{Abrahams-1979,Aleiner-2006,Altland-2006}.

As the QP hopping is increased the gaps approach $E=0$, which ``flattens'' the semimetal miniband until a non-zero value of the DOS is generated after a critical QP hopping strength. For $Q_L=2\pi F_{n-2}/F_n$ with $L=F_n$ we find that this occurs at $W_c=0.485 \pm 0.005$ by studying the $N_C$ dependence as shown in Fig.~\ref{Fig:rho0real}.
After the transition we find a low-energy peak centered about $E=0$ survives (which eventually develops structure at larger QP hopping strength), see Fig.~\ref{Fig:DOS0}.
We find that all of the states that make up the second (smaller) miniband  $=2(F_{n-6})^2$ for $Q_L/2\pi = F_{n-2}/L$ and $L=F_n$ in the semimetal phase become mixed in the metallic phase and are all \emph{contained} in the peak about zero energy in Fig.~\ref{Fig:DOS0} for $W=0.50$ and $0.52$.
This behavior only holds for the chiral model and does not necessarily occur for the case of a QP potential~\cite{FuPixley2018}.
The location of the transition $W_c$ is not universal and depends on the model details.

\begin{figure}
\includegraphics[width=.9\columnwidth]{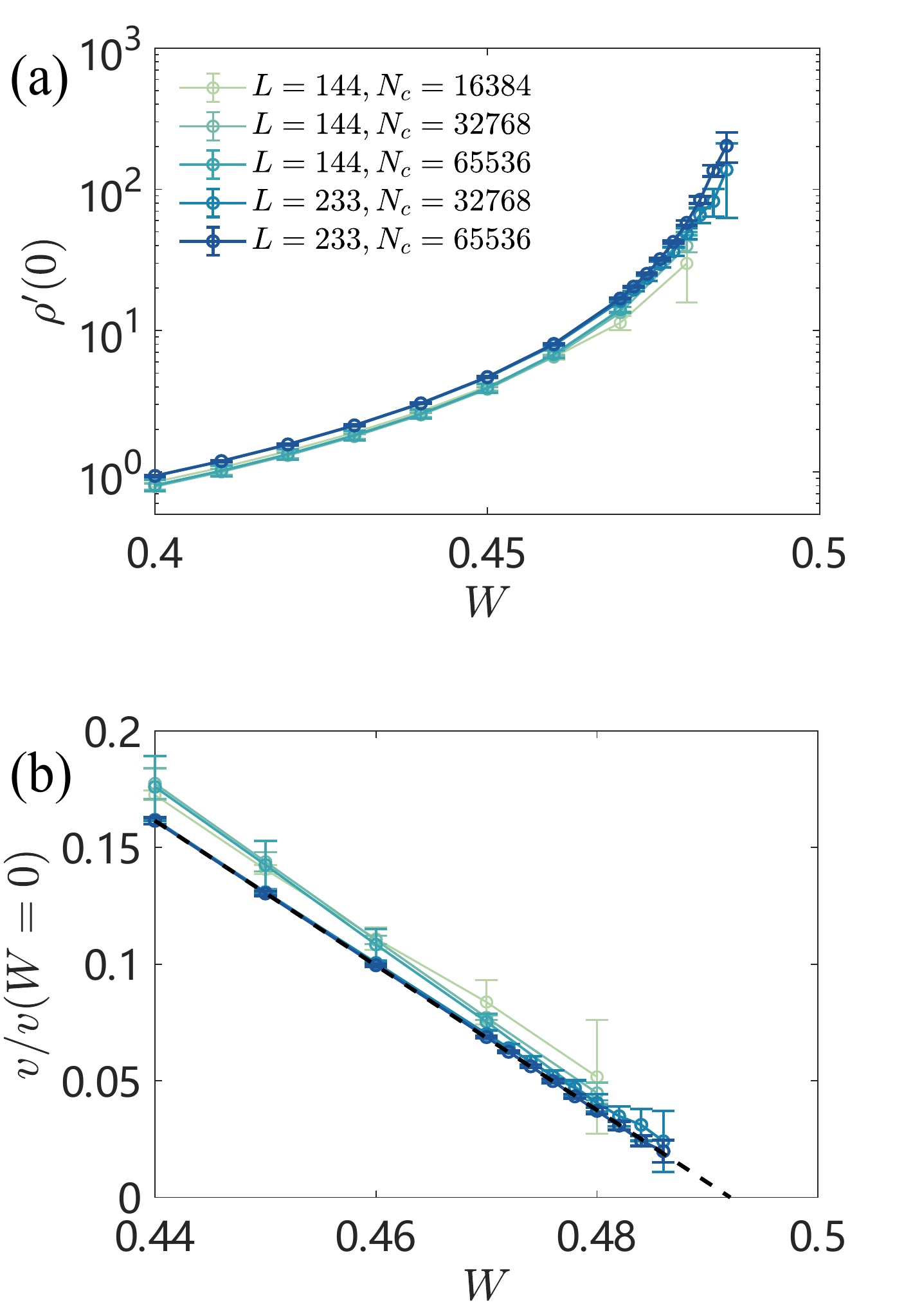}
\caption{The effective Dirac cone velocity extracted from the scaling of the  low-energy DOS $\rho(E) \sim \rho'(0) |E|$ [formally we compute $\rho'(0^+)$]. (a) The slope $ \rho'(0)$ vs $W$ for various combination of $N_c$ and $L$. We find that $\rho'(0)$ rises steeply, strongly suggesting a divergence and a non-analytic DOS at the transition. We extract $\rho'(0)$ from a fit to the scaling of the low-energy DDOS $\rho(E) \sim \rho'(0) |E|$.
    (b) Velocity $v=1/\sqrt{\rho'(0)}$. The dashed line shows the linear fit of highest $N_c$ and $L$ we have. The linear scaling of $\rho'(0)^{-0.5}$ indicates $\rho'(E=0)\sim (W_c-W)^{-2}$, and predicts critical point $W 0.485 \pm 0.005$ that is consistent with our other analysis.}
\label{Fig:v}
\end{figure}

We find that the semimetal miniband is well described by $\rho(E) \approx \rho'(0) |E|$, with no change to the power law in energy as the quantum phase transition is approached.
The Fermi velocity of the Dirac cone $v$ is related to the DOS via $\rho'(0) \propto 1/v^2$. As the transition is approached from the semimetal side we find $\rho'(0)$ diverges like $\rho'(0)\sim (W_c-W)^{-\beta}$, with $\beta=2\pm 0.2$
, see Fig.~\ref{Fig:v}.  This signals that the DOS develops non-analytic behavior at the semimetal-to-metal transition. As a result the velocity of the Dirac cone goes to zero like $v \sim (W_c -W)$.
It is very interesting to compare this result with what we found in Ref.~\cite{FuPixley2018} for the case of a QP potential, which yielded $\beta = 1.8 \pm 0.4$, which suggests (rather remarkably) that this exponent seems to be independent of the symmetry class.

\begin{figure}[t!]
\includegraphics[width=.9\columnwidth]{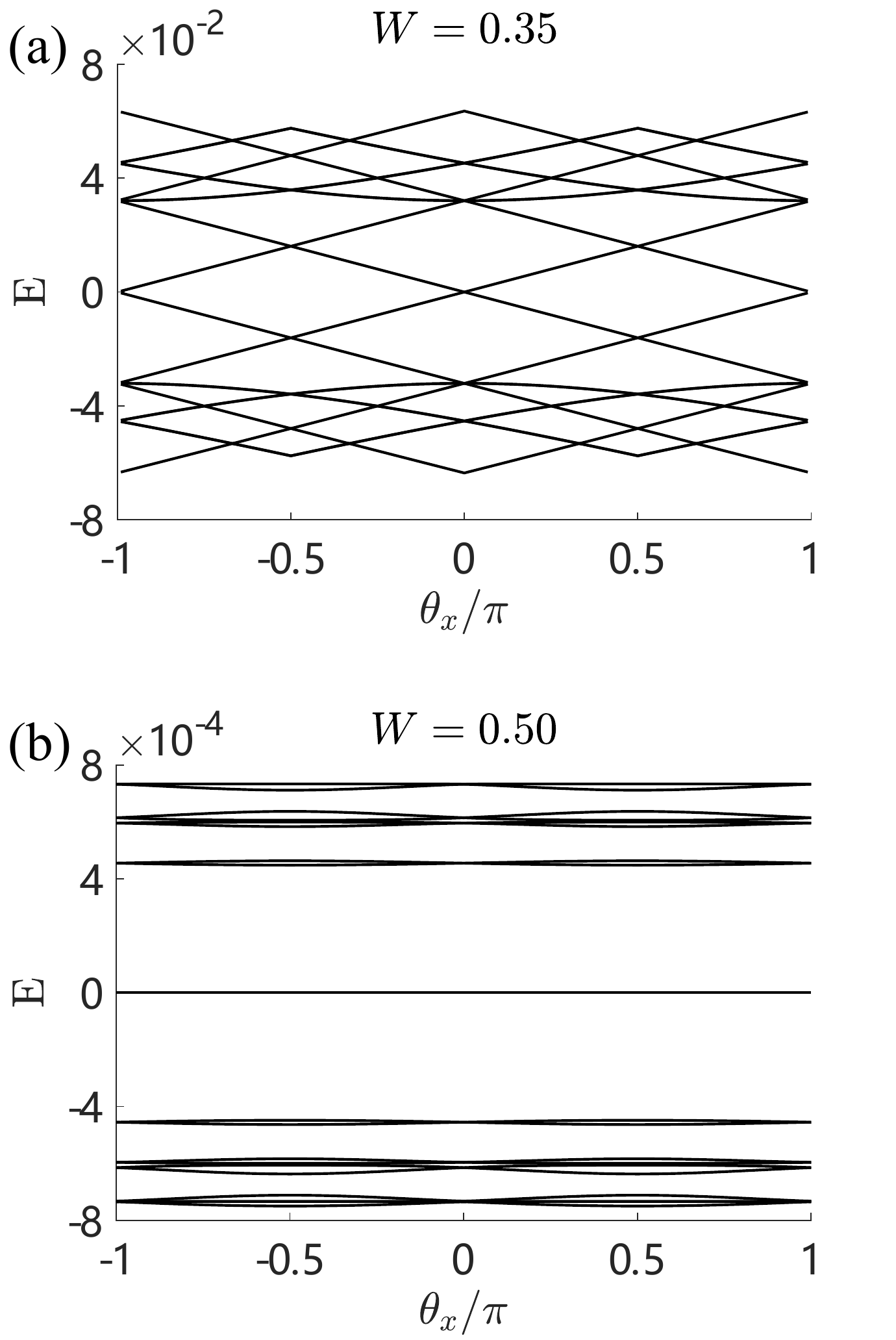}
\caption{The twist dispersion in the semimetal phase (a) and in the chiral metal (b), i.e. low-energy eigenvalues ($E$) as a function of a twist ($\theta_x$) in the boundary condition along the $x$-direction obtained by diagonalizing an $L=89$ sample. (a) For $W=0.35$ in the semimetal phase with clear Dirac points at $(0,0)$ and $(\pi, 0$). (b) Focusing on $W=0.50$  that is right after the semimetal to metal transition. We see the low-energy minibandwidth for $W=0.5$ has been substantially renormalized, the band in the center of the spectrum has a bandwidth that has been renormalized by a factor $ \sim 10^{-8} $ from its unperturbed value, which is an even stronger effect then has been seen previously~\cite{FuPixley2018}.}
\label{Fig:twist}
\end{figure}

The suppression of the velocity for $0<Q<\pi$ can also be captured analytically using perturbation theory in the QP hopping strength, borrowing techniques originally applied to twisted bilayer graphene \cite{BistritzerMacDonald2011,FuPixley2018}. Using this framework and going to second order in the QP hopping strength we find (see Appendix~\ref{App:Velocity})
\begin{align}
\frac{v}{2 J_0} &=\frac{1 - \frac{W^2}{4 J_0^2} [1+ 2\sec(Q/2)]}{1 + \frac{W^2}{4 J_0^2}\sec(Q/2)^2}.
\label{eqn:v}
\end{align}
This yields a vanishing velocity, i.e. a magic-angle condition $v=0$, for $W=W_c^{\rm (v)} \equiv 2/\sqrt{5	+ 2 \sec(Q/2)}$ which we compare to the numerical calculation of the DOS at zero energy in Fig.~\ref{Fig:varQ}(a).
In the regime near $Q = \pi$, where the $W_c$ is small and perturbation theory is controlled, both methods agree well.

These results strongly suggest that the semimetal-to-metal transition generates flat bands due to the vanishing velocity.
To clearly demonstrate the presence of flat bands, we  study how the low-energy eigenvalues evolve as a function of the twist in the boundary condition. 
To twist the boundaries we apply a gauge transformation that is equivalent to replacing the hopping terms $J_0 + J_{\mu}({\bf r}) \rightarrow e^{i\theta_{\mu}/L}[J_0 + J_{\mu}({\bf r})]$ for a twist $\theta_{\mu}$ in the $\mu$ direction. 
We use this as a measure of the low-energy dispersion in the mini (twist) Brillouin zone of size ($2\pi/L$). This is mathematically equivalent of tiling an infinite system with supercells of size $L\times L$ and finding the corresponding band structure (much akin to tiling graphene with moir\'e unit cells).
As shown in Fig.~\ref{Fig:twist}(a), we clearly see the presence of the Dirac cones at $(0,0)$ and $(\pi,0)$ for weak QP hopping. These bands become incredibly flat in the metallic phase, as shown in Fig.~\ref{Fig:twist}(b), which confirms both the qualitative expectation from the perturbative analysis and our approach of extracting the velocity from the scaling of the density of states. The flattening effect is substantial in the chiral model and suppresses the minibandwidth orders of magnitude more from the magic-angle transition driven by a quasiperiodic potential~\cite{FuPixley2018}. Interestingly, incredibly flat bands have also been seen in the so-called chiral model of twisted bilayer graphene~\cite{Tarnopolsky-2019}, and we find a similar effect here in this much simpler model that also possess a chiral symmetry. Thus, we conclude that the particle-hole symmetry leads to a significant enhancement of miniband renormalization effects.

\subsubsection{Wavefunction delocalization in momentum space}

\begin{figure}[b]
	\centering
	\includegraphics[width=0.33\textwidth,angle=-90]{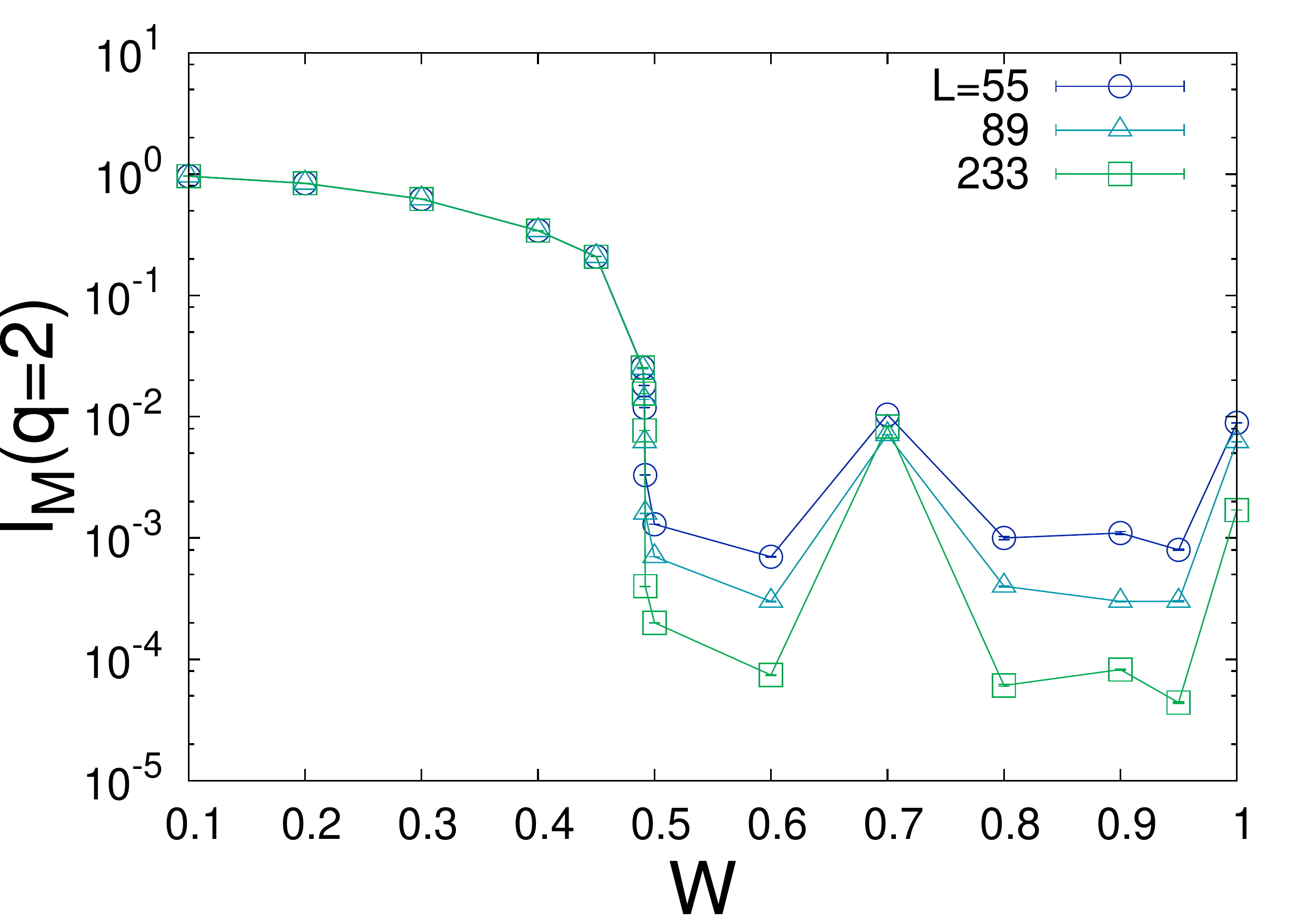}
	\caption{The $q=2$ inverse participation ratio in momentum space $\mathcal{I}_M(q=2)$ as a function of $W$ for various system sizes $L$. In the semimetal regime the momentum-space IPR is $L$-independent and becomes $L$-dependent in the chiral metal phase due to the wavefunction delocalizing in momentum space. At $W=0.7$, the momentum-space wavefunctions are still delocalized (see Fig.~\ref{Fig:MSW_W07}) even though the IPR data seems to be only weakly depending on the sizes.
	All the statistical errorbars in this plot are smaller than the symbols.}
	\label{Fig:iprq=2}
\end{figure}

\begin{figure*}[t]
	\includegraphics[width=0.9\textwidth]{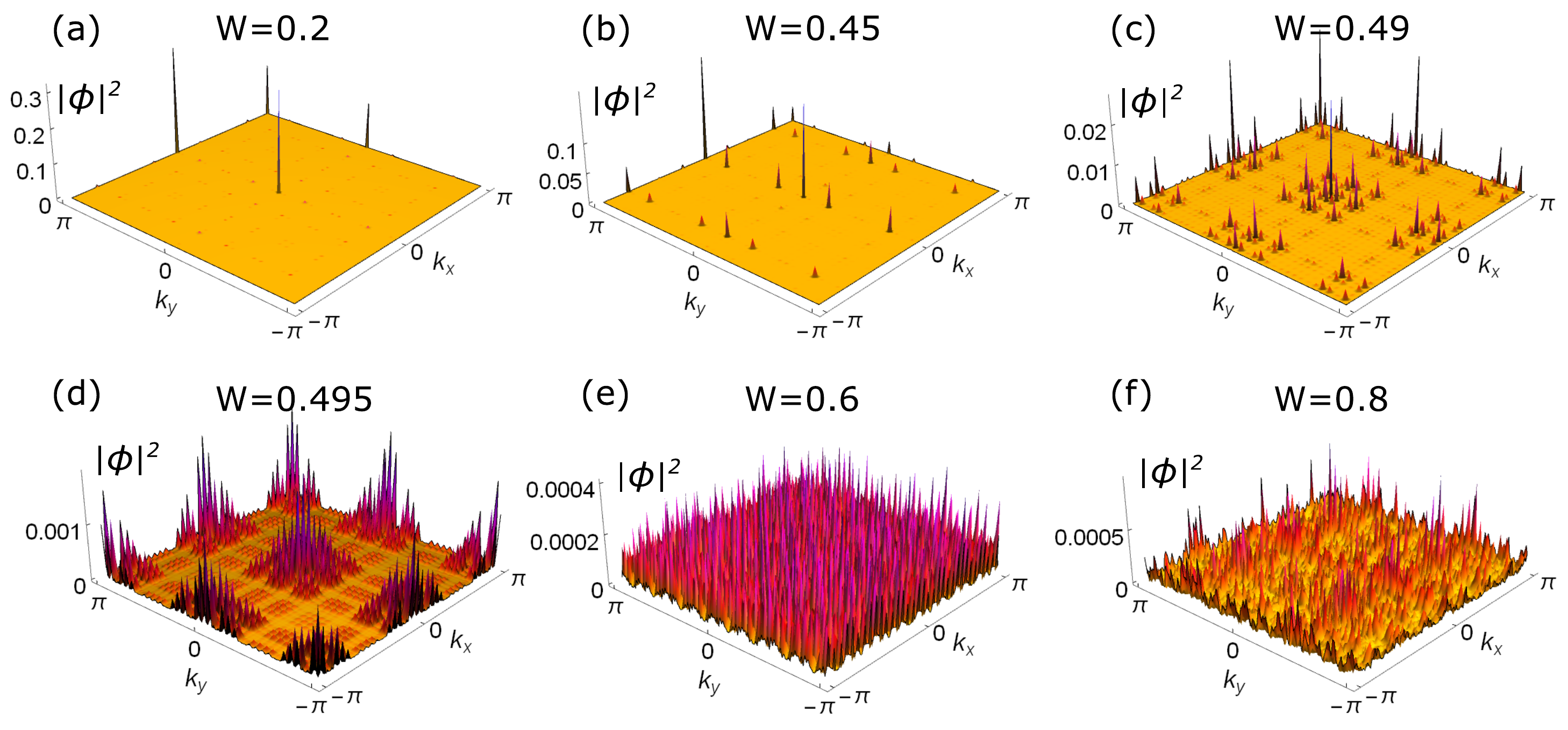}
	\caption{Probability distributions of zero energy wavefunctions in momentum space with $L=144$ and different values of $W$. (a)-(b): The wavefunctions contain well-defined ballistic peaks at $(k_x,k_y)=(0,0)$, $(0,\pi)$, $(\pi,0)$, and $(\pi,\pi)$. A few of satellite peaks can be seen in (b) while the major ballistic peaks are still well resolved from the figures. (c): The wavefunction is close to the critical point; The ballistic peaks can still be resolved. Meanwhile, the satellite peaks start to form regions instead of a few well-separated points.
	(d)-(f): The ballistic peaks are no longer sharply defined due to the hybridization with the satellite peaks which arise from scattering off QP potentials. In (f), the momentum-space wavefunction looks very much like a conventional delocalized state.
	The critical value is close to $W=0.49$.}
	\label{Fig:ZE_MSW}
\end{figure*}

We now connect the structure of the eigenvalues that we have probed through the DOS with the structure of the wavefunction.
A complementary way to understand the transition is to study how the
zero-energy  plane-wave eigenstates are perturbed by the QP hopping.
For the case of two-dimensional/three-dimensional Dirac/Weyl cones subject to a QP scalar potential it has been shown that the generation of a non-zero DOS coincides with a momentum-space delocalization transition \cite{Pixley2018,FuPixley2018}, which can be seen in the momentum-space IPR ($\mathcal{I}_M$) for $q=2$. Similar results for the current model are shown in Figs.~\ref{Fig1}(b) and \ref{Fig:iprq=2}.
In the absence of the QP hopping, the wavefunction at zero energy is composed of the Fourier modes at the Dirac points $(k_x,k_y)=(0,0)$, $(0,\pi)$, $(\pi,0)$, and $(\pi,\pi)$. Generically, the zero-energy states are linear combinations of these four plane waves.
Therefore, the probability distributions (integrating over the internal degrees of freedom) of the momentum-space wavefunction contains four peaks, which we call ``ballistic peaks.'' If we now translate the multifractal nomenclature to the present problem, we see that these ballistic peaks give rise to a frozen wavefunction. We note that the momentum-space wavefunction here has peaks at the Dirac points regardless of the QP potential (as long as it is weak). On the other hand, the real-space frozen wavefunctions, as realized in the the random vector potential Dirac model \cite{Ludwig1994,Castillo1997}, have peaks randomly distributed depending on the disorder realization.

\begin{figure}[t]
	\includegraphics[width=0.375\textwidth]{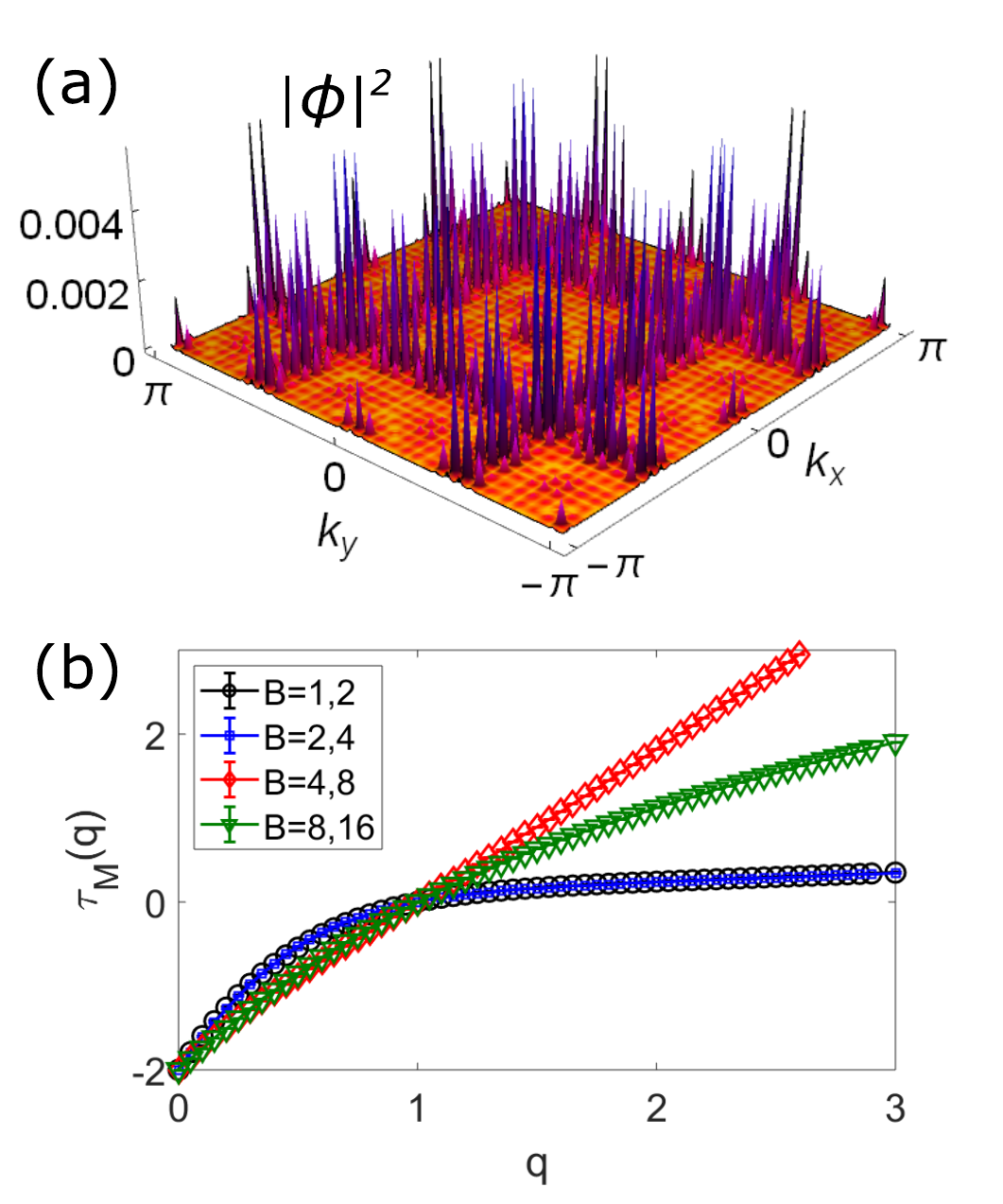}
	\caption{Zero-energy momentum-space wavefunction with $W=0.7$. (a) The probability distribution. The wavefunction is made of sparse peaks and is still delocalized in momentum space. (b) The multifractal spectra $\tau_M(q)$. Each data is averaged over 100 realizations. For smaller binning sizes ($B=1,2$ and $B=2,4$), the $\tau_M(q)$ show strongly multifractal (but still unfreezing) behavior. Note that $\tau_M(q=2)$ is not zero for all the binning sizes.
	}
	\label{Fig:MSW_W07}
\end{figure}

To support the argument of perturbing stable ballistic peaks, we plot the momentum-space wavefunctions in Fig.~\ref{Fig:ZE_MSW}. In Fig.~\ref{Fig:ZE_MSW} (a), the momentum-space wavefunction is essentially composed of the four ballistic peaks.
Generically, the QP hopping decreases the ballistic peaks via ``hopping'' in momentum space and generates other satellite peaks which arise
due to the coupling of the QP wavevectors $(\pm Q_L,0)$ and $(0,\pm Q_L)$. Those satellite peaks have weights related to the order of scattering off of the QP hopping.
While there are infinitely many such peaks in the thermodynamic limit, the wave function is weighted subextensively among them (akin to how a localized state dies off exponentially from a central localized site).
In finite system sizes and sufficiently weak $W$, only a finite number (smaller than $L^2$) of satellite peaks dominate, as shown in Fig.~\ref{Fig:ZE_MSW} (b).
For $W< 0.49$, where $W=0.49$ is close to the critical point, the ballistic peaks remain sharply defined even in the presence of the satellite peaks, and this structure can be captured perturbatively.
The weight of the wavefunction on the satellite peaks increases when driving W to a larger value, similar to a localized wavefunction as we approach a delocalization transition.
For $W>0.49$,
the ballistic peaks hybridize with extensively many satellite peaks,
the wavefunction is “delocalized” in momentum space, as displayed in Figs.~\ref{Fig:ZE_MSW} (d), (e), and (f).
Throughout this transition, the wave function is delocalized in real space; however, it acquires a definitive structure that we explain qualitatively in terms of topological zero modes in Sec.~\ref{sec:cm}.
This state is delocalized in both real- and momentum- space, in contrast to the wavefunctions with $W<0.49$ which are ballistic and composed of a measure-zero set of momenta.
The hybridization of an extensive number of momenta most likely creates extensive degenerate zero energy states, causing a finite DOS. And indeed, we witness numerically [see Fig.~\ref{Fig1}(b)] that the unfreezing transition in the momentum-space wavefunction coincides with the semimetal to metal transition in the DOS.

\begin{figure*}[t]
	\includegraphics[width=0.95\textwidth]{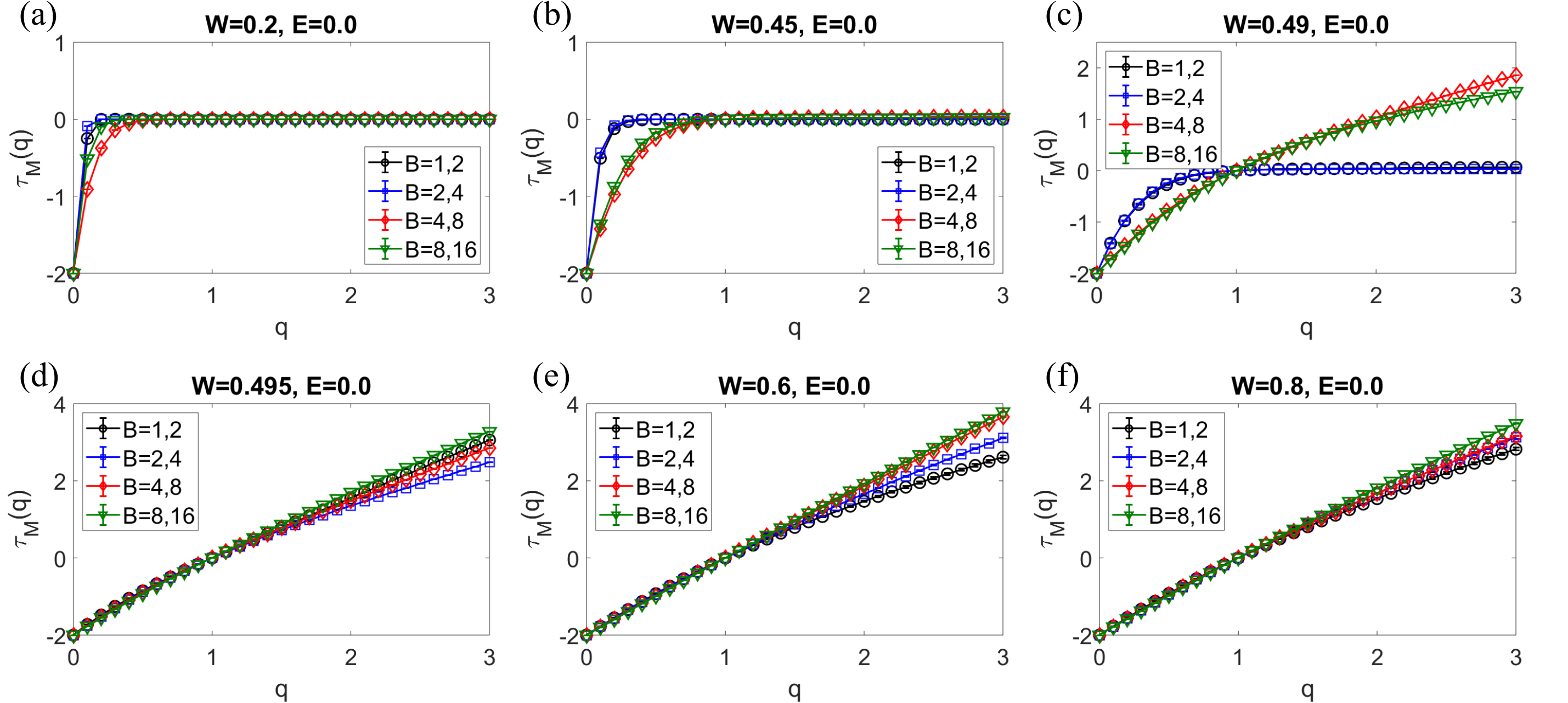}
	\caption{Multifractal spectrum of the zero-energy momentum-space wavefunction with different $W$ for $L=144$. Each $\tau_M(q)$ is obtained via numerical extrapolation of two different values of the binning size $B$. Each data is averaged over 100 realizations. (a)-(b) All the $\tau_M(q)$ spectra show freezing behavior. (c) $\tau_M(q)$ spectra extracted from larger binning sizes ($B=4,8$ and $B=8,16$) start to show unfreezing behavior. While the spectra from $B=1,2$ and $B=2,4$ are still frozen. This is very close to the critical value of $W$. (d)-(f) All the $\tau_M(q)$ spectra show unfreezing, weakly multifractal behavior.
	}
	\label{Fig:MSW_tauq}
\end{figure*}

To study the momentum-space wavefunction quantitatively, we first compute the second momentum-space IPR $\mathcal{I}_M(q=2,B=1,N=L)$ [given by Eq.~(\ref{Eq:Pm_tauq})] for different system sizes ($L=55,89,233$).
In Fig.~\ref{Fig:iprq=2}, the IPR with $q=2$ in different system sizes are essentially $L$-independent for $W<0.49$. For $W>0.49$, the IPR becomes size-dependent, an indication that the wave function is composed of an extensive number of momentum states. Similar results can be obtained for $L=34,144,610$. Note that, while it looks like $W=0.7$ is close to being localized in momentum space, this is not the case as we demonstrate in Fig.~\ref{Fig:MSW_W07}.
For even numbers, the Dirac nodes gap out at order $L/2$ in perturbation theory, so while the trend of the IPR is the same as for odd numbers, it quantitatively differs.
Correspondingly, we compute the $\tau_M(q)$ spectrum \cite{Evers2008_RMP} for $L=144$ by varying the binning size $B$ in every realization as shown in Figs.~\ref{Fig:MSW_W07} (b) and \ref{Fig:MSW_tauq}. This analysis directly answers if the wavefunctions are governed by well-localized peaks. For $W<0.49$, the wavefunctions show freezing which is characterized by $\tau_M(q)=0$ for all $q\ge 1$. We note that a single localized peak results in a spectrum with $\tau_M(q)=0$ for all $q>0$.
The frozen spectrum indicates that the dominating regions in the probability distribution of a wavefunction are characterized by a measure-zero set of peaks.
For $W>0.49$, the well-defined ballistic peaks are broadened with finite widths due to hybridization with the satellite peaks. We find that the $\tau_M(q)$ spectrum is
weakly ``multifractal.''
For instance, with $W=0.495$, the $\tau_M(q)\approx 2(q-1)-0.34q(q-1)$ for $|q|<1$.
These results are summarized in Fig.~\ref{Fig:MSW_tauq}.
The ballistic peaks are no longer sharply defined as their weights strongly depend on the binning size $B$.
The location of the semimetal to metal transition obtained from the wavefunction diagnostic is in excellent agreement with the semimetal to metal transition in the DOS.
As a comparison, we also plot the real-space wavefunctions with the associated parameters in Fig.~\ref{Fig:ZE_RSW}.
We also emphasize that the present transition is not related to the freezing transition \cite{Castillo1997,Carpentier2001,Motrunich2002,Horovitz2002,Mudry2003,Chou2014} in the context of highly random delocalized systems. Here, we simply use the multifractal analysis to explore the intricate structures in the momentum-space wavefunctions due to the QP hopping.

\subsubsection{A theory for the chiral metal phase in terms of topological zero modes}
\label{sec:cm}

\begin{figure*}[t!]
	\includegraphics[width=0.95\textwidth]{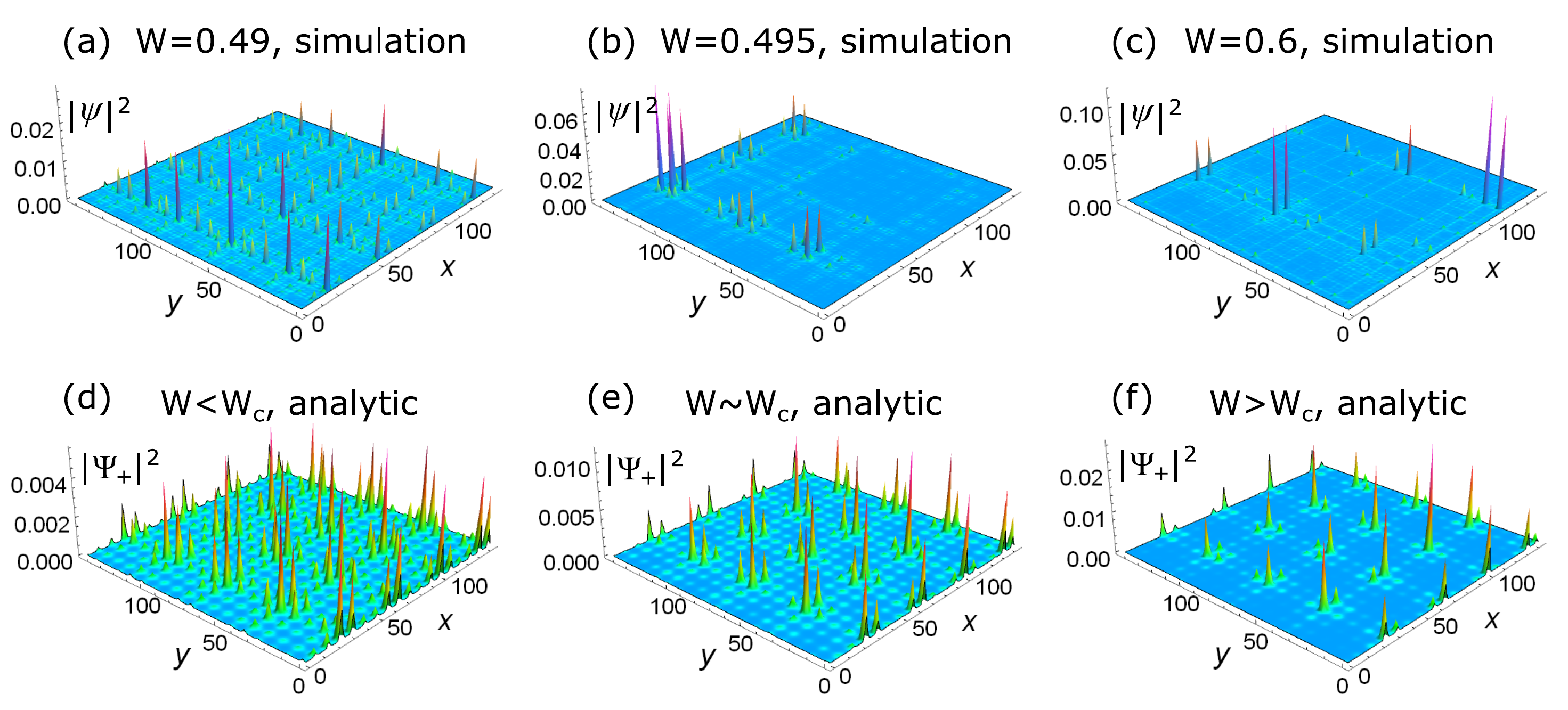}
	\caption{Probability distributions of zero-energy wavefunctions in real space with $L=144$ and different values of $W$ comparing the exact numerical calculations (top row) with the analytic results (bottom row) for the wavefunctions of the chiral metal, in Eq.~\eqref{eq:ZeroMode}.  (a) and (d): The wavefunctions are plane waves. (b) and (e): The model is close to the critical point of the semimetal to metal transition and the wavefunction looks like a periodic array of localized peaks.
		(c) and (f): The wavefunctions are delocalized but possess intricate structure that agrees qualitatively well with the analytic prediction.
		The critical value obtained from numerics is close to $W=0.49$.  Despite the analytical treatment overestimating the position of the semimetal to metal transition by a factor of 2, it leads to qualitatively similar behavior near the transition. As a result for the analytic results we show $W=0.83$ in (d), $W=0.87$ in (e), and $W=0.91$ in (f).
	}
	\label{Fig:ZE_RSW}
\end{figure*}

For $W>W_c(Q)$, we have seen how the low-energy eigenstates delocalize in momentum space, which induces well-defined patterns in the  real-space structure of the wavefunction (see Fig.~\ref{Fig:ZE_RSW}). There are a few key features that are unique to this chiral model and were not observed for a QP potential in Ref.~\cite{FuPixley2018}. Firstly, the low-energy excitations minibandwidth has been substantially renormalized reducing it by a factor of $ \sim 10^{-8}$, which is a much larger effect then we observed for a QP potential~\cite{FuPixley2018}, see Fig.~\ref{Fig:twist}. Second, we do not find any reentrant semimetal phase, for the chiral model, once the system has undergone a transition to the metallic phase, it remains there. This suggests that the metallic phase in the chiral limit should have a unique description that relies on the chiral symmetry. In the following, we will show that the our model possesses a band of quasizero modes which are intimately linked to the chiral symmetry. These solutions to an effective Dirac equation are bound states due to a sign changing Dirac mass induced by the QP hopping. For $W<W_c(Q)$ these bound state solutions strongly overlap: They are not well-defined local eigenstates, therefore they hybridize with the continuum of plane waves and hence do not play a role in the low-energy behavior. On the other hand for larger $W> W_c(Q)$, these zero mode bound states become sufficiently sharp to be stable. This produces a finite DOS at zero energy and a non-trivial structure in the wavefunction that agrees well with our numerical results in the metallic phase.
Since it exists only due to the chiral symmetry (e.g. they do not occur in the QP potential model in Ref.~\cite{FuPixley2018}) we dub this phase the chiral metal.

To mathematically derive the above statements, we invoke a perturbative inclusion of the incommensurate modulation on top of a continuum model.
In view of the stability of the semimetallic phase below the ``magic-angle'' semimetal-to-metal transition. Therefore, the physics near the center of the band may be treated in the continuum approximation leading to Dirac Hamiltonians subjected to certain background ``Higgs'' fields (i.e. a spatially dependent mass fields \cite{Jackiw1976,Jackiw1981}). In Appendix~\ref{app:LowEnergy}, we explicitly derive such effective Hamiltonians, which take the form
$
H = \sum_\pm h_{\pm} \frac{\mathbf 1 \pm \tau_y}{2}
$
with ($v_0 = 2 J_0$)
\begin{equation}
h_\pm = v_0 \slashed p \lambda_z
+ V(x) \lambda_y \pm V(y) \lambda_x. \label{eq:HDirac4}
\end{equation}
Here $\slashed p = p_x \sigma_x + p_y \sigma_y$ and the original basis in Eq. (2) has been rotated for convenience;
to account for all four Dirac nodes, we require  more sets of Pauli matrices, $\tau_\mu$ works within blocks of the same helicities $(0,0)$ and $(\pi,\pi)$ [or $(0,\pi)$ and $(\pi,0)$], while $\lambda_\mu$ connect these blocks.
In this basis, the chiral symmetry is represented by $\lbrace \sigma_z \lambda_z,H \rbrace = 0$ and time reversal symmetry implies $H = \sigma_y \lambda_z H^T \sigma_y \lambda_z$. Both constrain the structure of the effective Hamiltonian.
The dominant contributions for the model at $Q = 2\pi [2/(\sqrt{5}+1)]^2$ are
\begin{eqnarray}
V(x) &=& V_1 \sin((\pi-Q)x) + V_4 \sin((4Q- 3\pi)x), \label{eq:VofX}
\end{eqnarray}
with $V_1 = 2 W$, $V_4 =W^4/[J_0^3 \prod_{l = 1}^3 (2 \sin(lQ))]$.
Since, in the chiral model $\gamma_1 = \sigma_{x} \lambda_z, \gamma_2 = \sigma_{y} \lambda_z,  \gamma_3 = \lambda_y, \gamma_4 = \lambda_x$ form a Clifford algebra, zero modes (as in other magic-angle systems, such as twisted bilayer graphene \cite{Longzhang2018}) may be readily found analytically at the vortex like nodes of $(V(x),V(y))$. The zero modes of $h_\pm$ have the form
\begin{equation}
\Psi_\pm(\v x) = \mathcal N e^{-  \sum_{i = 1,4}\frac{2V_i }{v_0 q_i}[\sin^2(\frac{q_i x}{2}) \lambda_x \sigma_x \mp\sin^2(\frac{q_i y}{2})\lambda_y \sigma_y]} \Phi_\pm, \label{eq:ZeroMode}
\end{equation}
with $q_1 = \pi - Q$, $q_4 = 4 Q - 3\pi$, $\Phi_+ = (1,0,0,1)$, and $\Phi_- = (0,1,1,0)$ such that the eigenvalues of $\sigma_x \lambda_x$ and $\mp \sigma_y \lambda_y$ are both 1.
The solution of Eq.~\eqref{eq:ZeroMode} is plotted in Fig.~\ref{Fig:ZE_RSW} along with the numerical solutions.
These bound states are irregularly localized at distances set by $2\pi/{q_{1,4}}$ and their decay length is given by $\sqrt{v_0/(q_{1,4}V_{1,4})}$. Therefore, a simplest estimate (keeping only $q_1$ and $V_1$) suggests that bound states become stable for $W \gtrsim W_c^{\rm (0\; modes)} \equiv 1/\sqrt{1 + \text{const.} \times (\pi - Q)^{-2}}$, in good agreement for $Q$ close to $\pi$ (apart from the numerical constants) with the $W_c$ obtained of Eq.~\eqref{eqn:v}.

We conclude with three remarks: First, we repeat that this non-perturbative analysis is based on the continuum Dirac Hamiltonian which is clearly only justified for sufficiently low $W$ and inapplicable deep in the metallic phase. Second, we highlight that the bound state picture explains the observation of the sparse real-space structure of the eigenstates for $W \gtrsim W_c$, see Fig.~\ref{Fig:ZE_RSW}.
Finally, in order to analyze the importance of symmetries, we also applied the same method to a non-chiral model with a QP potential (from Ref.~\cite{FuPixley2018}) and to the model with complex hopping (from Appendix~\ref{sec:cqph}). In both cases additional mass terms appear in Eq.~\eqref{eq:HDirac4}, which breaks the topologically protected depletion of the gap inside a vortex configuration of $[V(x),V(y)]$. As a consequence, topological bound state solutions are absent in these cases.

\subsubsection{Real-space Anderson localization and structure of the mobility edges}
\label{sec:AndersonLocReal}

Real-space Anderson localization in disordered systems of orthogonal and unitary chiral
classes are special, because the zero energy state is robust against localization~\cite{Gade1991,Motrunich2002,Konig-2012}, and tend to form a line of critical fixed points between Anderson localized states at finite energy~\cite{Abrahams-1979}.
This model \cite{fn1} is fundamentally distinct from its random counterpart because the QP hopping is, in some sense, infinitely correlated
and generic localization at $E \neq 0$ no longer occurs.
It is therefore non-trivial to determine the localization phase diagram in the present model at finite energies. To do so we compare the typical and average DOS [see Eq.~\eqref{eqn:tdos}]. Anderson localized eigenstates necessarily have a typical DOS that goes to zero for increasing KPM expansion order (or system size), and we compare with the average DOS to differentiate between a hard gap (with no states) and localized states. We also use Lanczos diagonalization to examine the localization properties directly via wavefunctions.

\begin{figure*}
\begin{minipage}{\columnwidth}
\centering
	\includegraphics[width=0.68\columnwidth,angle=-90]{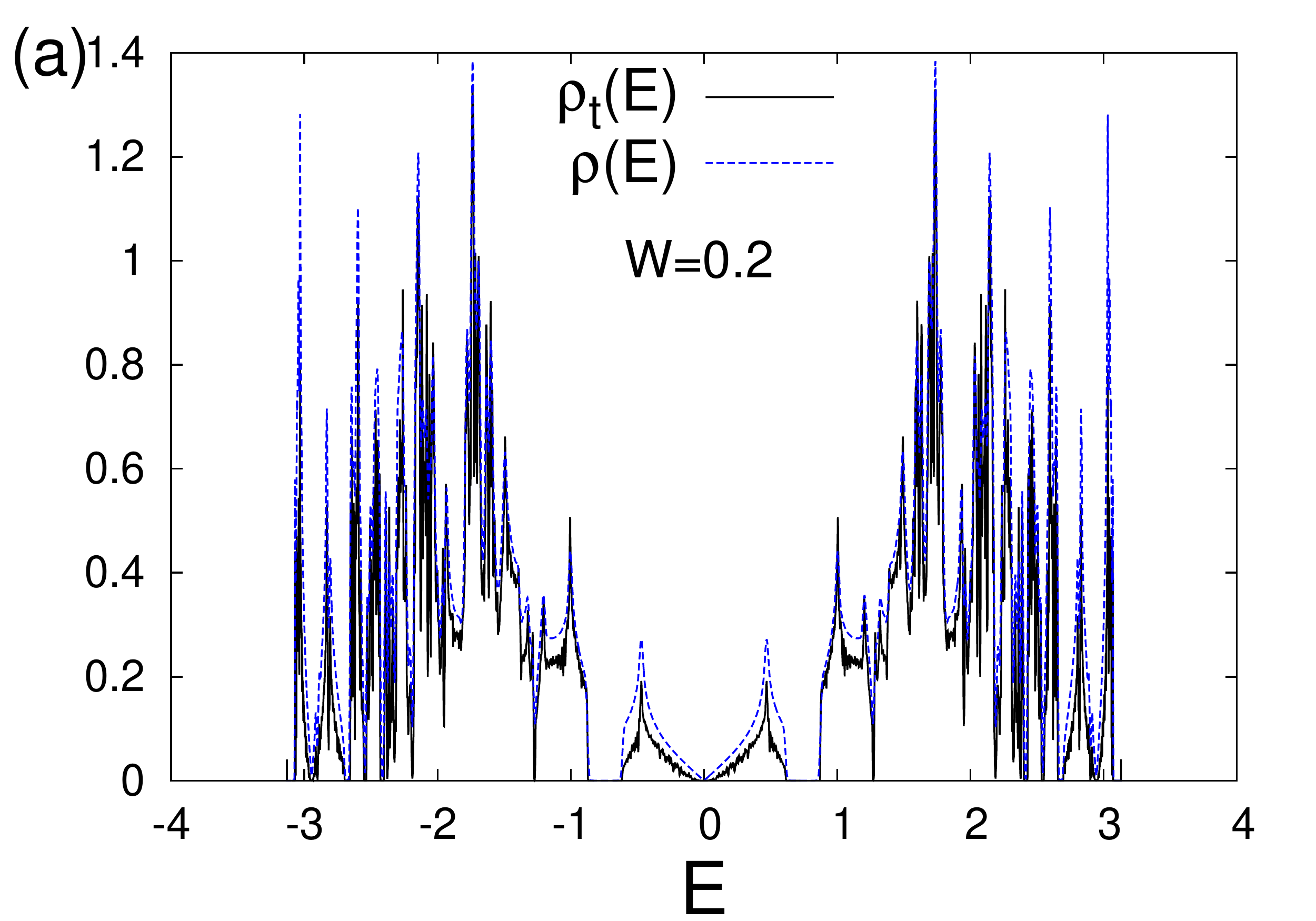}
\end{minipage}%
\begin{minipage}{\columnwidth}
\includegraphics[width=0.68\columnwidth,angle=-90]{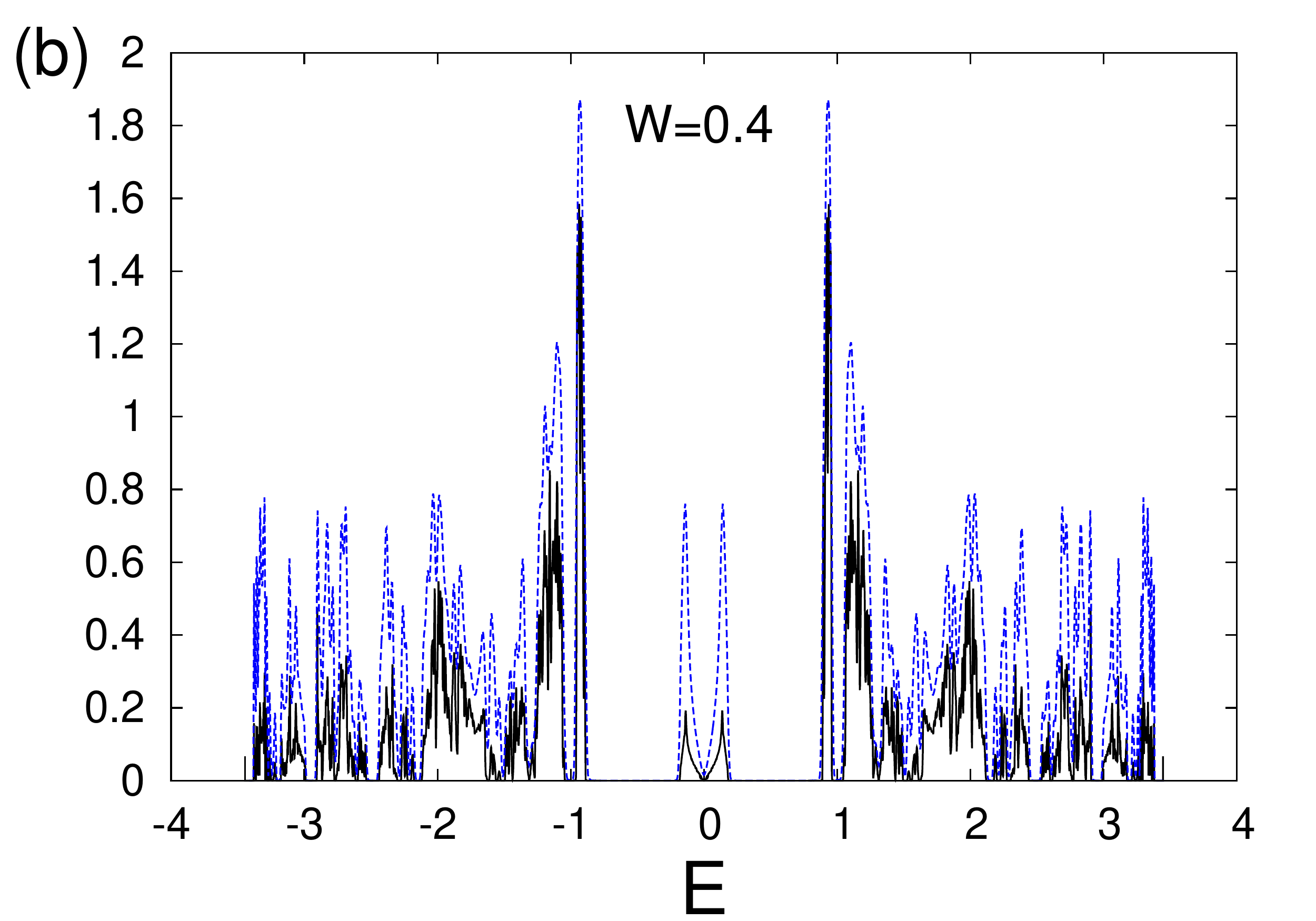}
\end{minipage}
\begin{minipage}{\columnwidth}
	\includegraphics[width=0.68\columnwidth,angle=-90]{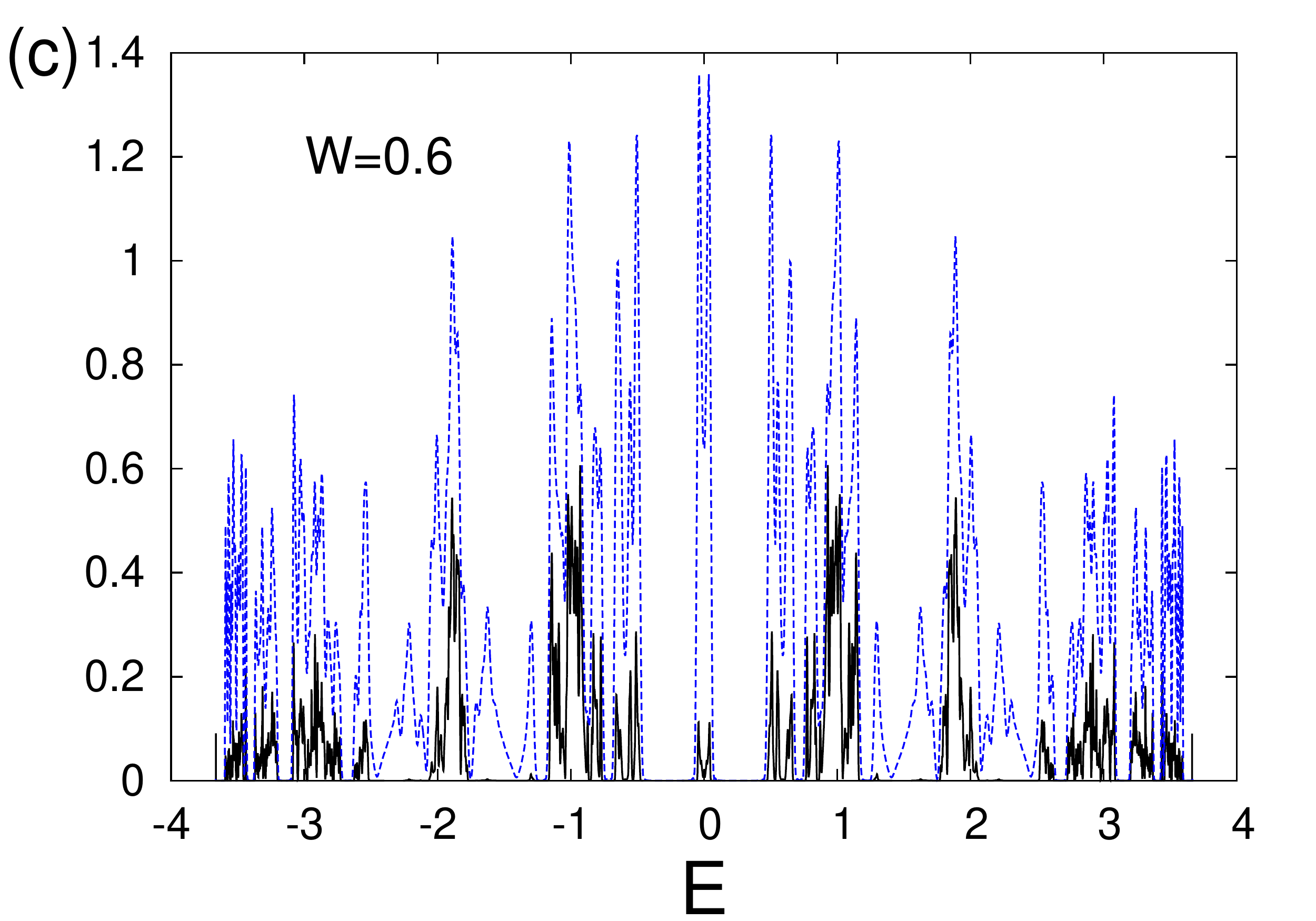}
\end{minipage}%
\begin{minipage}{\columnwidth}
	\includegraphics[width=0.68\columnwidth,angle=-90]{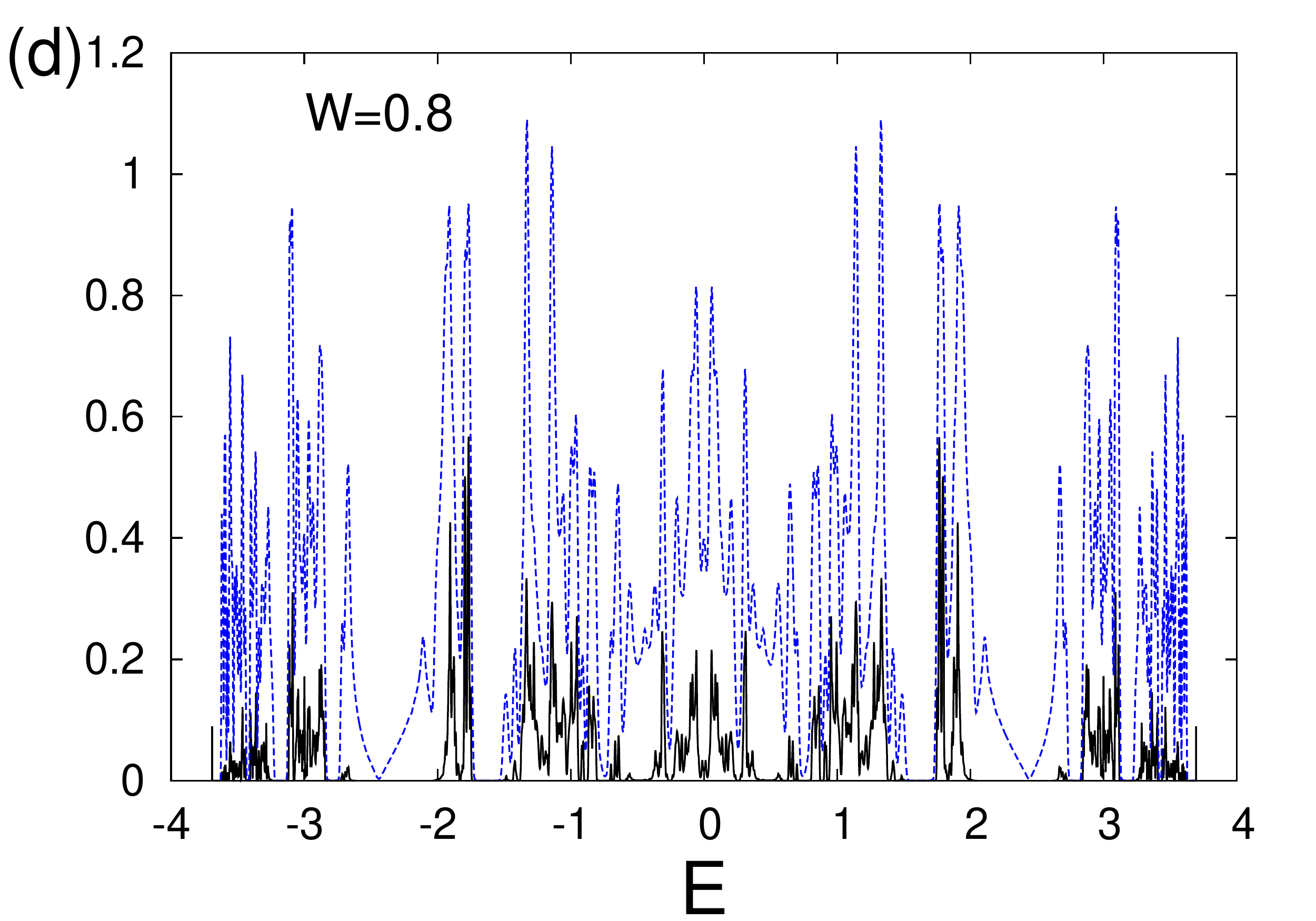}
\end{minipage}
\begin{minipage}{\columnwidth}
	\includegraphics[width=0.68\columnwidth,angle=-90]{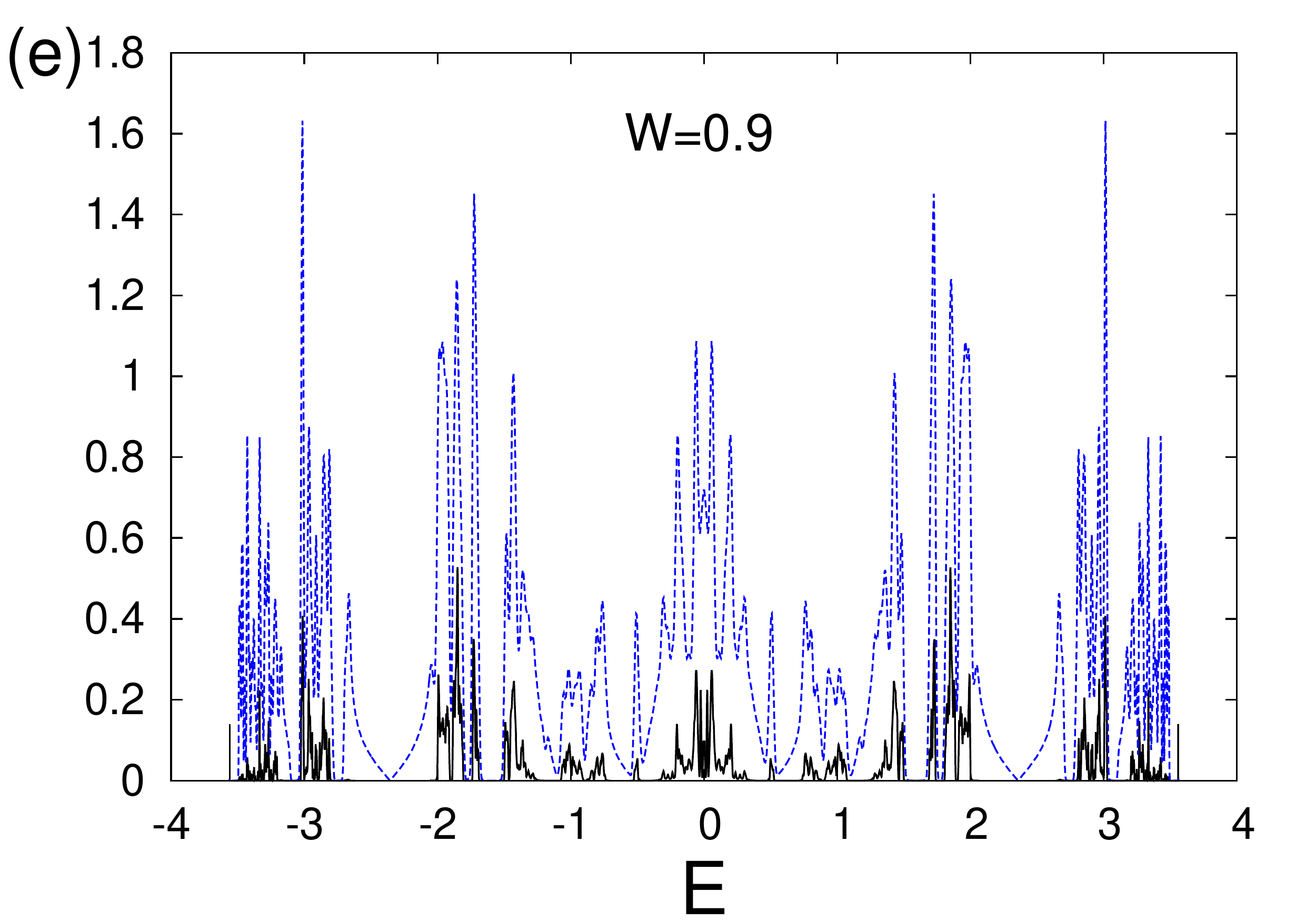}
\end{minipage}
\begin{minipage}{\columnwidth}
	\includegraphics[width=0.68\columnwidth,angle=-90]{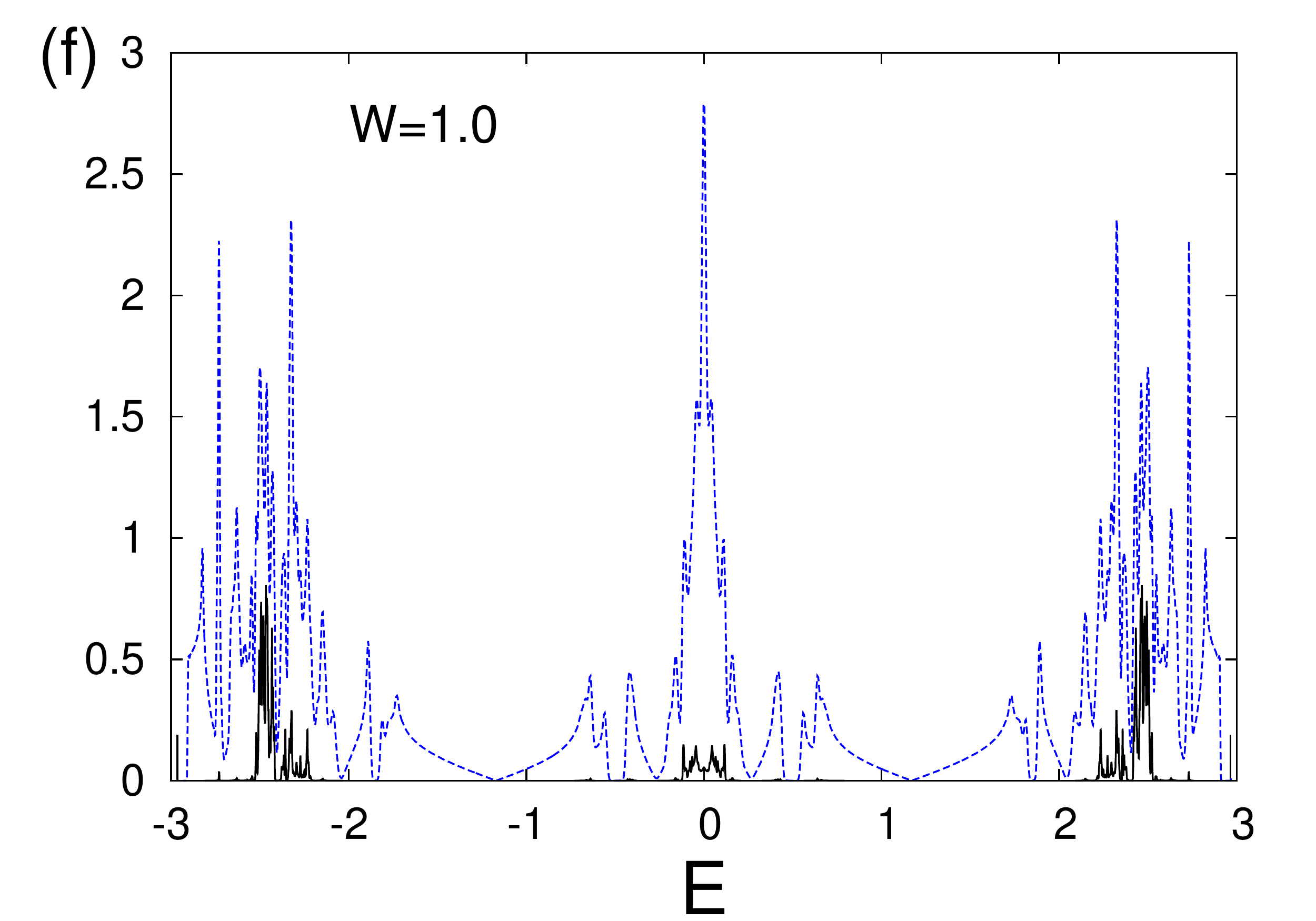}
\end{minipage}
	\caption{Localization properties obtained through the typical DOS. Typical DOS are in black solid lines, and average DOS are in blue dashed lines (to distinguish hard gaps and localized states) for $L=144$ and $N_C=2^{14}$ [(a) $W=0.2$; (b) $W=0.4$; (c)$W=0.6$; (d) $W=0.8$; (e) $W=0.9$ and (f) $W=1.0$].}
	\label{Fig:rhot_all}
\end{figure*}

\begin{figure*}[t]
	\includegraphics[width=0.9\textwidth]{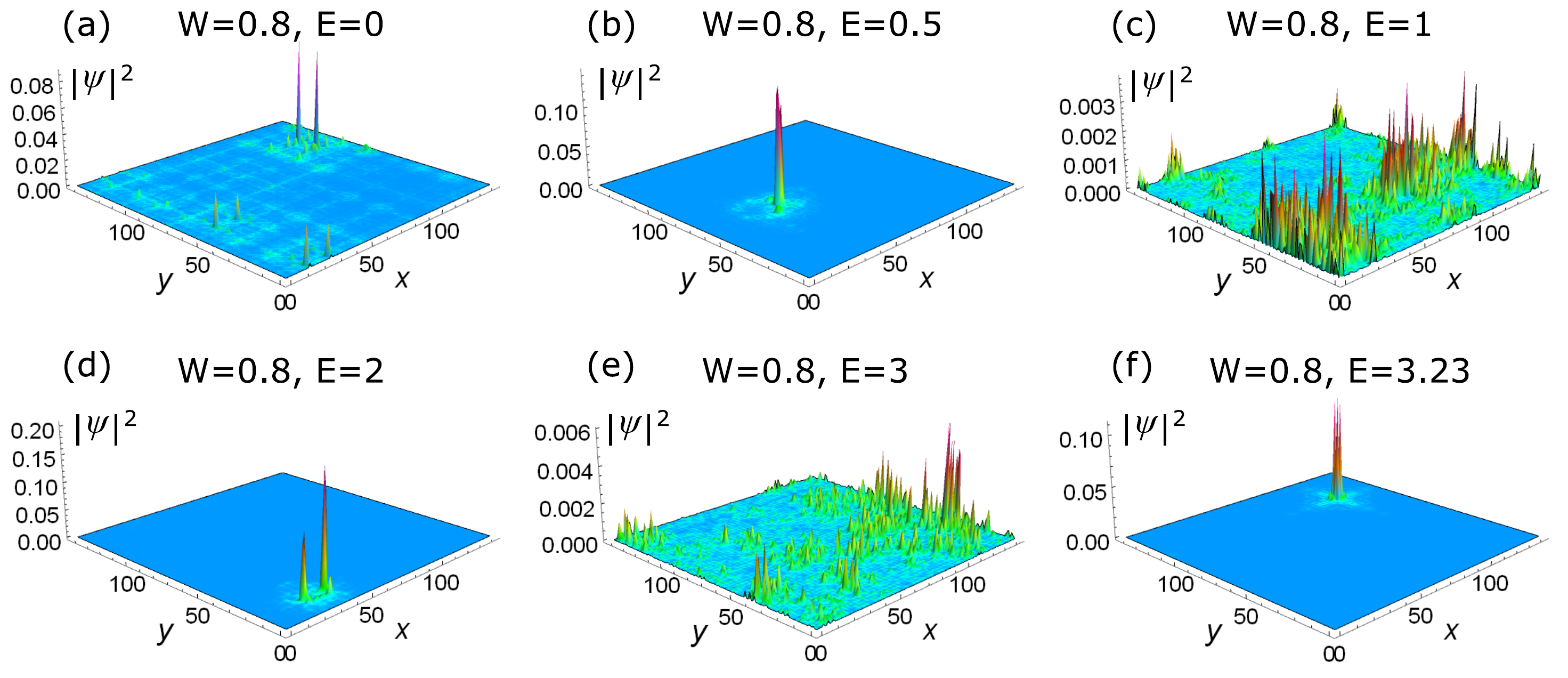}
	\caption{Real-space wavefunctions at various energies corresponding to $W=0.8$ and $L=144$. (a), (c), and (e) are delcoalized wavefunctions; (b), (d), and (f) are localized wavefunctions. This confirms the multiple mobility edges observed in the typical DOS in Fig.~\ref{Fig:rhot_all}.
	}
	\label{Fig:RSW_W08}
\end{figure*}

As shown in Fig.~\ref{Fig:rhot_all}, we find that the finite energy eigenstates are not localized for weak QP hopping strength. For QP hopping strengths beyond the semimetal to metal phase transition we find semimetal minibands develop at \emph{finite energy} with a linearly vanishing DOS that is shifted away from $E=0$ and the edges of the these minibands have  Van Hove-like peaks in the average DOS.  Interestingly, the typical DOS shows that these finite energy semimetal minibands are Anderson localized
As a result, for a single value of $W$ there can be various mobility edges in the system and the region separating localized and delocalized states does not monotonically vary as we tune $W$.
Looking directly at wavefunctions, we confirm the non-monotonic localization behavior and multiple mobility edges in Fig.~\ref{Fig:rhot_all}.
For example, wavefunctions for $W=0.8$ and $L=144$ at different energies are plotted in Fig.~\ref{Fig:RSW_W08}.
The results clearly show the same non-monotonic localization properties as a function of energy and are consistent with the typical DOS diagnostics.

Upon increasing the QP hopping strength further, the number of localized states increases but even for pure QP hopping ($W=1.0$) we still find a finite number of delocalized states. In particular, the low-energy states that contribute to the diverging DOS do not appear to localize.

\subsection{Strong quasiperiodic hopping}

We now turn to the properties of the QP hopping model in the limit of large $W$, where our parametrization of the model gives a purely QP hopping model for $W=1$, see Eq.~\eqref{eqn:J0}. A striking feature of random chiral class models is the presence of a divergence in the low-energy DOS \cite{Gade1991,Gade1993,Motrunich2002,Mudry2003,Evers2008_RMP}, but this behavior is strongly dependent on the type of model chosen. In random hopping models the precise form of this divergence is modified due to Griffith effects~\cite{Motrunich2002}.  This is naturally a very interesting problem to compare with the QP hopping model since we know \textit{a priori} it has no rare region effects.
However, observing anything beyond just a power-law divergence is notoriously difficult numerically and therefore that is not our goal here. Instead, we aim to demonstrate the existence of a divergence and not necessarily pinpoint its precise analytic form beyond the leading power-law dependence.
\subsubsection{Diverging low-energy density of states}

\begin{figure}[b]
	\centering
	\includegraphics[width=0.3\textwidth,angle=-90]{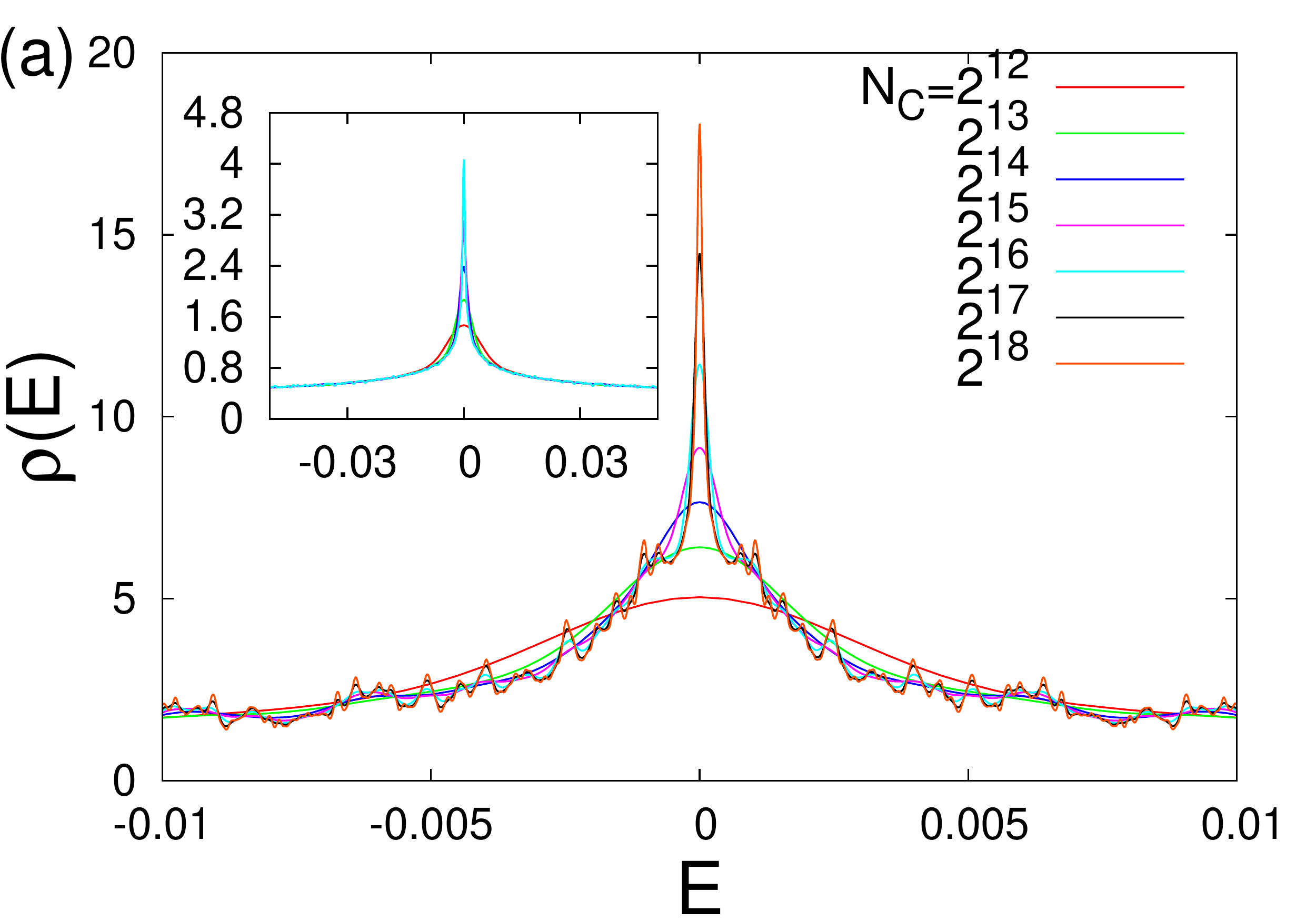}
	\includegraphics[width=0.3\textwidth,angle=-90]{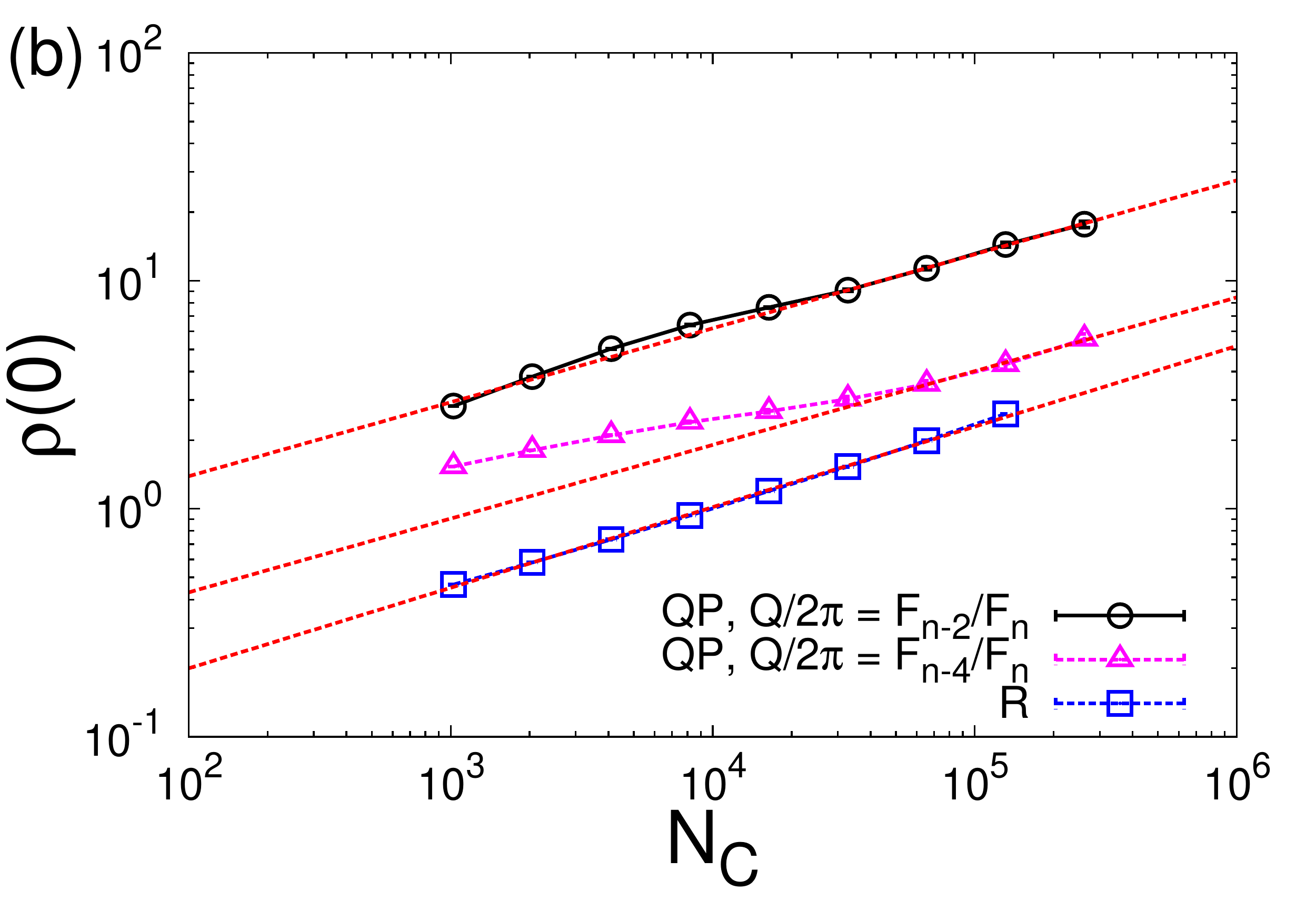}
	\caption{Divergence of the low-energy DOS for $W=1$ (i.e. pure QP hopping). (a) $N_C$-dependence near zero energy for a very large system size $L=987$ and $Q_L=2 \pi F_{n-2}/F_n$. (Inset) Similar results for the randomized version of the model (letting the phase be random at each site) with $L=233$ for $N_C= 2^{12}, 2^{13}, 2^{14}$, $L=377$ for $N_C=2^{15}$ and $L=610$ for $N_C=2^{16}$, note that the divergence is similar between the two. (b) Divergence of the low-energy DOS for $W=1$ in the pure QP limit comparing two different quasiperiodic wavevectors and the random (R) hopping model with the KPM expansion order that acts like a low-energy scale that rounds out the divergence of the DOS. Fits to the power law form are shown as red dashed lines.
	}
	\label{Fig:Div}
\end{figure}

Focusing on the pure QP limit $W=1$, we compute the DOS using KPM on very large system sizes ($L=987$) such that any low-energy divergence of the DOS is not affected by the mean level spacing on finite size systems. Any low-energy divergence in the DOS will be rounded out to due the extrinsic effects of finite system size and KPM expansion order. By going to $L=987$ we are able to reach large enough system sizes so that all of the (artificial) rounding is due to the KPM expansion order i.e. a finite $N_C$ \cite{fn2}. We now reach one of our main results, as shown in Fig.~\ref{Fig:Div}, we find a clear divergence of the low-energy DOS in the pure QP hopping model (rounded by the finite KPM expansion order $N_C$).
Since we are working at such large system sizes we can use the rounding of the divergence in the DOS by $N_C$ to our advantage:  in order to accurately compute the power-law divergence in the DOS $\rho(E) \sim 1/|E|^{x_{QP}}$,
we use the fact that the KPM expansion order is related to an infrared energy scale $N_C \sim 1/\delta E$ that implies the ansatz
\begin{equation}
\rho(E=0) \sim (N_C)^{x_{QP}}.
\end{equation}
As shown in Fig.~\ref{Fig:Div}, we find that $x_{QP}\approx 0.32$ for $Q = 2\pi F_{n-2}/F_n$ and  $Q = 2\pi F_{n-4}/F_n$, which is consistent with the divergence and value of $x_{QP}$ being $Q$-independent for irrational $Q$. Thus, we conclude that randomness is not necessary to create a low-energy divergence in the DOS. Using $\rho(E)\sim |E| ^{d/z-1}$ this leads to the estimate $z \approx 3$ for $W=1$.

\begin{figure}[t]
	\centering
	\includegraphics[width=0.3\textwidth,angle=-90]{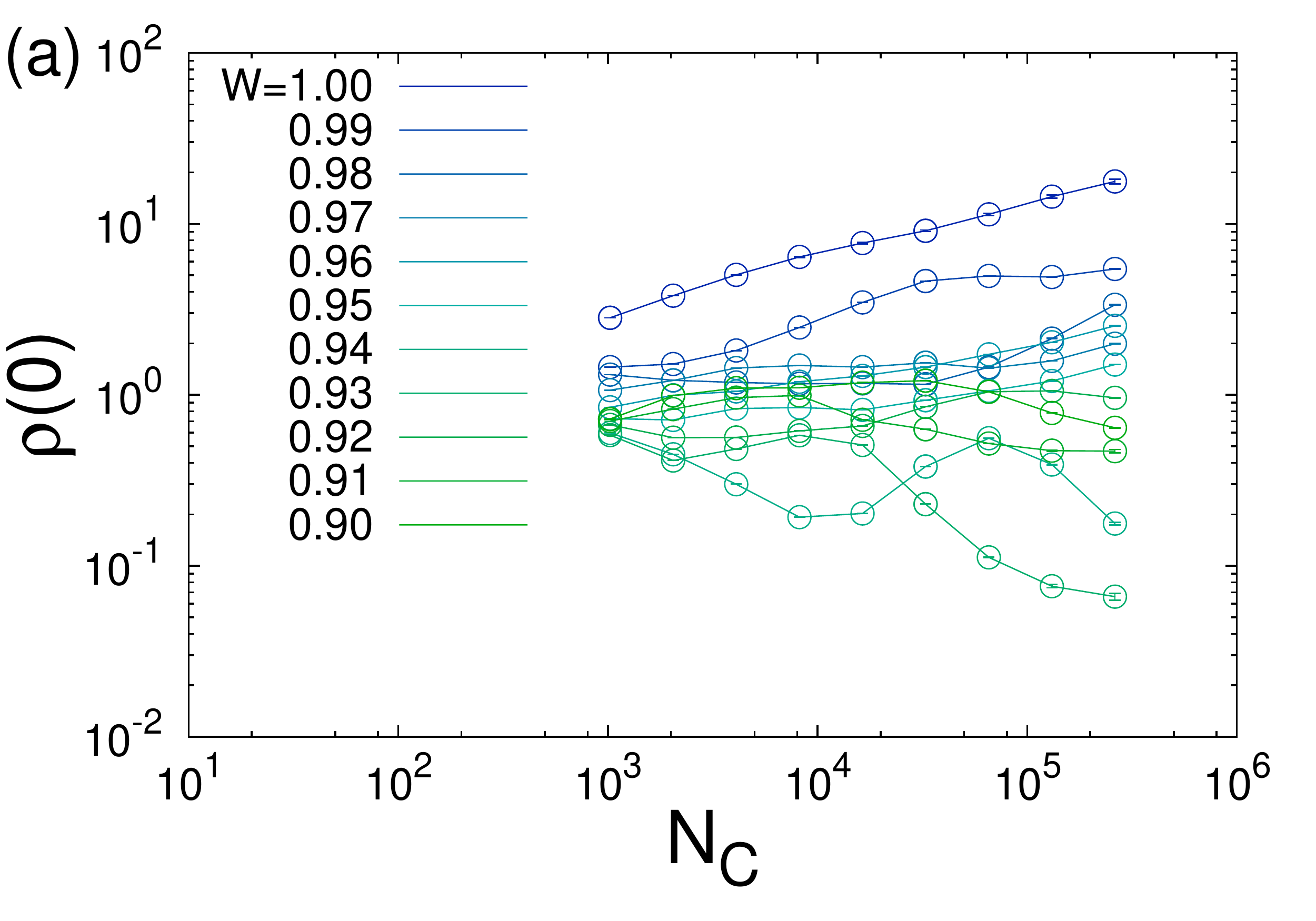}
	\includegraphics[width=0.3\textwidth,angle=-90]{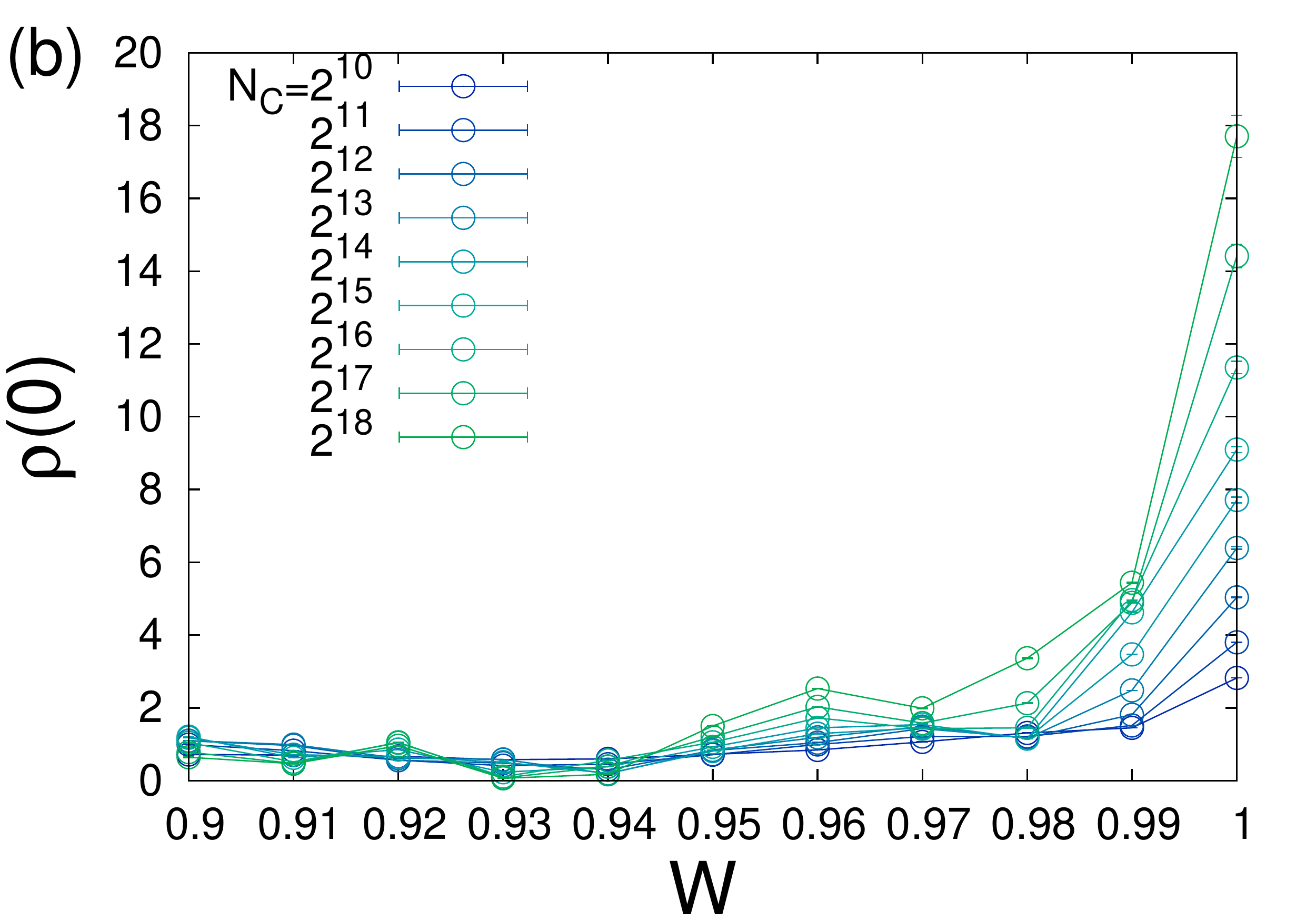}
	\caption{The onset of a divergence in the DOS at zero energy $\rho(0)$ versus (a) $N_C$ and (b) $W$ close to $W=1$ and $L=610$. We see a trend towards an increasing $\rho(0)$ for $W > 0.95$,  but there is no clear sign of divergence in the data other then at $W=1$.}
	\label{fig:rho0W1}
	\end{figure}

It is interesting to compare this result with the corresponding randomized version of the model, which has phases that are random across each bond [i.e. the $\phi_{\nu}$ in Eq.~(\eqref{eqn:hop}) are replaced by $\phi_{\nu}({\bf r})$ and sampled between $(0,2\pi)$ at each site]. We find the nature of the divergence of the DOS goes like
$
\rho(E=0) \sim (N_C)^{x_R}
$
with $x_R\approx0.35$.
Thus, we  find that the low-energy divergence in the QP hopping model agrees well with that of the random model to within our numerical accuracy. Since these two problems share the same distribution of hopping strengths at each bond, with the distinction being that the phases ($\phi_{\nu}$) are correlated across the entire sample for the QP model.
Note that this distribution is $Q$-independent and is given by the distribution of $\cos(x)+\cos(y)$ for $x,y\in[0,2\pi]$, which is consistent with $x_{QP}$ being $Q$-independent as we have already found.
In this way, our results on $x_{QP}$ and $x_R$ implies that the nature of the low-energy divergence, is dictated by the distribution and not whether the models possess rare regions. We note that other numerical studies have also seen just a simple power-law divergence in related (but not equivalent) disordered models \cite{Motrunich2002}.

\begin{figure}
	\centering
	\includegraphics[width=0.4\textwidth]{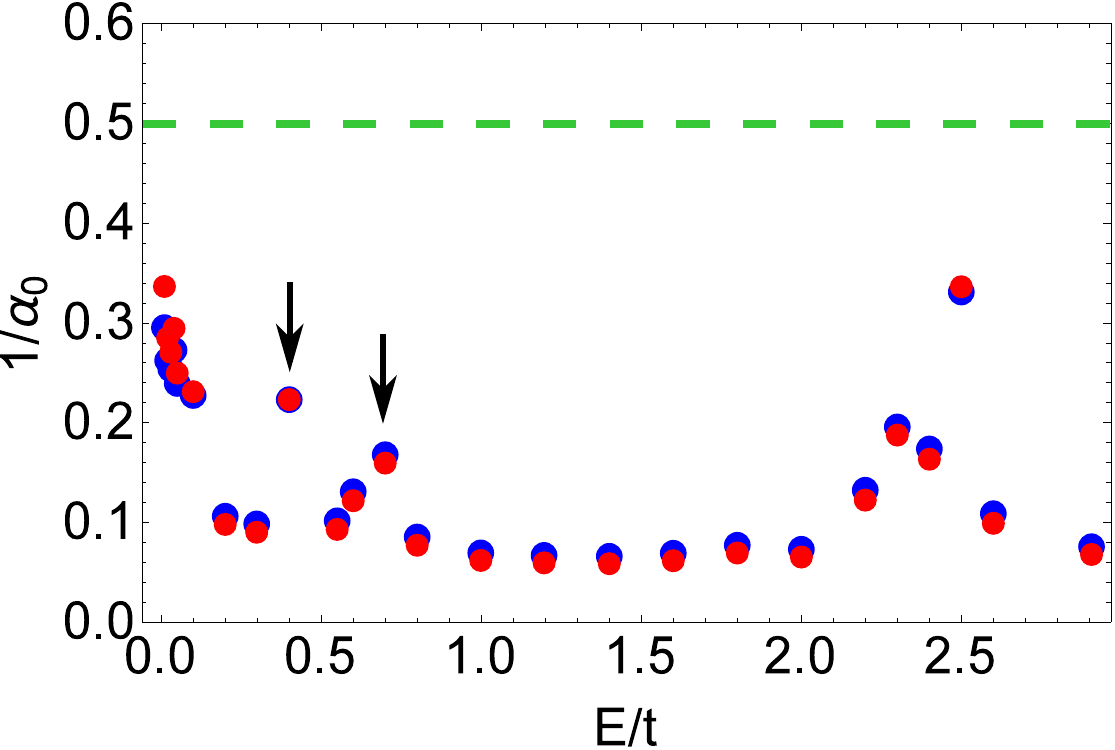}
	\caption{Inverse multifractal exponent $\alpha_0$ as a function of energy for $W=1$ and $L=144$. The green dashed line indicate the plane wave value $1/\alpha_0=0.5$. Localized states in the thermodynamic limit give $1/\alpha_0\rightarrow 0$. The results demonstrate non-monotonic dependence as a function of energy. Blue dots indicate the data extracting from $\psi(\vex{x})$ ($b=1$); red dots correspond to the data extracting from binned wavefunctions with resolution length $b=2$.
	The black arrows indicate the states consisted of double identical peaks. The corresponding typical DOS values are very small but non-zero in Fig.~\ref{Fig:rhot_all}.
	}
	\label{Fig:alpha0_W1}
\end{figure}

The low-energy divergence of the DOS for the pure QP limit of the model poses a natural question: is there a phase with a divergent low-energy DOS or is it only an isolated point as a function of $W$? As shown in Fig.~\ref{fig:rho0W1}, for KPM expansion orders up to $N_C=2^{18}$ and $L=610$ we do not find a clear sign of a divergence at $W<1$ in the data for $\rho(0)$ versus $N_C$, but we do find that the DOS is showing trends to a divergence at the largest expansion orders for $W \gtrsim 0.95$.
Thus, our data suggests that the point $W=1$ is fundamentally distinct from the phases of the model with $W<1$, i.e. any finite bare hopping $(J_0>0$) appears to be sufficient to suppress this divergence. As we show in Appendix~\ref{sec:cqph}, if we instead consider complex QP hopping matrix elements then the low-energy divergence goes away. As we discuss in Sec.~\ref{sec:discussion}, we attribute the divergence in the low-energy DOS to the hopping vanishing along lines in real space which induces an extensive number of zero modes.

\subsubsection{Real-space wavefunctions at $W=1$}

\begin{figure}
	\centering
	\includegraphics[width=0.49\textwidth]{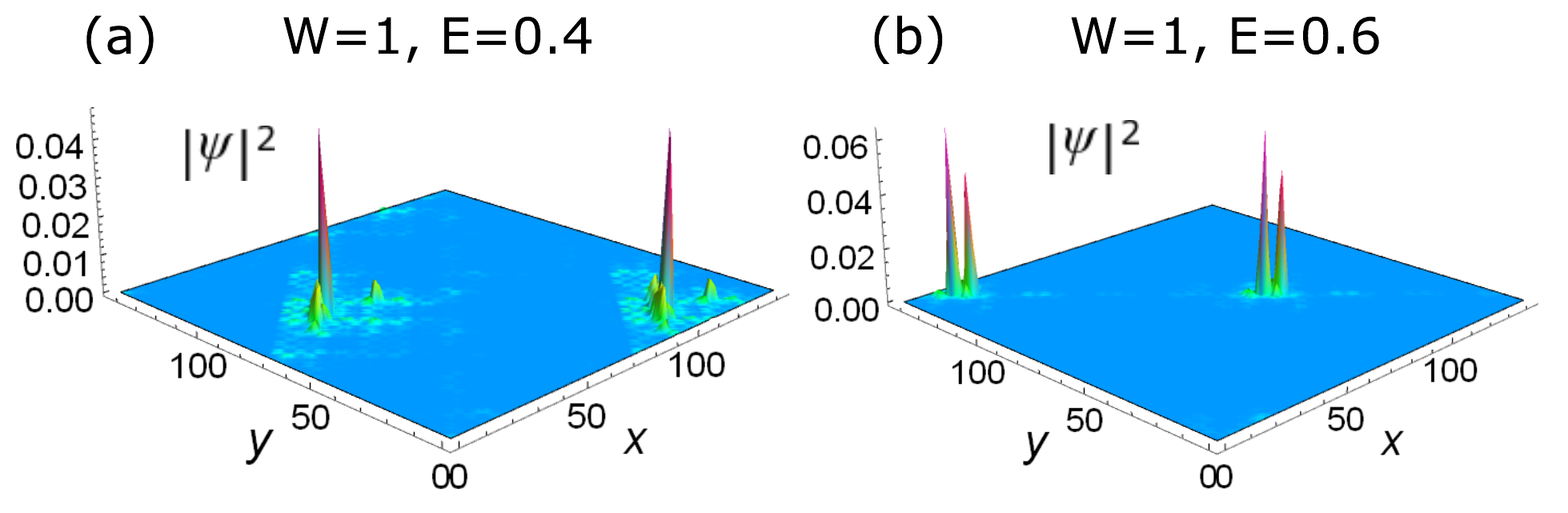}
	\caption{Real-space wavefunctions that show double peaks structure for $W=1$ and certain finite energies [(a) $E=0.4$; (b) $E=0.6$].
		These two wavefunctions correspond to the data in Fig.~\ref{Fig:alpha0_W1} indicated by the black arrows. They are not the conventional localized or frozen wavefunctions that are found in the disordered systems. Such an unconventional feature is probably due to the quasiperiodicity.
	}
	\label{Fig:DP}
\end{figure}

Here, we focus on the pure QP hopping case ($W=1$). As plotted in Fig.~\ref{Fig:rhot_all}(f), both low ($|E| \ll 1$) and finite energy ($|E|\approx 2.2-2.5$) delocalized states still appear in the pure QP hopping limit.
This is very different from the expectation from the disordered problem where all finite-energy states are localized. Therefore, it is important to confirm the detailed features of the finite-energy localized states.

We compute the multifractal exponent $\alpha_0$ (Ref.~\cite{Chhabra1989}, see Appendix.~\ref{App:alpha0}) as an indicator of localization. For a uniformly distributed plane wave, $\alpha_0=d=2$. For a localized state, $\alpha_0\rightarrow \infty$. As shown in Fig.~\ref{Fig:alpha0_W1}, the values of $\alpha_0$ show non-monotonic dependence as a function of energy. We found strongly multifractal delocalized states (intermediate $\alpha_0$ values) in certain finite energies. Importantly, the low-energy states remain delocalized within every measure we have considered so far. In addition, we identify a few delocalized states within the region where the typical DOS is small but finite (near $E\approx 0.5$). Those finite energy wavefunctions consist of two similar peaks with arbitrary separation in $L=144$ as shown in Fig.~\ref{Fig:DP}. We attribute this feature to the QP hopping rather than the (chiral) symmetry of the present model.
Similar features are also presents for larger system sizes ($L=610$), but the associated energy region becomes narrower.
We can not conclude if such states are due to a finite-size effect in the current study.

\begin{figure}
	\centering
	\includegraphics[width=0.45\textwidth]{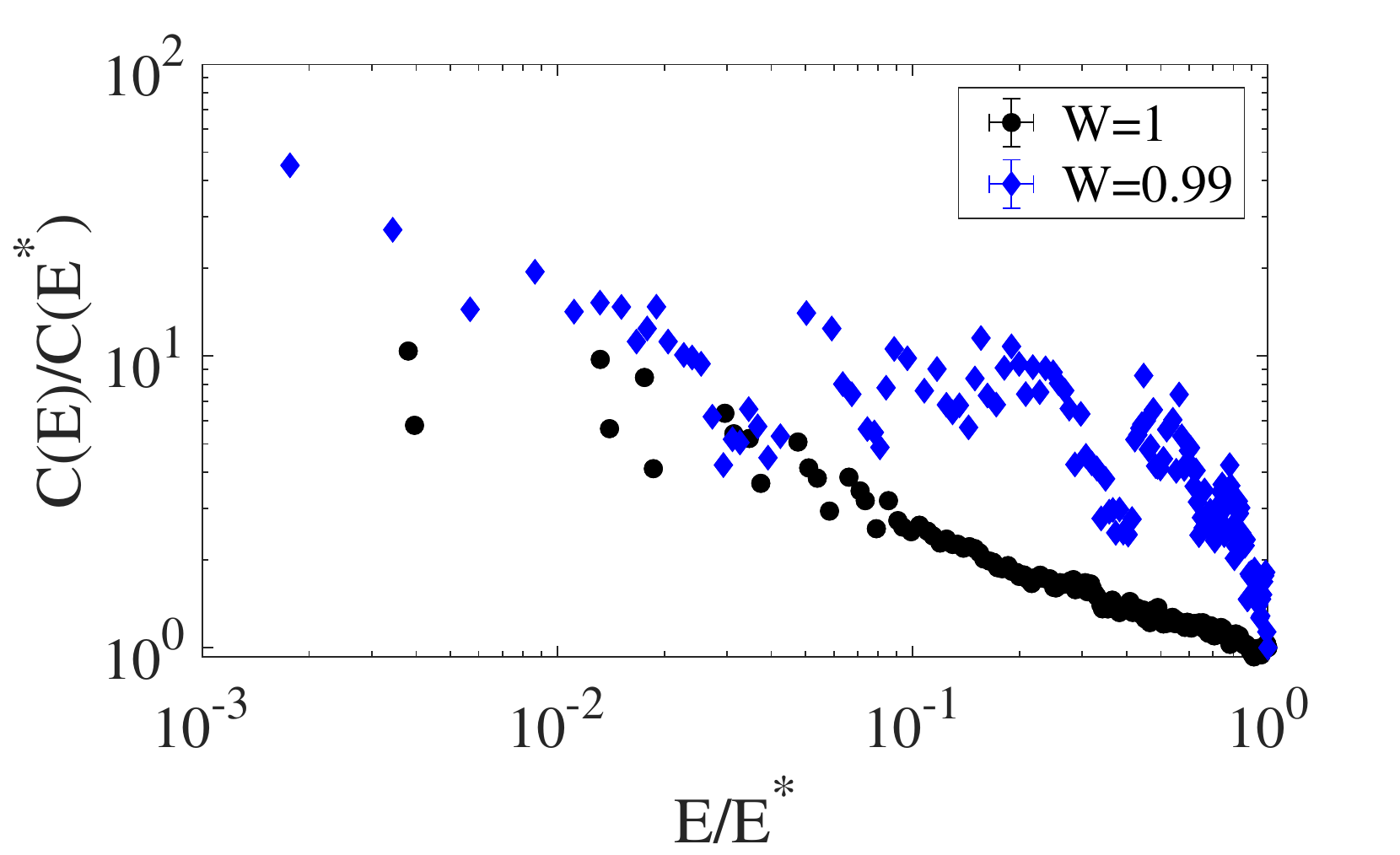}
	\caption{Two-wavefunction correlation [given by Eq.~(\ref{Eq:C_E}) with $E_0\approx0$] as a function of energy ($E$). We take 300 lowest positive energy states of $L=144$ per realization and compute the probability overlap of two wavefunctions in the same realization. The data is averaged over 400 realizations. $E^*=0.01$ for $W=0.99$; $E^*=0.0025$ for $W=1$.
	We rescale all the data points with the rightmost point. In the pure QP hopping limit ($W=1$), the two wavefunction correlation shows a clear power law scaling. For $W=0.99$, the low-energy wavefunctions lose clear power law overlapping features.
	}
	\label{Fig:Ch_Sc}
\end{figure}

We also study the low-energy wavefunctions in a fixed realization. The low-energy wavefunctions are strongly multifractal for $L=144$ and $L=610$.
We compute the two-wavefunction correlation $C(E)$ [given by Eq.~(\ref{Eq:C_E})] to quantify the degrees of probability amplitude overlap. The numerical results of $L=144$ with $W=1$ and $W=0.99$ ($W/J_0\approx7$) are plotted in Fig.~\ref{Fig:Ch_Sc}.
The finite overlap of the wavefunctions with adjacent energies signals the metallic rather than localized behavior and is consistent with our intuitive argument about the hybridizing subregion states. Remarkably, the pure QP hopping ($W=1$) limit gives a power-law behavior,
$
C(E)\sim E^{-\mu},
$
where $\mu\approx 0.48$ for $L=144$. In the disordered problems with a power-law low-energy DOS, the exponent $\mu$ is given by  $\mu=[d-\tau_R(2)]/z$. In the QP hopping model, we are not aware of any scaling argument that supports such a relation. If we assume $\mu=[d-\tau_R(2)]/z$ and compute the $\tau_R(2)$ numerically, the dynamic exponent extracted this way is $z^*\approx 2$, different from the dynamic exponent from low-energy DOS. The discrepancy might come from (a) the sampled energies are not low enough in $C(E)$ or (b) the relation $\mu=[d-\tau_R(2)]/z$ does not hold in this QP hopping model.

The presence of power law correlations in the wavefunctions implies a multifractal enhancement of the interactions \cite{Feigelman2007,Foster2012,BurmistrovMirlin2012,Foster2014}. Unlike the for plane wave states, these multifractal wavefunctions have an intricate spatial probability distribution.
The existence of correlations in energy indicates that the probability distributions of wavefunctions at adjacent energies have significant overlaps. Therefore, we  expect this potentially produces an enhancement of correlated effects for certain types of four-fermion interactions.
In disordered systems, the multifractal enhancement of interactions is related to the wavefunction multifractality directly due to quantum-critical scaling.
The relevance of the four-fermion interaction ($U$) is determined by~\cite{Foster2014}
$dU/dl=x_1-x_2^{(U)}$,
where $x_1=d-z$ is the local DOS exponent and $x_2^{(U)}$ is the scaling dimension of the four fermion operator. In the clean case, the relevance is determined by $x_1$ alone since $x_2^{(U)}=2x_1$. For disorder systems, $x_2^{(U)}\ge x_2$ where $x_2=\tau_R(2)-2(1-x_1)$ is the scaling exponent for the second moment of the local DOS operator after the disorder average has been performed. Nevertheless, it is not currently clear if one can apply the above results to the present QP hopping model at $W=1$; if we do, they imply a strong multifractal enhancement of some short-range interactions (e.g., the density-density interaction).

On the other hand, we do not observe power law correlation in our finite size data for $W=0.99$. This indicates that the power law correlation is a special feature in the pure QP hopping limit.
More quantitative tests (e.g., much larger system sizes) are required to pin down the precise mechanism.

\subsubsection{Wavepacket Dynamics}

\begin{figure}[h!]
	\centering
	\includegraphics[width=0.9\linewidth, angle =0]{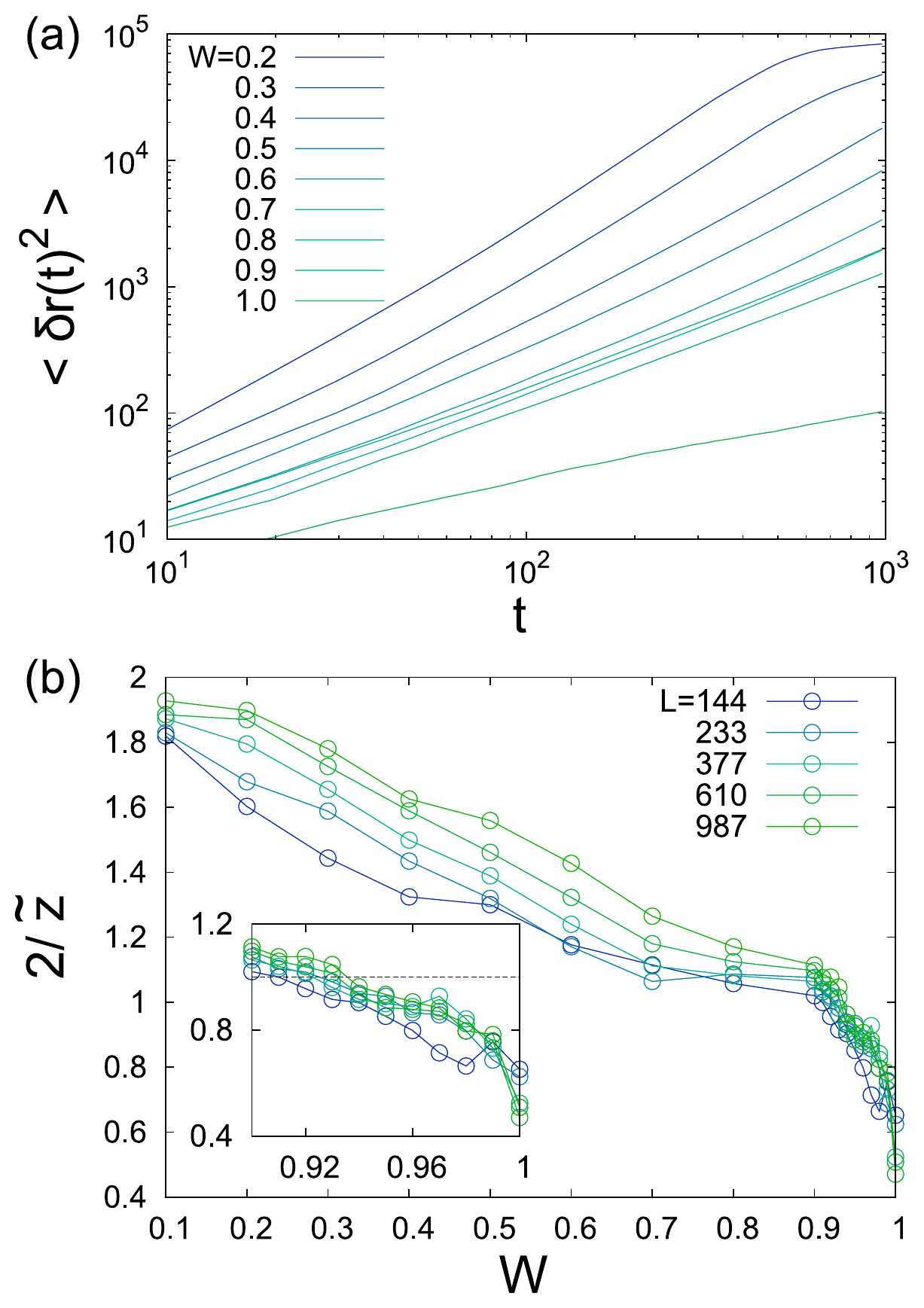}
	\caption{Wave packet dynamics, we initialize the wavefunction to be localized to a single site and evolve it under $H$. (a) Spread of the wavepacket as a function of time $t$ on a log-log scale with $L=987$ and $N_C=2^{13}$ we never see a clear diffusive phase ($z=2$). (b) Extracted dynamic exponent $z$ from $\langle \delta r(t)^2 \rangle\sim t^{2/\tilde z}$ (inset) zoom in near $W=1$ with a dashed line to mark diffusion $2/\tilde z =1$. Note that the wave packet dynamics is not sensitive to the semimetal to metal transition at $E=0$.}
	\label{Fig:rvst}
\end{figure}

Lastly, we now study the wavepacket dynamics in the QP hopping model using an expansion of the time evolution operator in terms of Chebyshev polynomials.
We are interested in
the spread of the wavepacket $\langle\delta r(t)^2\rangle$ in the long-time limit, see Eq.~\eqref{eqn:r2}.
We initialize the state in an up-spin state localized to one lattice site.
Then, we use Eq.~\eqref{eqn:zwp} to extract estimates of an averaged dynamic exponent $\tilde z$ via $\langle \delta r(t)^2 \rangle \sim t^{2/\tilde z}$ as shown in Fig.~\ref{Fig:rvst} for the largest system size $L=987$ considered.  Despite the wave packet dynamics not being energy resolved, for moderate QP strength when a mobility edge is present in the spectrum, the localized states will not contribute and therefore the long-time limit of the wavepacket spreading probes contributions to transport from the ``quickest'' parts of the spectrum. Thus, in the limit of a large QP potential
wavepackets are a good way to probe dynamical transport properties, despite not being energy resolved.

As shown in Fig.~\ref{Fig:rvst} we do not see any clearly diffusive regime in the model (consistent with other QP studies in two-dimensions~\cite{Devakul2017,FuPixley2018}). Instead $2/\tilde z$ smoothly decreases from 2 (for ballistic transport) as a function of the QP hopping strength and the transport looks super-diffusive $1<\tilde z<2$ and $2/\tilde z$ passes through 1 at $W \approx 0.95$. For $W>0.95$ we find $\tilde z>2$ and the transport appears sub-diffusive, approaching $\tilde z \approx 4$ in the pure QP hopping limit.

It is an interesting finding that for the low-energy DOS to diverge requires $z>2$, and our current estimate for $\tilde z$ from the wavepackets yields $\tilde z >2 $ for $W \gtrsim 0.95$. However, the DOS does not appear to have any divergence in this regime (see Fig.~\ref{fig:rho0W1}), which suggests that this feature is due to $\tilde z$ not being energy resolved.
From this perspective, we contrast this estimate of $z$ with that of the divergence in the DOS. From the power law divergence at $W=1$ we estimate from the DOS $z \approx 3$, which is close but does not completely match the wave packet estimate ($\tilde z \approx 4$). However, this is not entirely surprising since  the wave packet estimate gets contributions from states across the spectrum at finite energies
(which possess both finite-energy delocalized and localized states as shown in Fig.~\ref{Fig:rhot_all}),
whereas the DOS is energy resolved and only probes the states near $E=0$.
The presence of finite-energy localized states will slow down the energy averaged transport and give an enhanced value of $\tilde{z}$.
These results suggest that the energy averaged transport properties are sub-diffusive over a range of $W$, while the low-energy states only develop sub-diffusion at $W=1$.

\section{Experimental Realization}
\label{sec:expt}

In this section we present a way to realize Eq.~\eqref{eqn:H} in a cold atomic setup and discuss how to probe the phase diagram. In addition, we also briefly  discuss how the model in Eq.~\eqref{eqn:H}  can be implemented using metamaterials.

We closely follow Ref.~\cite{WuPan2016}, where two-dimensional spin-orbit coupling in ultracold atomic bosonic systems was proposed and experimentally tested.
The continuum version of Eq.~\eqref{eqn:H} has the following form (we consider 2 internal degrees of freedom per atom)
\begin{equation}
H = \frac{\hat p^2}{2m} + V_{\rm latt}(\v x) + \mathcal M_x(\v x) \sigma_x + \mathcal M_y(\v x) \sigma_y. \label{eq:ContSOC}
\end{equation}

The limit of interest is a deep optical potential $V_{\rm latt}(\v x)$, in which spin preserving hopping is suppressed. However, an appropriately designed $\mathcal M_{x,y}(\v x)$ assists spin flip hopping in a certain direction and generates the Hamiltonian of interest.

To realize Eq.~\eqref{eq:ContSOC}, we follow the recent implementation of two-dimensional SOC in Ref.~\cite{WuPan2016}. However,
in contrast to that work, we tune the angle of incidence of the Raman beam and detune the system sufficiently strongly such that the Raman laser (called $E_{2x,z}$ in Ref.~\cite{WuPan2016}) has a wavelength $2\pi/k_2$ which differs from twice the lattice constant $2\pi/k_0$. Then, tuning the optical path such that $\delta \varphi_L = \pi/2$ and $\varphi_L = 0$, we find that $\mathcal M_{x} \propto [\cos(k_0 x) \cos(k_2 y) -\cos(k_0 y) \sin(k_2 x)]$ (and analogously for $x \leftrightarrow y$). For $k_0$ and $k_2$ incommensurate, spin-flip hopping acquires a QP modulation, which in the tight binding limit leads to a Hamiltonian akin to Eq.~\eqref{eqn:H}.

In such a setup, experimental verification of the semimetal to metal transition (where the kinetic energy is quenched, i.e. the ``magic-angle'' effect) as well as a probe of the divergent DOS at $W = 1$ may be achieved using radiofrequency spectroscopy~\cite{ChenLevin2009}.
Within such an experiment, the magic-angle effect of quenched kinetic energy can be observed by means of momentum resolved radiofrequency spectroscopy. As a complementary approach, band mapping techniques~\cite{GreinerEsslinger2001,KoehlEsslinger2005}, allow one to reconstruct the miniband structure experimentally.

Alternatively, metamaterial setups can also realize our model with current experimental techniques. For example, using an array of connected electrical resonators with a suitable choice of the intrinsic frequency and connecting capacitance, one can construct a circuit equivalent to the tight-binding model we have studied here and the overall absorption spectrum is analogous to the DOS \cite{peterson2018resonator,Kollar2019} and thus allows one to probe the semimetal-to-metal transition we have explored here. The spatial distribution of the eigenmodes of resonance can also verify our results regarding localization. Besides resonators, photonic \cite{khanikaev2017photonic} and phononic~\cite{nash2015phononic} systems are also nicely tunable and we also expect that they can be used to engineer the  Hamiltonian in Eq~\eqref{eqn:H} in a majority of the parameter space.

\section{Discussion and Conclusion}
\label{sec:discussion}

We have analyzed the properties of a two-dimensional Dirac semimetal with quasiperiodicity that respects chiral symmetry. The quasiperiodicity takes the form of a QP hopping on a tight-binding model. As shown in Fig.~\ref{Fig1}(a), 
the low-energy states demonstrates a semimetal phase with Dirac cones in the band structure, a chiral metal phase with non-trivial real space structure in the wavefunctions, as well as the pure QP hopping limit $W=1$ [see the paramaterization of $J_0$ in Eq.~\eqref{eqn:J0}], which is critical exhibiting sub-diffusive dynamics. 
A clear demonstration of the semimetal to metal EPT, in the DOS [see Eq.~\eqref{eqn:dos}] and the inverse participation ratio (IPR) in momentum space [see Eq.~\eqref{Eq:Pm_tauq}], is shown in Fig.~\ref{Fig1}(b). The momentum-space IPR (indicating a delocalization in the momentum basis) vanishes in a continuous fashion concomitantly with the onset of the zero-energy DOS, which demonstrates the nature of this phase transition in the structure of the eigenstates and eigenvalues, respectively. In Fig.~\ref{Fig1}(c) we show the diverging DOS in the pure QP hopping limit and we find that the low-energy eigenstates in this regime exhibit quantum-critical Chalker scaling.

First, we demonstrate the stability of the two-dimensional semimetal phase to QP hopping. We find that the QP hopping introduces gaps at finite energy that create a low-energy semimetal miniband that retains the scaling $\rho(E) \sim |E|$. The semimetal phase persists until a critical, $Q$-dependent, potential strength $W_c$ where a semimetal to metal transition takes place. 
At this transition the Dirac velocity vanishes in a universal fashion and the low-energy bands become flat, which should strongly enhance correlation effects and has been dubbed magic-angle transitions in analogy to twisted bilayer graphene at the magic-angle \cite{tbg1,tbg2}. 
Concomitantly, the single-particle wavefunctions delocalize in momentum space. Interestingly, we find that the velocity vanishes with a critical exponent that is in excellent agreement with models that have a QP potential and are lacking chiral symmetry. While these results suggest that the chiral symmetry does not play a role in the critical properties of the semimetal to metal transition, they do have a strong effect on the structure of the phase diagram and the minibandwidth renormalization (being about 4 orders of magnitude smaller then for a QP potential~\cite{FuPixley2018}). 
For example, we find that the metallic phase does not undergo an additional transition back to a reentrant semimetal phase, which occurs in a wide multitude of other models~\cite{FuPixley2018}.
In the metallic phase, we find that the low-energy eigenstates are weakly multifractal in momentum space and wavepacket dynamics are super-diffusive over a large region of the phase diagram ($W<0.95$). Using the chiral symmetry of the model, we characterize this transition and the formation of the low-energy DOS as a band of topological zero modes that form due to bound zero-energy states that arise from a sign-changing Dirac mass \cite{Jackiw1976,Jackiw1981}. If we consider values of $Q$ that are commensurate but are close to the irrational values we have investigated here, then  the single particle phase transition will be rounded into a cross over, which will result in a small but non-vanishing velocity and the momentum-space wavefunctions that do not truly delocalize.

We also investigate the effects of strong quasiperiodicity and therefore determine the real-space Anderson localization properties of this model. We demonstrate that the model exhibits a sequence of real space Anderson localization-delocalization transitions as a function of energy and thus the system hosts multiple mobility edges. Interestingly, the low-energy eigenstates evade exponential localization and appear to remain critical even for maximal QP hopping strength ($W=1$). These results are markedly distinct from disordered systems, where all the finite-energy eigenstates would be localized for the models with real and complex random hopping terms. We verify this non-trivial structure of the phase diagram characterizing real space localization by using a combination of typical density of states and wavefunction analysis.

In the pure QP hopping limit ($W=1$), the system exhibits a diverging DOS at zero energy, see Fig.~\ref{Fig1}(c). We provide evidence that this power-law divergence is universal, for irrational $Q$. The low-energy states that make up this divergence are not exponentially localized, and instead appear strongly multifractal, i.e. critical.
Using wavepacket dynamics we have shown that the majority of the chiral metal phase is super-diffusive and crosses over to sub-diffusion near $W \approx 0.95$. These results are consistent with that the low-energy states are not localized. The slow sub-diffusive wavepacket dynamics gives a dynamical exponent $z\approx 3$.
In addition, we find power-law scaling as a function of energy for almost two decades in the two-wavefunction correlation [see Eq.~\eqref{Eq:C_E}] (in the $W=1$ limit). This provides strong numerical evidence of Chalker scaling without randomness~\cite{Chalker1988,Chalker1990}.
Interestingly, we find Chalker scaling does not clearly hold in the limit of the pure complex quasiperiodic hopping (not shown), demonstrating that the strong correlations between wavefunctions seem to rely on the low-energy diverging DOS in the limit of real quasiperiodic hopping.

\begin{figure}
	\centering
	\includegraphics[width=0.475\textwidth]{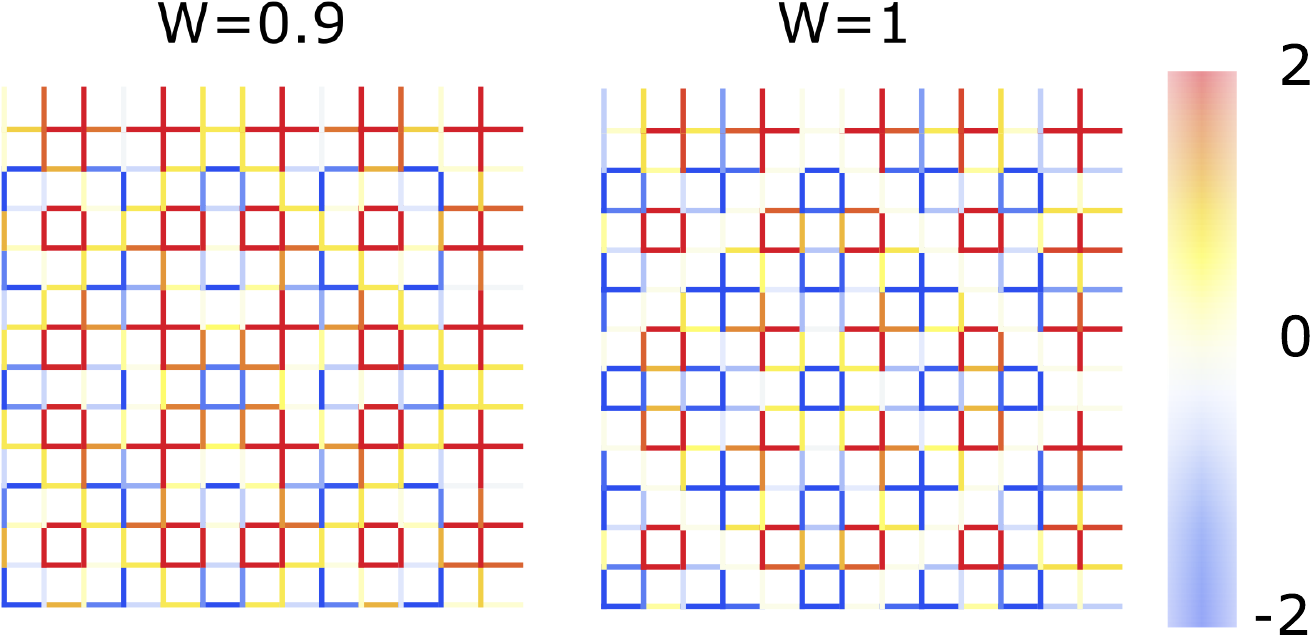}
	\caption{
	The hopping configurations of QP hopping models with $L=13$. (Right) The hopping configuration of the QP hopping model with $W=1$. The QP pattern generates nearly zero lines of bonds which effectively separate the system into many subsystems. (Left) The hopping configuration of the QP hopping model with $W=0.9$. The system is typically well connected as a whole.
	}
	\label{Fig:hopping}
\end{figure}

One remaining important question is to understand the origin of the diverging low-energy DOS for $W=1$. We provide evidence that this is a result of local sub regions with an imbalance $N_A \neq N_B$ of sublattice sites. This induces a pile up of an extensive number of zero modes due to the QP hopping elements vanishing along certain lines in real space.
In Fig.~\ref{Fig:hopping} we plot the configuration of hopping matrix elements in the pure QP hopping model ($W=1$) and strong QP hopping ($W=0.9$).
The pure QP hopping case shows nearly zero hopping lines which  effectively cut the system into many subsystems. Those nearly zero lines roughly track the zeros of the QP hopping, which are obtained by solving $\cos(2\pi Q_L x^*+\phi_x)+\cos(2\pi Q_L y^*+\phi_y)=0$ for $x^*$ and $y^*$.
It is apparent that there are several virtually disconnected subregions in which $N_A - N_B \neq 0$. Those are an imperative origin of zero modes by means of a poor-man's index theorem (rectangular matrices have a non-zero kernel)\cite{InuiAbrahams1994,WeikEvers2016}. To add additional support to this picture we have also studied a model with complex QP hopping amplitudes. This model is chosen to have no lines of vanishing hopping strength as in Fig.~\ref{Fig:hopping}, since the bonds' norms can now never vanish. As shown in Appendix~\ref{sec:cqph}, we find that the complex QP hopping model has no diverging DOS for pure complex hopping. In addition, we also find that this model does not exhibit Chalker scaling. These results lend support to the above argument but are not conclusive and therefore we leave the question of the origin of the pile up of zero energy states at $W=1$ to future work.

Lastly, our work demonstrates two separate routes to inducing strong correlations in quasiperiodic semimetals. The first is due to magic-angle transitions, where the Dirac cone velocity vanishes at an EPT. The second route is due to Chalker scaling in the limit of pure QP hopping. The presence of power-law correlations in the wavefunctions potentially implies a multifractal enhancement of the interactions~\cite{Feigelman2007,Foster2012,BurmistrovMirlin2012,Foster2014}.
Our work provides a clear cut example of how this can occur in the absence of randomness.

\section*{Acknowledgment}

We thank Sarang Gopalakrishnan, David Huse, Alexander  Mirlin, Rahul Nandkishore, and Zhentao Wang for useful discussions. In particular, we thank Matthew Foster for suggesting to us to look into Chalker scaling as well as for numerous insightful discussions.
Y.-Z.C. was sponsored in part by the Army Research Office and was accomplished under Grant Number W911NF-17-1-0482 and by a Simons Investigator award from the Simons Foundation to Leo Radzihovsky.
J.H.P. and J.H.W. performed part of this work at the Aspen Center for Physics, which is supported by NSF Grant No. PHY-1607611, and J.H.P. at the Kavli Institute for Theoretical Physics, which is supported by NSF Grant No. PHY-1748958.
E.J.K acknowledges support by the U.S. Department of Energy (DOE), Office of Basic
Energy Sciences (BES), under Award No. DE-FG02-
99ER45790.
The authors acknowledge the Beowulf cluster at the Department of Physics and Astronomy of Rutgers University and the Office of Advanced Research Computing (OARC) at Rutgers, The State University of New Jersey (http://oarc.rutgers.edu) for providing access to the Amarel cluster and associated research computing resources that have contributed to the results reported here.
The views and conclusions contained in this document are those of the authors and should not be interpreted as representing the official policies, either expressed or implied, of the Army Research Office or the U.S. Government. The U.S. Government is authorized to reproduce and distribute reprints for Government purposes notwithstanding any copyright notation herein.

\appendix

\section{Complex quasiperiodic hopping model}
\label{sec:cqph}
As a comparison to real QP hopping, we also introduce a complex QP hopping model. The complex QP hopping model is realized by introducing complex hopping matrix elements to replace Eq.~\eqref{eqn:hop} with
\begin{equation}
J_{\mu}(\vex{r}) =W\sum_{\nu=x,y}\exp \left[iQ_L \left(r_{\nu}+\hat{\mu}\cdot\hat{\nu}/2\right)+i\phi_{\nu}\right],
\label{eqn:hop_complex}
\end{equation}
where $J_{\mu}(\vex{r})$ is the QP hopping amplitude between site $\vex{r}$ and $\vex{r}+\hat\mu$. In the pure complex QP hopping limit, the bonds are non-zero almost everywhere in contrast to the ``nearly cut lines'' as plotted in Fig.~\ref{Fig:hopping} for real QP hopping with $W=1$. Therefore, we expect that the low-energy physics in the complex QP hopping model with $W=1$ should be distinct from the real hopping model with $W=1$ discussed in the main text.

\begin{figure}[b]
	\centering
	\includegraphics[width=0.3\textwidth,angle=-90]{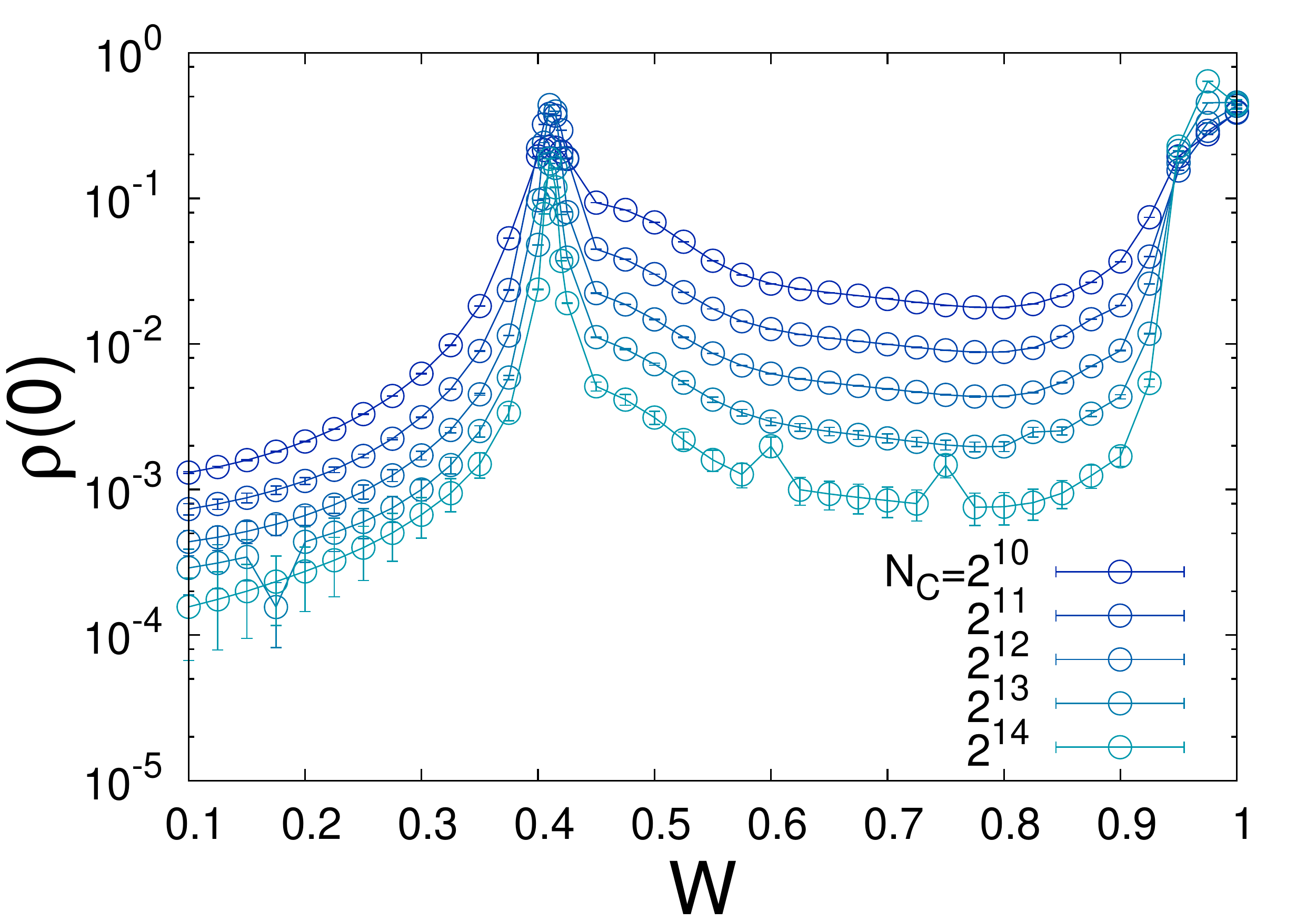}
	\caption{The zero energy density of states as a function of $W$ in the complex hopping model for various KPM expansion orders $N_C$ and a system size of $L=233$. We find a small metallic phase near $W\approx 0.41$. }
	\label{Fig:rho0complex}
\end{figure}

The zero energy density of states in this model is shown in Fig.~\ref{Fig:rho0complex} for various KPM expansion orders. We find a small metallic phase near $W \approx 0.41$, which transitions back into a reentrant semimetal phase, which is distinct from the case of real QP hopping (see Fig.~\ref{Fig1}). The metallic phases are clear from where the data is (roughly) $N_C$ independent. We also find a second semimetal to metal transition at a larger $W\approx 0.93$, and the zero energy DOS does not look divergent at $W=1$.

\begin{figure}[b]
	\centering
	\includegraphics[width=0.3\textwidth,angle=-90]{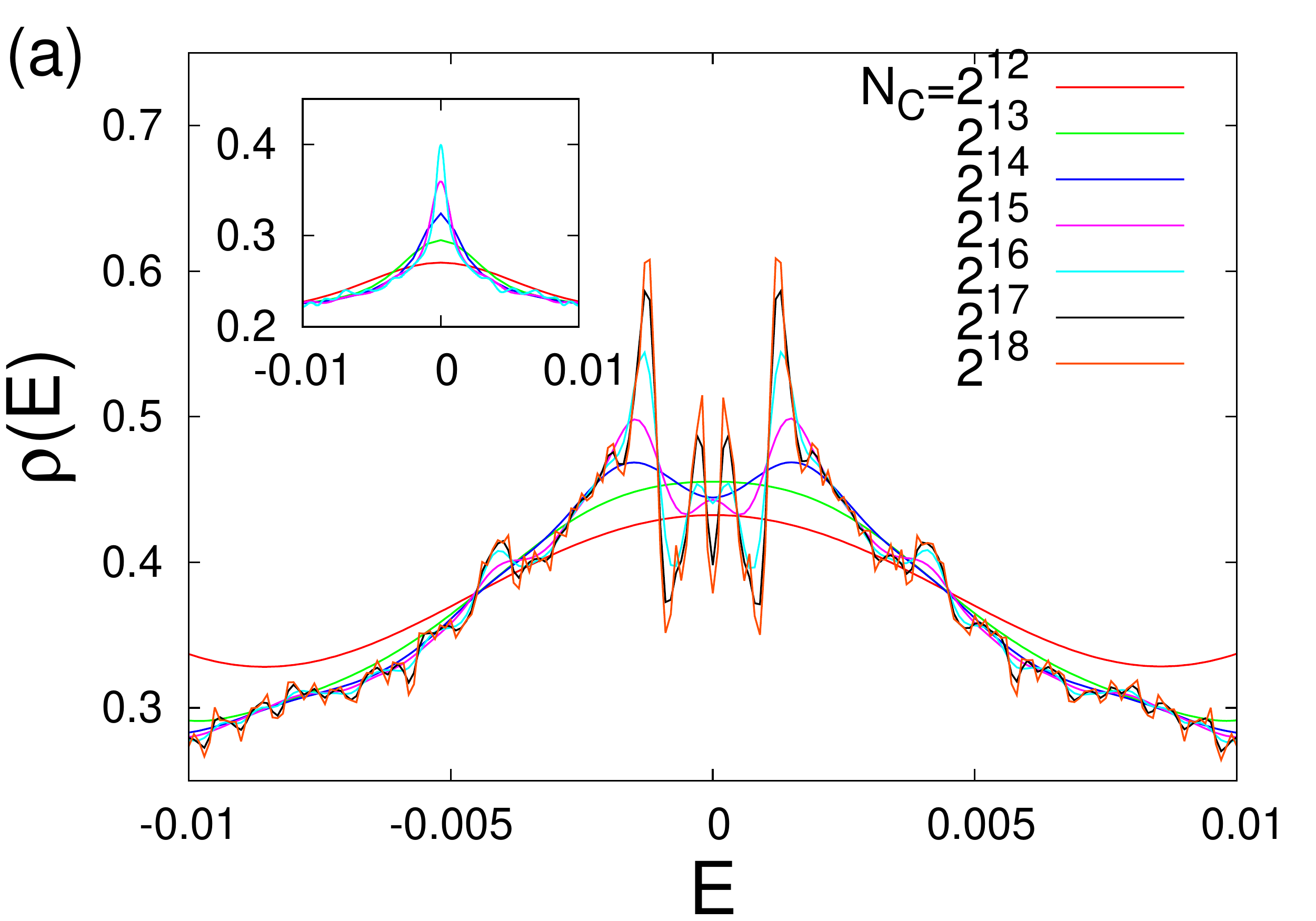}
	\includegraphics[width=0.3\textwidth,angle=-90]{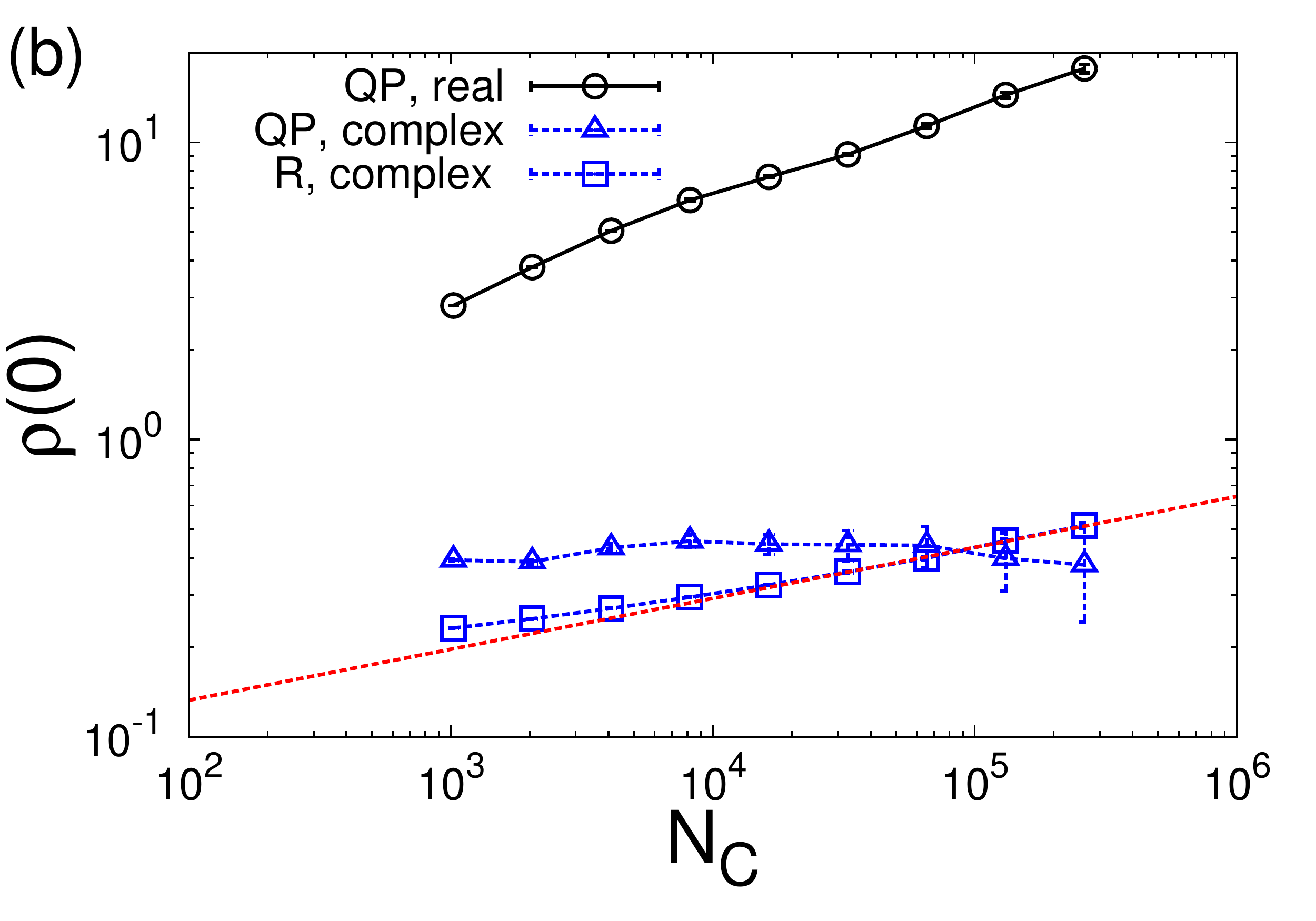}
	\caption{The lack of a divergence of the low-energy DOS for $W=1$ i.e. pure QP complex hopping model. (a) $N_C$ dependence near zero energy for a  system size $L=233$ and $Q_L=2 \pi F_{n-2}/F_n$. (Inset) Similar results for the randomized version of the model (letting the phase be random at each site) with $L=233$, displaying a clear divergence unlike the complex QP hopping model. (b) The zero energy DOS for $W=1$ in the pure QP limit comparing real and complex QP hopping with the complex random (R) hopping model. The complex QP hopping model clearly has no low-energy divergence in the DOS. The KPM expansion order that acts like a low-energy scale that rounds out the divergence of the DOS in the random model and the red line is a fit to the power law form $\rho(E=0) \sim (N_C)^{y_R}$. }
	\label{Fig:Div-complex}
\end{figure}

Interestingly, we find that the existence of a reentrant phase is consistent with our zero mode analysis. As mentioned in Sec.~\ref{sec:cm} for complex QP hopping, similar to the case of a QP potential, the zero mode solution is not topologically protected. This implies that for the case of real QP hopping, the model cannot return to the semimetal phase due to a stable proliferation of overlapping zero modes.  Whereas in the complex QP hopping model there is no band of zero modes and thus the model can in principle return to the semimetallic phase as in the case of the QP potential model.

\begin{figure}
	\centering
	\includegraphics[width=0.45\textwidth]{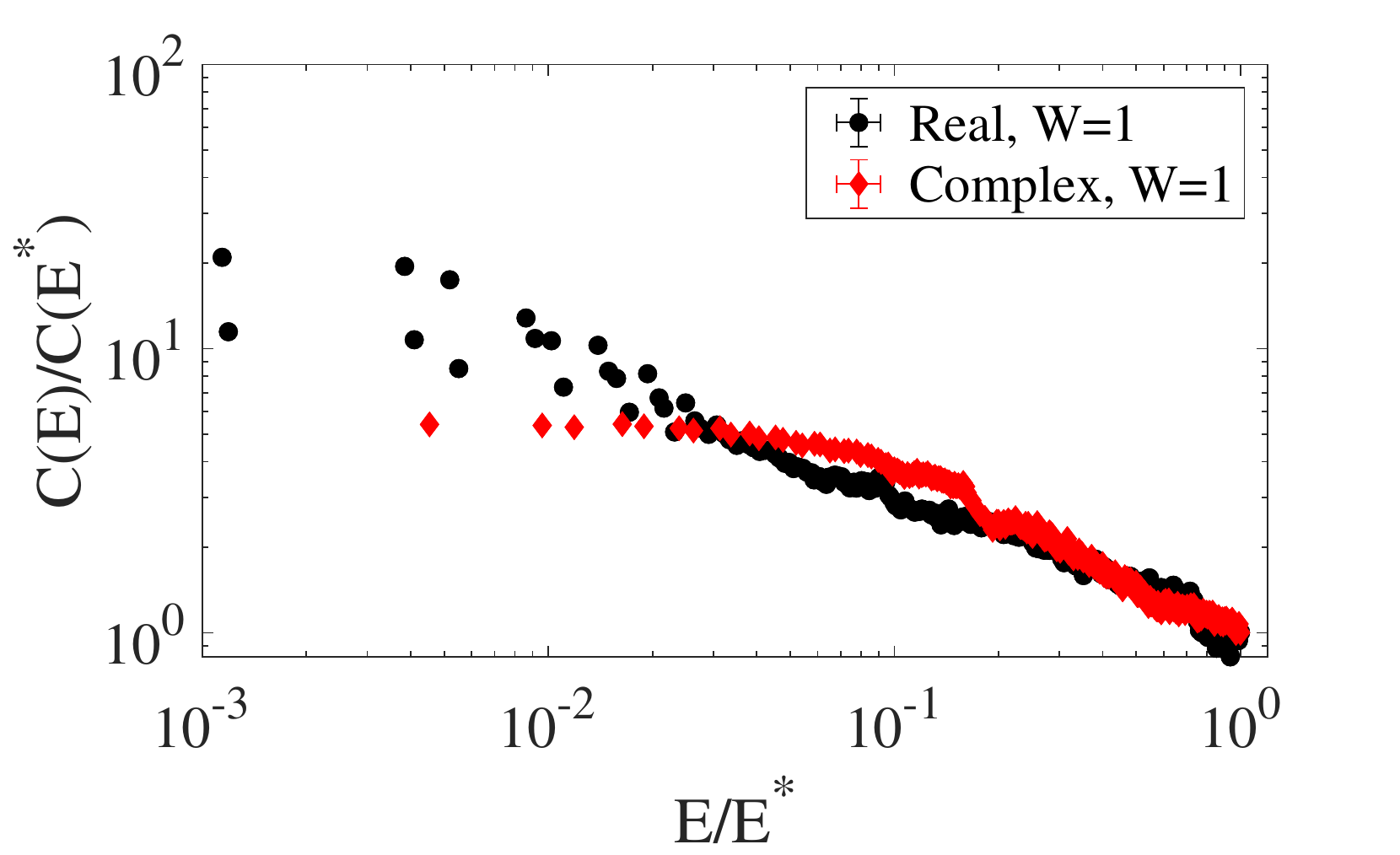}
	\caption{Two-wavefunction correlation [given by Eq.~(\ref{Eq:C_E}) with $E_0=0$] as a function of energy ($E$) for the QP hopping model and for the complex QP hopping model with $W=1$. We take 600 lowest positive energy states of $L=144$ per realization and compute the probability overlap of two wavefunctions in the same realization. The data is averaged over 400 realizations. $E^*\approx 0.01$ for the real hopping model; $E^*\approx 0.1$ for the complex hopping model. We rescale all the data points with the rightmost point. The real QP hopping model shows a power law scaling persisting around two decades. The complex QP hopping model (red diamond) does not exhibit a clear power law scaling behavior.
	}
	\label{Fig:Ch_Sc_RvsC}
\end{figure}

Lastly we turn to the pure complex QP hopping model, i.e. at $W=1$. As shown in Fig.~\ref{Fig:Div-complex}, we find that the complex QP hopping model does not have a low-energy divergence. However, if we randomize the model, by letting the $\phi_{\nu}$ in Eq.~\eqref{eqn:hop_complex} be random at each site then we find that the divergence returns as we would expect for the random model~\cite{Motrunich2002}.
The random model has the low-energy divergence given by $\rho(0) \sim (N_C)^{y_R}$ with $y_R \approx 0.17$, which is half of the value of the random real hopping model. Thus, the complex QP hopping model is an example of a system that has no broken bonds since their norm is always non-zero and the low-energy DOS does not diverge, whereas its random counterpart has a DOS that does diverge. To complete this analysis we test for Chalker scaling from Eq.~\eqref{Eq:C_E} in the complex QP hopping model at $W=1$ as shown in Fig.~\ref{Fig:Ch_Sc_RvsC}. While the regime of power law scaling extends over about two decades of energy in the real QP hopping model we do not find clear evidence of a power-law scaling with energy in the complex QP hopping model. Thus we conclude that the complex QP hopping model does not have Chalker scaling.

\section{Analytical calculations}
\label{app:LowEnergy}
\subsection{Perturbative Velocity Renormalization}
\label{App:Velocity}

To second order in perturbation theory, it is sufficient to consider the truncated effective Hamiltonian
\begin{equation}
H_{\rm eff} =  \left (\begin{array}{ccccc}
h_0 & W_{x,+} & W_{x,-} & W_{y,+} & W_{y,-} \\
W_{x,+} & h_{x,+} & 0 & 0 & 0 \\
W_{x,-} & 0 & h_{x,-} & 0 & 0 \\
W_{y,+} & 0 & 0 & h_{y,+} & 0\\
W_{y,-} & 0 & 0 & 0 & h_{y,-}
\end{array} \right )
\end{equation}

We introduced the notation $h_0 = 2J_0 [\sin(k_x) \sigma_x + \sin(k_y) \sigma_y]$, $h_{x,\pm} = 2J_0 [\sin(k_x \pm Q) \sigma_x + \sin(k_y) \sigma_y]$, $h_{y,\pm} = 2J_0 [\sin(k_x) \sigma_x + \sin(k_y \pm Q) \sigma_y]$, and
$W_{x,\pm} = W[ \sigma_x \sin(k_x \pm Q/2) +\sigma_y \sin(k_y)]$ and $W_{y,\pm} = W [\sigma_x \sin(k_x)  +\sigma_y \sin(k_y \pm Q/2) ]$.

The perturbative calculation of the self energy near the $\Gamma$ point leads to
\begin{equation}
\Sigma \simeq \frac{W^2}{J_0^2} \left [-\frac{E}{2[1+\cos(Q)]} - J_0 \slashed{p} \frac{1 + 2 \sec(Q/2)}{2} \right ].
\end{equation}
The velocity renormalization, Eq.~\eqref{eqn:v}, immediately follows.

\subsection{Topological bound states in the effective low-energy theory.}

Here we map the problem to the dominant low-energy physics near the Dirac nodes. The translationally invariant Hamiltonian
may be expanded and, in first quantization, takes the form
\begin{equation}
H_0 = \left (\begin{array}{cccc}
v_0 \slashed p & 0 & 0 & 0 \\
0 & -v_0 \slashed p & 0 & 0 \\
0 & 0 & -v_0 \slashed p^* & 0 \\
0 & 0 & 0 & v_0 \slashed p^*
\end{array} \right ) \label{eq:HDirac1}
\end{equation}
where $v_0 = 2J_0$,  each element is a two-by-two matrix with $\slashed p=p_x\sigma_x+p_y\sigma_y$, and each column represents a different Dirac point in momentum space: the $\Gamma$ $(0,0)$, $M$ $(\pi,\pi)$, $X$ $(\pi,0)$, and $Y$ $(\pi, 0)$ points.
For $Q = 2\pi [2/(\sqrt{5}+1)]^2$, the most important low-energy processes are have momentum transfer $Q$ (close to $\pi$) and $4Q$ (close to $3\pi$), both of which connect Dirac points either vertically or horizontally (diagonal coupling is included by higher order processes in the Hamiltonian that is about to be derived. $2Q$ and $3Q$ processes are virtual processes that we integrate out.

The off diagonal components of the self energy introduce
\begin{equation}
V =  \left (\begin{array}{cccc}
0 & 0 & -i V(x) \sigma_x & - i V(y) \sigma_y \\
0 & 0 &  i  V(y) \sigma_y & i V(x) \sigma_x \\
i V(x) \sigma_x & -i V(y) \sigma_y  & 0 & 0 \\
i V(y) \sigma_y & -i V(x) \sigma_x  & 0 & 0 \\
\end{array} \right ). \label{eq:PotDirac1}
\end{equation}
The function $V(x)$ is defined in Eq.~\eqref{eq:VofX}
There are also terms of higher-order in gradients that we omitted (i.e. terms with both $p$ and $x$ dependence). Note that the chiral symmetry $\lbrace H, \sigma_z \rbrace =0$ is preserved.

It is instructive to rotate the Hamiltonian by means of $U = \text{diag}(\mathbf 1, - i \sigma_z, \sigma_x, \sigma_y)$ so that the effective low-energy Hamiltonian may be written as
\begin{widetext}
\begin{equation}{
H = \left (\begin{array}{cccc}
v_0 \slashed p & 0 & 0 & 0 \\
0 & v_0 \slashed p & 0 & 0 \\
0 & 0 & -v_0 \slashed p & 0 \\
0 & 0 & 0 & -v_0 \slashed p
\end{array} \right ) +  \left (\begin{array}{cccc}
0 & 0 & -i V(x)  & - i V(y)\\
0 & 0 &  i  V(y) & -i V(x)  \\
i V(x) & -i V(y)  & 0 & 0 \\
i V(y) & i V(x)   & 0 & 0 \\
\end{array} \right ). \label{eq:HDirac}}
\end{equation}
\end{widetext}
We remind ourselves of the matrix structure of this $8\times 8$ matrix: The diagonal kinetic parts reflect $\Gamma, M, X,Y$ points (in this order).
We can compactly write
\begin{equation}
H = v_0 \slashed p \lambda_z + V(x) \lambda_y + V(y) \tau_y \lambda_x.
\end{equation}
Here, $\tau$ are Pauli matrices within $\Gamma - M$ (or $X-Y$) blocks of equal winding in Eq.~\eqref{eq:HDirac}, while $\lambda$ matrices connect these blocks. Since only $\tau_y$ and $\mathbf 1_\tau$ appear, we may diagonalize in $\tau$ (i.e. choose wave functions with equal weight at, e.g. $\Gamma$ and $M$ points).
This leads to the direct sum of two $4\times 4$ Hamiltonians presented in Eq.~\eqref{eq:HDirac4},
i.e. the approximate low-energy theory is the theory of two two-dimensional Dirac electrons coupled by two incommensurate to one another off-diagonal terms.

The involved matrices $\gamma_1 = \sigma_{x} \lambda_z, \gamma_2 = \sigma_{y} \lambda_z,  \gamma_3 = \lambda_y, \gamma_4 = \lambda_x$ form a Clifford algebra. In particular, it therefore follows that the zero energy wave function $h_\pm \Psi_{0}(\v x) = 0$ can be found via the usual Ansatz
\begin{equation}
\Psi_0(\v x)  = e^{- [\int^x dx' V(x') \sigma_x \lambda_x \mp \int^y dy' V(y') \sigma_y \lambda_y]/v_0}\Phi.
\end{equation}
Here, the position independent four spinor $\Phi$ is constraint by the normalizability condition (ultimately, by the wish of having maximum weight at $V(x)V(y) = 0$.) For example, focusing on the node at $x = y = 0$ and the model at $Q = 2\pi[2/(\sqrt{5}+1)]^2$, we obtain
\begin{equation}
\Psi_0(\v x) \simeq e^{- \frac{V_1 q_1 + V_4 q_4}{2 v} (x^2 \sigma_x \lambda_x \mp y^2\sigma_y \lambda_y)}\Phi
\end{equation}
 We defined $q_1 = \pi - Q$ and $q_4 = 4 Q- 3\pi$. Normalizability then implies $\Phi_+ \propto (1,0,0,1)$ ($\Phi_- \propto (0,1,1,0)$) for $h_+$ ($h_-$), such that the eigenvalues of $\sigma_x \lambda_x$ and $\mp \sigma_y \lambda_y$ are both 1. Keeping the whole system, this leads to a wave function given in Eq.~\eqref{eq:ZeroMode}.

\section{Multifractal Exponent $\alpha_0$}\label{App:alpha0}

Here, we define the multifractal exponent $\alpha_0$ which is employed for characterizing the localization properties in the main text. The $\alpha_0$ can be computed via numerical Legendre transformation of $\tau(q)$. Instead, we use the method by Chhabra and Jensen \cite{Chhabra1989} to compute $\alpha_0$. For a real-space wavefunction $\psi(\vex{x})$, we define \cite{Chhabra1989}
\begin{align}
\mu^{(q)}_{\vex{x}}=&\frac{|\psi(\vex{x})|^{2q}}{\sum_{\vex{x}}|\psi(\vex{x})|^{2q}},\\
\label{Eq:f_q}f_q=&\frac{\sum_{\vex{x}}\mu^{(q)}_{\vex{x}}\ln\mu^{(q)}_{\vex{x}}}{-\ln(L^2) },\\
\label{Eq:alpha_q}\alpha_q=&\frac{\sum_{\vex{x}}\mu^{(q)}_{\vex{x}}\ln|\psi(\vex{x})|^2}{-\ln(L^2) },
\end{align}
where $f_q$ and $\alpha_q$ form the singularity multifractal spectrum.
The multifractal exponent $\alpha_0$ corresponds to $q=0$ in Eq.~(\ref{Eq:alpha_q}). In the $f(\alpha)$ spectrum, the most probably value of the probability density is given by $|\psi(\vex{x})|\sim L^{-\alpha_0}$. For plane wave states, $\alpha_0=d$ due to the uniform distributing nature. For a localized state, $\alpha_0\rightarrow\infty$ as all the probability densities are vanishingly small except the localized peak.

One can extend the above definition with the binned wavefunction $\psi_b(\vex{X}_j)$ [defined in Sec.~\ref{Sec:eigenstates}] in order to test the robustness of the results.

\section{Quadrupole topological insulator at commensurate limits of the model}
\label{app:QTI}

\begin{figure}
    \centering
    \includegraphics[width=.95\columnwidth]{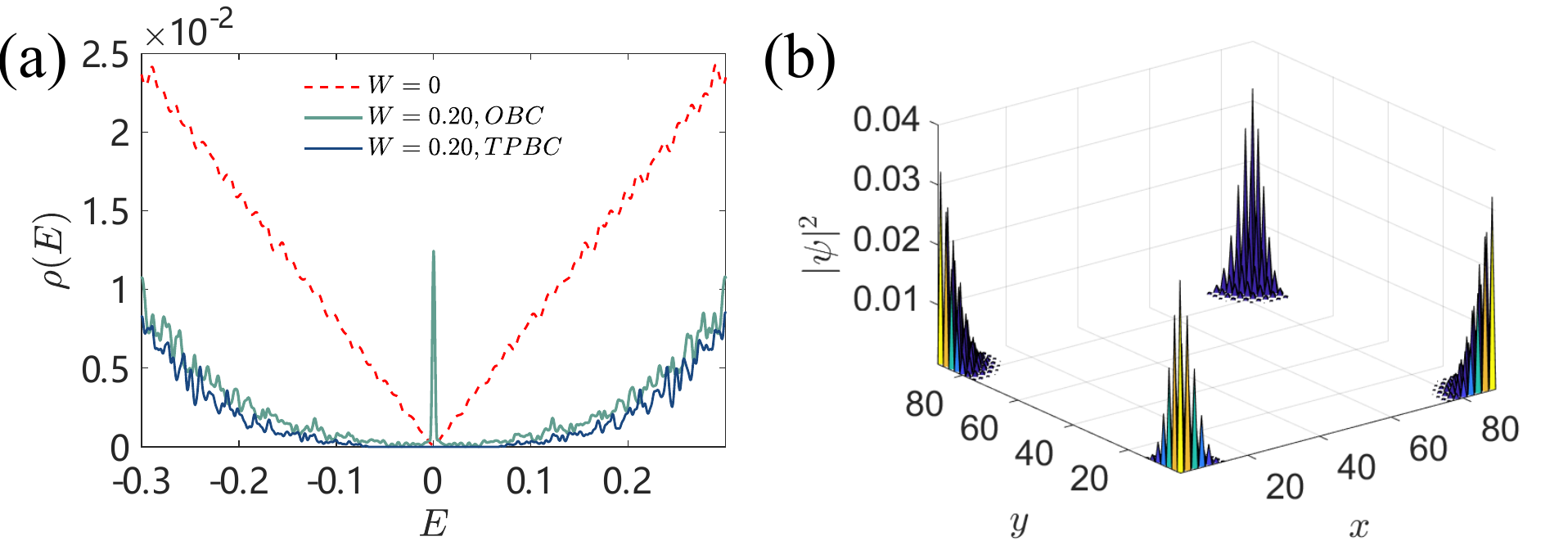}
    \caption{(a) Density of state by energy, with $Q_L=\pi$, in twisted periodic boundary condition (TPBC) and open boundary condition (OBC). Both boundary conditions show bulk gap, while OBC allows the topological corner states. The system size is $L=144$, and $Q_L=2\pi (72/L)$. $N_C=8192$ for KPM calculations. (b) real space wave function
    at $Q_L=\pi$ and $W=0.4$. System size is $L=89$. }
    \label{fig:HOTI}
\end{figure}

As we already discussed in the main text, for $Q=\pi$, the model in Eq.~\eqref{eqn:H} is a quadrupole topological insulator~\cite{benalcazar2017HOTI}. In this case, the model can be separated into two copies of decoupled $\pi$ flux model by alternating spin. For each copy, four lattice sites on the corners of a plaquette form a unit cell when $Q=\pi$. We label them from the left-bottom corner as $|1\uparrow\rangle$, $|3\downarrow\rangle$, $|4\uparrow\rangle$ and $|2\downarrow\rangle$ counterclockwise (and opposite spin labels for the other copy). The Bloch Hamiltonian is given by $h(\vex{k}) = W(\cos(k_x)\tau_x\sigma_0 -\sin(k_x)\tau_y\sigma_x - \cos(k_y)\tau_y\sigma_y-\sin(k_y)\tau_y\sigma_x) +E_0(\vex{k})\tau_z\sigma_0$, where $\sigma, \tau$ are Pauli matrices that act on the degrees of freedom within a unit cell, with identical/opposite spin respectively.
The dispersion with $W=0$ is $E_0({\bf k}) = \pm 2 J_0 \sqrt{\sin k_x^2 + \sin k_y^2}$. 

For $W>0$, we see a hard gap near $E=0$. When $L$ is odd with twisted periodic boundary condition, or $L$ is even with open boundary condition, a small peak is seen at $E=0$ [Fig.~\ref{fig:HOTI}(a)]. When $L$ is even and taking closed boundary condition, the corner state do not show up. The corner states survive twisted periodic boundary condition when $L$ is odd because the unit cell has size $2\times 2$, and hence a strip of half unit cells opens the boundary. The peak includes two states, independent of what $L$ is chosen to calculate the DOS, indicating a topological nature of such a peak. The wavefunction data shown in Fig.~\ref{fig:HOTI} (b) also indicates that the system is in a quadrupole TI phase since the zero-energy wavefunction concentrates near the corners.


\begin{thebibliography}{77}%
	\makeatletter
	\providecommand \@ifxundefined [1]{%
		\@ifx{#1\undefined}
	}%
	\providecommand \@ifnum [1]{%
		\ifnum #1\expandafter \@firstoftwo
		\else \expandafter \@secondoftwo
		\fi
	}%
	\providecommand \@ifx [1]{%
		\ifx #1\expandafter \@firstoftwo
		\else \expandafter \@secondoftwo
		\fi
	}%
	\providecommand \natexlab [1]{#1}%
	\providecommand \enquote  [1]{``#1''}%
	\providecommand \bibnamefont  [1]{#1}%
	\providecommand \bibfnamefont [1]{#1}%
	\providecommand \citenamefont [1]{#1}%
	\providecommand \href@noop [0]{\@secondoftwo}%
	\providecommand \href [0]{\begingroup \@sanitize@url \@href}%
	\providecommand \@href[1]{\@@startlink{#1}\@@href}%
	\providecommand \@@href[1]{\endgroup#1\@@endlink}%
	\providecommand \@sanitize@url [0]{\catcode `\\12\catcode `\$12\catcode
		`\&12\catcode `\#12\catcode `\^12\catcode `\_12\catcode `\%12\relax}%
	\providecommand \@@startlink[1]{}%
	\providecommand \@@endlink[0]{}%
	\providecommand \url  [0]{\begingroup\@sanitize@url \@url }%
	\providecommand \@url [1]{\endgroup\@href {#1}{\urlprefix }}%
	\providecommand \urlprefix  [0]{URL }%
	\providecommand \Eprint [0]{\href }%
	\providecommand \doibase [0]{http://dx.doi.org/}%
	\providecommand \selectlanguage [0]{\@gobble}%
	\providecommand \bibinfo  [0]{\@secondoftwo}%
	\providecommand \bibfield  [0]{\@secondoftwo}%
	\providecommand \translation [1]{[#1]}%
	\providecommand \BibitemOpen [0]{}%
	\providecommand \bibitemStop [0]{}%
	\providecommand \bibitemNoStop [0]{.\EOS\space}%
	\providecommand \EOS [0]{\spacefactor3000\relax}%
	\providecommand \BibitemShut  [1]{\csname bibitem#1\endcsname}%
	\let\auto@bib@innerbib\@empty
	\bibitem [{\citenamefont {Sachdev}(2007)}]{Sachdev-book}%
	\BibitemOpen
	\bibfield  {author} {\bibinfo {author} {\bibfnamefont {S.}~\bibnamefont
			{Sachdev}},\ }\href@noop {} {\emph {\bibinfo {title} {Quantum phase
				transitions}}}\ (\bibinfo  {publisher} {Wiley Online Library},\ \bibinfo
	{year} {2007})\BibitemShut {NoStop}%
	\bibitem [{\citenamefont {Goldenfeld}(1992)}]{Goldenfeld-book}%
	\BibitemOpen
	\bibfield  {author} {\bibinfo {author} {\bibfnamefont {N.}~\bibnamefont
			{Goldenfeld}},\ }\href@noop {} {\emph {\bibinfo {title} {Lectures on phase
				transitions and the renormalization group}}}\ (\bibinfo  {publisher}
	{Addison-Wesley, Advanced Book Program, Reading},\ \bibinfo {year}
	{1992})\BibitemShut {NoStop}%
	\bibitem [{\citenamefont {Anderson}(1958)}]{Anderson1958}%
	\BibitemOpen
	\bibfield  {author} {\bibinfo {author} {\bibfnamefont {P.~W.}\ \bibnamefont
			{Anderson}},\ }\href {\doibase 10.1103/PhysRev.109.1492} {\bibfield
		{journal} {\bibinfo  {journal} {Phys. Rev.}\ }\textbf {\bibinfo {volume}
			{109}},\ \bibinfo {pages} {1492} (\bibinfo {year} {1958})}\BibitemShut
	{NoStop}%
	\bibitem [{\citenamefont {Abrahams}\ \emph {et~al.}(1979)\citenamefont
		{Abrahams}, \citenamefont {Anderson}, \citenamefont {Licciardello},\ and\
		\citenamefont {Ramakrishnan}}]{Abrahams-1979}%
	\BibitemOpen
	\bibfield  {author} {\bibinfo {author} {\bibfnamefont {E.}~\bibnamefont
			{Abrahams}}, \bibinfo {author} {\bibfnamefont {P.~W.}\ \bibnamefont
			{Anderson}}, \bibinfo {author} {\bibfnamefont {D.~C.}\ \bibnamefont
			{Licciardello}}, \ and\ \bibinfo {author} {\bibfnamefont {T.~V.}\
			\bibnamefont {Ramakrishnan}},\ }\href {\doibase 10.1103/PhysRevLett.42.673}
	{\bibfield  {journal} {\bibinfo  {journal} {Phys. Rev. Lett.}\ }\textbf
		{\bibinfo {volume} {42}},\ \bibinfo {pages} {673} (\bibinfo {year}
		{1979})}\BibitemShut {NoStop}%
	\bibitem [{\citenamefont {Lee}\ and\ \citenamefont
		{Ramakrishnan}(1985)}]{Lee-1985}%
	\BibitemOpen
	\bibfield  {author} {\bibinfo {author} {\bibfnamefont {P.~A.}\ \bibnamefont
			{Lee}}\ and\ \bibinfo {author} {\bibfnamefont {T.~V.}\ \bibnamefont
			{Ramakrishnan}},\ }\href {\doibase 10.1103/RevModPhys.57.287} {\bibfield
		{journal} {\bibinfo  {journal} {Rev. Mod. Phys.}\ }\textbf {\bibinfo {volume}
			{57}},\ \bibinfo {pages} {287} (\bibinfo {year} {1985})}\BibitemShut
	{NoStop}%
	\bibitem [{\citenamefont {Evers}\ and\ \citenamefont
		{Mirlin}(2008)}]{Evers2008_RMP}%
	\BibitemOpen
	\bibfield  {author} {\bibinfo {author} {\bibfnamefont {F.}~\bibnamefont
			{Evers}}\ and\ \bibinfo {author} {\bibfnamefont {A.~D.}\ \bibnamefont
			{Mirlin}},\ }\href {\doibase 10.1103/RevModPhys.80.1355} {\bibfield
		{journal} {\bibinfo  {journal} {Rev. Mod. Phys.}\ }\textbf {\bibinfo {volume}
			{80}},\ \bibinfo {pages} {1355} (\bibinfo {year} {2008})}\BibitemShut
	{NoStop}%
	\bibitem [{\citenamefont {Basko}\ \emph {et~al.}(2006)\citenamefont {Basko},
		\citenamefont {Aleiner},\ and\ \citenamefont {Altshuler}}]{Basko2006}%
	\BibitemOpen
	\bibfield  {author} {\bibinfo {author} {\bibfnamefont {D.}~\bibnamefont
			{Basko}}, \bibinfo {author} {\bibfnamefont {I.}~\bibnamefont {Aleiner}}, \
		and\ \bibinfo {author} {\bibfnamefont {B.}~\bibnamefont {Altshuler}},\ }\href
	{\doibase http://dx.doi.org/10.1016/j.aop.2005.11.014} {\bibfield  {journal}
		{\bibinfo  {journal} {Annals of Physics}\ }\textbf {\bibinfo {volume}
			{321}},\ \bibinfo {pages} {1126 } (\bibinfo {year} {2006})}\BibitemShut
	{NoStop}%
	\bibitem [{\citenamefont {Gornyi}\ \emph {et~al.}(2005)\citenamefont {Gornyi},
		\citenamefont {Mirlin},\ and\ \citenamefont {Polyakov}}]{Gornyi2005}%
	\BibitemOpen
	\bibfield  {author} {\bibinfo {author} {\bibfnamefont {I.~V.}\ \bibnamefont
			{Gornyi}}, \bibinfo {author} {\bibfnamefont {A.~D.}\ \bibnamefont {Mirlin}},
		\ and\ \bibinfo {author} {\bibfnamefont {D.~G.}\ \bibnamefont {Polyakov}},\
	}\href {\doibase 10.1103/PhysRevLett.95.206603} {\bibfield  {journal}
		{\bibinfo  {journal} {Phys. Rev. Lett.}\ }\textbf {\bibinfo {volume} {95}},\
		\bibinfo {pages} {206603} (\bibinfo {year} {2005})}\BibitemShut {NoStop}%
	\bibitem [{\citenamefont {Nandkishore}\ and\ \citenamefont
		{Huse}(2015)}]{Nandkishore2015}%
	\BibitemOpen
	\bibfield  {author} {\bibinfo {author} {\bibfnamefont {R.}~\bibnamefont
			{Nandkishore}}\ and\ \bibinfo {author} {\bibfnamefont {D.~A.}\ \bibnamefont
			{Huse}},\ }\href@noop {} {\bibfield  {journal} {\bibinfo  {journal} {Annual
				Review of Condensed Matter Physics}\ }\textbf {\bibinfo {volume} {6}},\
		\bibinfo {pages} {15} (\bibinfo {year} {2015})}\BibitemShut {NoStop}%
	\bibitem [{\citenamefont {Abanin}\ \emph {et~al.}(2019)\citenamefont {Abanin},
		\citenamefont {Altman}, \citenamefont {Bloch},\ and\ \citenamefont
		{Serbyn}}]{Abanin-2019}%
	\BibitemOpen
	\bibfield  {author} {\bibinfo {author} {\bibfnamefont {D.~A.}\ \bibnamefont
			{Abanin}}, \bibinfo {author} {\bibfnamefont {E.}~\bibnamefont {Altman}},
		\bibinfo {author} {\bibfnamefont {I.}~\bibnamefont {Bloch}}, \ and\ \bibinfo
		{author} {\bibfnamefont {M.}~\bibnamefont {Serbyn}},\ }\href {\doibase
		10.1103/RevModPhys.91.021001} {\bibfield  {journal} {\bibinfo  {journal}
			{Rev. Mod. Phys.}\ }\textbf {\bibinfo {volume} {91}},\ \bibinfo {pages}
		{021001} (\bibinfo {year} {2019})}\BibitemShut {NoStop}%
	\bibitem [{\citenamefont {Dyson}(1953)}]{Dyson1953}%
	\BibitemOpen
	\bibfield  {author} {\bibinfo {author} {\bibfnamefont {F.~J.}\ \bibnamefont
			{Dyson}},\ }\href {\doibase 10.1103/PhysRev.92.1331} {\bibfield  {journal}
		{\bibinfo  {journal} {Phys. Rev.}\ }\textbf {\bibinfo {volume} {92}},\
		\bibinfo {pages} {1331} (\bibinfo {year} {1953})}\BibitemShut {NoStop}%
	\bibitem [{\citenamefont {Gade}\ and\ \citenamefont {Wegner}(1991)}]{Gade1991}%
	\BibitemOpen
	\bibfield  {author} {\bibinfo {author} {\bibfnamefont {R.}~\bibnamefont
			{Gade}}\ and\ \bibinfo {author} {\bibfnamefont {F.}~\bibnamefont {Wegner}},\
	}\href {\doibase http://dx.doi.org/10.1016/0550-3213(91)90401-I} {\bibfield
		{journal} {\bibinfo  {journal} {Nucl. Phys. B}\ }\textbf {\bibinfo {volume}
			{360}},\ \bibinfo {pages} {213 } (\bibinfo {year} {1991})}\BibitemShut
	{NoStop}%
	\bibitem [{\citenamefont {Gade}(1993)}]{Gade1993}%
	\BibitemOpen
	\bibfield  {author} {\bibinfo {author} {\bibfnamefont {R.}~\bibnamefont
			{Gade}},\ }\href {\doibase http://dx.doi.org/10.1016/0550-3213(93)90601-K}
	{\bibfield  {journal} {\bibinfo  {journal} {Nucl. Phys. B}\ }\textbf
		{\bibinfo {volume} {398}},\ \bibinfo {pages} {499 } (\bibinfo {year}
		{1993})}\BibitemShut {NoStop}%
	\bibitem [{\citenamefont {Motrunich}\ \emph {et~al.}(2002)\citenamefont
		{Motrunich}, \citenamefont {Damle},\ and\ \citenamefont
		{Huse}}]{Motrunich2002}%
	\BibitemOpen
	\bibfield  {author} {\bibinfo {author} {\bibfnamefont {O.}~\bibnamefont
			{Motrunich}}, \bibinfo {author} {\bibfnamefont {K.}~\bibnamefont {Damle}}, \
		and\ \bibinfo {author} {\bibfnamefont {D.~A.}\ \bibnamefont {Huse}},\ }\href
	{\doibase 10.1103/PhysRevB.65.064206} {\bibfield  {journal} {\bibinfo
			{journal} {Phys. Rev. B}\ }\textbf {\bibinfo {volume} {65}},\ \bibinfo
		{pages} {064206} (\bibinfo {year} {2002})}\BibitemShut {NoStop}%
	\bibitem [{\citenamefont {Mudry}\ \emph {et~al.}(2003)\citenamefont {Mudry},
		\citenamefont {Ryu},\ and\ \citenamefont {Furusaki}}]{Mudry2003}%
	\BibitemOpen
	\bibfield  {author} {\bibinfo {author} {\bibfnamefont {C.}~\bibnamefont
			{Mudry}}, \bibinfo {author} {\bibfnamefont {S.}~\bibnamefont {Ryu}}, \ and\
		\bibinfo {author} {\bibfnamefont {A.}~\bibnamefont {Furusaki}},\ }\href
	{\doibase 10.1103/PhysRevB.67.064202} {\bibfield  {journal} {\bibinfo
			{journal} {Phys. Rev. B}\ }\textbf {\bibinfo {volume} {67}},\ \bibinfo
		{pages} {064202} (\bibinfo {year} {2003})}\BibitemShut {NoStop}%
	\bibitem [{\citenamefont {H{\"a}fner}\ \emph {et~al.}(2014)\citenamefont
		{H{\"a}fner}, \citenamefont {Schindler}, \citenamefont {Weik}, \citenamefont
		{Mayer}, \citenamefont {Balakrishnan}, \citenamefont {Narayanan},
		\citenamefont {Bera},\ and\ \citenamefont {Evers}}]{HafnerEvers2014}%
	\BibitemOpen
	\bibfield  {author} {\bibinfo {author} {\bibfnamefont {V.}~\bibnamefont
			{H{\"a}fner}}, \bibinfo {author} {\bibfnamefont {J.}~\bibnamefont
			{Schindler}}, \bibinfo {author} {\bibfnamefont {N.}~\bibnamefont {Weik}},
		\bibinfo {author} {\bibfnamefont {T.}~\bibnamefont {Mayer}}, \bibinfo
		{author} {\bibfnamefont {S.}~\bibnamefont {Balakrishnan}}, \bibinfo {author}
		{\bibfnamefont {R.}~\bibnamefont {Narayanan}}, \bibinfo {author}
		{\bibfnamefont {S.}~\bibnamefont {Bera}}, \ and\ \bibinfo {author}
		{\bibfnamefont {F.}~\bibnamefont {Evers}},\ }\href@noop {} {\bibfield
		{journal} {\bibinfo  {journal} {Physical review letters}\ }\textbf {\bibinfo
			{volume} {113}},\ \bibinfo {pages} {186802} (\bibinfo {year}
		{2014})}\BibitemShut {NoStop}%
	\bibitem [{\citenamefont {Ostrovsky}\ \emph {et~al.}(2014)\citenamefont
		{Ostrovsky}, \citenamefont {Protopopov}, \citenamefont {K\"onig},
		\citenamefont {Gornyi}, \citenamefont {Mirlin},\ and\ \citenamefont
		{Skvortsov}}]{Ostrovsky-2014}%
	\BibitemOpen
	\bibfield  {author} {\bibinfo {author} {\bibfnamefont {P.~M.}\ \bibnamefont
			{Ostrovsky}}, \bibinfo {author} {\bibfnamefont {I.~V.}\ \bibnamefont
			{Protopopov}}, \bibinfo {author} {\bibfnamefont {E.~J.}\ \bibnamefont
			{K\"onig}}, \bibinfo {author} {\bibfnamefont {I.~V.}\ \bibnamefont {Gornyi}},
		\bibinfo {author} {\bibfnamefont {A.~D.}\ \bibnamefont {Mirlin}}, \ and\
		\bibinfo {author} {\bibfnamefont {M.~A.}\ \bibnamefont {Skvortsov}},\ }\href
	{\doibase 10.1103/PhysRevLett.113.186803} {\bibfield  {journal} {\bibinfo
			{journal} {Phys. Rev. Lett.}\ }\textbf {\bibinfo {volume} {113}},\ \bibinfo
		{pages} {186803} (\bibinfo {year} {2014})}\BibitemShut {NoStop}%
	\bibitem [{\citenamefont {Ferreira}\ and\ \citenamefont
		{Mucciolo}(2015)}]{FerreiraMucciolo2015}%
	\BibitemOpen
	\bibfield  {author} {\bibinfo {author} {\bibfnamefont {A.}~\bibnamefont
			{Ferreira}}\ and\ \bibinfo {author} {\bibfnamefont {E.~R.}\ \bibnamefont
			{Mucciolo}},\ }\href@noop {} {\bibfield  {journal} {\bibinfo  {journal}
			{Physical review letters}\ }\textbf {\bibinfo {volume} {115}},\ \bibinfo
		{pages} {106601} (\bibinfo {year} {2015})}\BibitemShut {NoStop}%
	\bibitem [{\citenamefont {Weik}\ \emph {et~al.}(2016)\citenamefont {Weik},
		\citenamefont {Schindler}, \citenamefont {Bera}, \citenamefont {Solomon},\
		and\ \citenamefont {Evers}}]{WeikEvers2016}%
	\BibitemOpen
	\bibfield  {author} {\bibinfo {author} {\bibfnamefont {N.}~\bibnamefont
			{Weik}}, \bibinfo {author} {\bibfnamefont {J.}~\bibnamefont {Schindler}},
		\bibinfo {author} {\bibfnamefont {S.}~\bibnamefont {Bera}}, \bibinfo {author}
		{\bibfnamefont {G.~C.}\ \bibnamefont {Solomon}}, \ and\ \bibinfo {author}
		{\bibfnamefont {F.}~\bibnamefont {Evers}},\ }\href@noop {} {\bibfield
		{journal} {\bibinfo  {journal} {Physical Review B}\ }\textbf {\bibinfo
			{volume} {94}},\ \bibinfo {pages} {064204} (\bibinfo {year}
		{2016})}\BibitemShut {NoStop}%
	\bibitem [{\citenamefont {Sanyal}\ \emph {et~al.}(2016)\citenamefont {Sanyal},
		\citenamefont {Damle},\ and\ \citenamefont {Motrunich}}]{Sanyal2016}%
	\BibitemOpen
	\bibfield  {author} {\bibinfo {author} {\bibfnamefont {S.}~\bibnamefont
			{Sanyal}}, \bibinfo {author} {\bibfnamefont {K.}~\bibnamefont {Damle}}, \
		and\ \bibinfo {author} {\bibfnamefont {O.~I.}\ \bibnamefont {Motrunich}},\
	}\href {\doibase 10.1103/PhysRevLett.117.116806} {\bibfield  {journal}
		{\bibinfo  {journal} {Phys. Rev. Lett.}\ }\textbf {\bibinfo {volume} {117}},\
		\bibinfo {pages} {116806} (\bibinfo {year} {2016})}\BibitemShut {NoStop}%
	\bibitem [{\citenamefont {Aleiner}\ and\ \citenamefont
		{Efetov}(2006)}]{Aleiner-2006}%
	\BibitemOpen
	\bibfield  {author} {\bibinfo {author} {\bibfnamefont {I.~L.}\ \bibnamefont
			{Aleiner}}\ and\ \bibinfo {author} {\bibfnamefont {K.~B.}\ \bibnamefont
			{Efetov}},\ }\href {\doibase 10.1103/PhysRevLett.97.236801} {\bibfield
		{journal} {\bibinfo  {journal} {Phys. Rev. Lett.}\ }\textbf {\bibinfo
			{volume} {97}},\ \bibinfo {pages} {236801} (\bibinfo {year}
		{2006})}\BibitemShut {NoStop}%
	\bibitem [{\citenamefont {Altland}(2006)}]{Altland-2006}%
	\BibitemOpen
	\bibfield  {author} {\bibinfo {author} {\bibfnamefont {A.}~\bibnamefont
			{Altland}},\ }\href {\doibase 10.1103/PhysRevLett.97.236802} {\bibfield
		{journal} {\bibinfo  {journal} {Phys. Rev. Lett.}\ }\textbf {\bibinfo
			{volume} {97}},\ \bibinfo {pages} {236802} (\bibinfo {year}
		{2006})}\BibitemShut {NoStop}%
	\bibitem [{\citenamefont {Pixley}\ \emph
		{et~al.}(2016{\natexlab{a}})\citenamefont {Pixley}, \citenamefont {Huse},\
		and\ \citenamefont {Das~Sarma}}]{Pixley-2016}%
	\BibitemOpen
	\bibfield  {author} {\bibinfo {author} {\bibfnamefont {J.~H.}\ \bibnamefont
			{Pixley}}, \bibinfo {author} {\bibfnamefont {D.~A.}\ \bibnamefont {Huse}}, \
		and\ \bibinfo {author} {\bibfnamefont {S.}~\bibnamefont {Das~Sarma}},\ }\href
	{\doibase 10.1103/PhysRevX.6.021042} {\bibfield  {journal} {\bibinfo
			{journal} {Phys. Rev. X}\ }\textbf {\bibinfo {volume} {6}},\ \bibinfo {pages}
		{021042} (\bibinfo {year} {2016}{\natexlab{a}})}\BibitemShut {NoStop}%
	\bibitem [{\citenamefont {Pixley}\ \emph
		{et~al.}(2016{\natexlab{b}})\citenamefont {Pixley}, \citenamefont {Huse},\
		and\ \citenamefont {Das~Sarma}}]{PixleyBR-2016}%
	\BibitemOpen
	\bibfield  {author} {\bibinfo {author} {\bibfnamefont {J.~H.}\ \bibnamefont
			{Pixley}}, \bibinfo {author} {\bibfnamefont {D.~A.}\ \bibnamefont {Huse}}, \
		and\ \bibinfo {author} {\bibfnamefont {S.}~\bibnamefont {Das~Sarma}},\ }\href
	{\doibase 10.1103/PhysRevB.94.121107} {\bibfield  {journal} {\bibinfo
			{journal} {Phys. Rev. B}\ }\textbf {\bibinfo {volume} {94}},\ \bibinfo
		{pages} {121107} (\bibinfo {year} {2016}{\natexlab{b}})}\BibitemShut
	{NoStop}%
	\bibitem [{\citenamefont {Pixley}\ \emph {et~al.}(2017)\citenamefont {Pixley},
		\citenamefont {Chou}, \citenamefont {Goswami}, \citenamefont {Huse},
		\citenamefont {Nandkishore}, \citenamefont {Radzihovsky},\ and\ \citenamefont
		{Das~Sarma}}]{PixleyB-2017}%
	\BibitemOpen
	\bibfield  {author} {\bibinfo {author} {\bibfnamefont {J.~H.}\ \bibnamefont
			{Pixley}}, \bibinfo {author} {\bibfnamefont {Y.-Z.}\ \bibnamefont {Chou}},
		\bibinfo {author} {\bibfnamefont {P.}~\bibnamefont {Goswami}}, \bibinfo
		{author} {\bibfnamefont {D.~A.}\ \bibnamefont {Huse}}, \bibinfo {author}
		{\bibfnamefont {R.}~\bibnamefont {Nandkishore}}, \bibinfo {author}
		{\bibfnamefont {L.}~\bibnamefont {Radzihovsky}}, \ and\ \bibinfo {author}
		{\bibfnamefont {S.}~\bibnamefont {Das~Sarma}},\ }\href {\doibase
		10.1103/PhysRevB.95.235101} {\bibfield  {journal} {\bibinfo  {journal} {Phys.
				Rev. B}\ }\textbf {\bibinfo {volume} {95}},\ \bibinfo {pages} {235101}
		(\bibinfo {year} {2017})}\BibitemShut {NoStop}%
	\bibitem [{\citenamefont {Wilson}\ \emph {et~al.}(2017)\citenamefont {Wilson},
		\citenamefont {Pixley}, \citenamefont {Goswami},\ and\ \citenamefont
		{Das~Sarma}}]{Wilson-2017}%
	\BibitemOpen
	\bibfield  {author} {\bibinfo {author} {\bibfnamefont {J.~H.}\ \bibnamefont
			{Wilson}}, \bibinfo {author} {\bibfnamefont {J.~H.}\ \bibnamefont {Pixley}},
		\bibinfo {author} {\bibfnamefont {P.}~\bibnamefont {Goswami}}, \ and\
		\bibinfo {author} {\bibfnamefont {S.}~\bibnamefont {Das~Sarma}},\ }\href
	{\doibase 10.1103/PhysRevB.95.155122} {\bibfield  {journal} {\bibinfo
			{journal} {Phys. Rev. B}\ }\textbf {\bibinfo {volume} {95}},\ \bibinfo
		{pages} {155122} (\bibinfo {year} {2017})}\BibitemShut {NoStop}%
	\bibitem [{\citenamefont {Pixley}\ \emph {et~al.}(2018)\citenamefont {Pixley},
		\citenamefont {Wilson}, \citenamefont {Huse},\ and\ \citenamefont
		{Gopalakrishnan}}]{Pixley2018}%
	\BibitemOpen
	\bibfield  {author} {\bibinfo {author} {\bibfnamefont {J.~H.}\ \bibnamefont
			{Pixley}}, \bibinfo {author} {\bibfnamefont {J.~H.}\ \bibnamefont {Wilson}},
		\bibinfo {author} {\bibfnamefont {D.~A.}\ \bibnamefont {Huse}}, \ and\
		\bibinfo {author} {\bibfnamefont {S.}~\bibnamefont {Gopalakrishnan}},\ }\href
	{\doibase 10.1103/PhysRevLett.120.207604} {\bibfield  {journal} {\bibinfo
			{journal} {Phys. Rev. Lett.}\ }\textbf {\bibinfo {volume} {120}},\ \bibinfo
		{pages} {207604} (\bibinfo {year} {2018})}\BibitemShut {NoStop}%
	\bibitem [{\citenamefont {Fu}\ \emph {et~al.}(2018)\citenamefont {Fu},
		\citenamefont {K{\"o}nig}, \citenamefont {Wilson}, \citenamefont {Chou},\
		and\ \citenamefont {Pixley}}]{FuPixley2018}%
	\BibitemOpen
	\bibfield  {author} {\bibinfo {author} {\bibfnamefont {Y.}~\bibnamefont
			{Fu}}, \bibinfo {author} {\bibfnamefont {E.}~\bibnamefont {K{\"o}nig}},
		\bibinfo {author} {\bibfnamefont {J.}~\bibnamefont {Wilson}}, \bibinfo
		{author} {\bibfnamefont {Y.-Z.}\ \bibnamefont {Chou}}, \ and\ \bibinfo
		{author} {\bibfnamefont {J.}~\bibnamefont {Pixley}},\ }\href@noop {}
	{\bibfield  {journal} {\bibinfo  {journal} {arXiv preprint arXiv:1809.04604}\
		} (\bibinfo {year} {2018})}\BibitemShut {NoStop}%
	\bibitem [{\citenamefont {Mastropietro}(2020)}]{Mastropietro2020}%
	\BibitemOpen
	\bibfield  {author} {\bibinfo {author} {\bibfnamefont {V.}~\bibnamefont
			{Mastropietro}},\ }\href {https://arxiv.org/abs/2003.01499} {\bibfield
		{journal} {\bibinfo  {journal} {arXiv preprint arXiv:2003.01499}\ } (\bibinfo
		{year} {2020})}\BibitemShut {NoStop}%
	\bibitem [{\citenamefont {Li}\ \emph {et~al.}(2010)\citenamefont {Li},
		\citenamefont {Luican}, \citenamefont {Lopes~dos Santos}, \citenamefont
		{Castro~Neto}, \citenamefont {Reina}, \citenamefont {Kong},\ and\
		\citenamefont {Andrei}}]{li_observation_2010}%
	\BibitemOpen
	\bibfield  {author} {\bibinfo {author} {\bibfnamefont {G.}~\bibnamefont
			{Li}}, \bibinfo {author} {\bibfnamefont {A.}~\bibnamefont {Luican}}, \bibinfo
		{author} {\bibfnamefont {J.~M.~B.}\ \bibnamefont {Lopes~dos Santos}},
		\bibinfo {author} {\bibfnamefont {A.~H.}\ \bibnamefont {Castro~Neto}},
		\bibinfo {author} {\bibfnamefont {A.}~\bibnamefont {Reina}}, \bibinfo
		{author} {\bibfnamefont {J.}~\bibnamefont {Kong}}, \ and\ \bibinfo {author}
		{\bibfnamefont {E.~Y.}\ \bibnamefont {Andrei}},\ }\href {\doibase
		10.1038/nphys1463} {\bibfield  {journal} {\bibinfo  {journal} {Nature
				Physics}\ }\textbf {\bibinfo {volume} {6}},\ \bibinfo {pages} {109} (\bibinfo
		{year} {2010})}\BibitemShut {NoStop}%
	\bibitem [{\citenamefont {Bistritzer}\ and\ \citenamefont
		{MacDonald}(2011)}]{BistritzerMacDonald2011}%
	\BibitemOpen
	\bibfield  {author} {\bibinfo {author} {\bibfnamefont {R.}~\bibnamefont
			{Bistritzer}}\ and\ \bibinfo {author} {\bibfnamefont {A.~H.}\ \bibnamefont
			{MacDonald}},\ }\href@noop {} {\bibfield  {journal} {\bibinfo  {journal}
			{Proceedings of the National Academy of Sciences}\ }\textbf {\bibinfo
			{volume} {108}},\ \bibinfo {pages} {12233} (\bibinfo {year}
		{2011})}\BibitemShut {NoStop}%
	\bibitem [{\citenamefont {dos Santos}\ \emph {et~al.}(2012)\citenamefont {dos
			Santos}, \citenamefont {Peres},\ and\ \citenamefont
		{Neto}}]{DosSantosNeto2012}%
	\BibitemOpen
	\bibfield  {author} {\bibinfo {author} {\bibfnamefont {J.~L.}\ \bibnamefont
			{dos Santos}}, \bibinfo {author} {\bibfnamefont {N.}~\bibnamefont {Peres}}, \
		and\ \bibinfo {author} {\bibfnamefont {A.~C.}\ \bibnamefont {Neto}},\
	}\href@noop {} {\bibfield  {journal} {\bibinfo  {journal} {Phys. Rev. B}\
		}\textbf {\bibinfo {volume} {86}},\ \bibinfo {pages} {155449} (\bibinfo
		{year} {2012})}\BibitemShut {NoStop}%
	\bibitem [{\citenamefont {Tarruell}\ \emph {et~al.}(2012)\citenamefont
		{Tarruell}, \citenamefont {Greif}, \citenamefont {Uehlinger}, \citenamefont
		{Jotzu},\ and\ \citenamefont {Esslinger}}]{Tarruell-2012}%
	\BibitemOpen
	\bibfield  {author} {\bibinfo {author} {\bibfnamefont {L.}~\bibnamefont
			{Tarruell}}, \bibinfo {author} {\bibfnamefont {D.}~\bibnamefont {Greif}},
		\bibinfo {author} {\bibfnamefont {T.}~\bibnamefont {Uehlinger}}, \bibinfo
		{author} {\bibfnamefont {G.}~\bibnamefont {Jotzu}}, \ and\ \bibinfo {author}
		{\bibfnamefont {T.}~\bibnamefont {Esslinger}},\ }\href@noop {} {\bibfield
		{journal} {\bibinfo  {journal} {Nature}\ }\textbf {\bibinfo {volume} {483}},\
		\bibinfo {pages} {302} (\bibinfo {year} {2012})}\BibitemShut {NoStop}%
	\bibitem [{\citenamefont {Aidelsburger}\ \emph {et~al.}(2015)\citenamefont
		{Aidelsburger}, \citenamefont {Lohse}, \citenamefont {Schweizer},
		\citenamefont {Atala}, \citenamefont {Barreiro}, \citenamefont {Nascimbene},
		\citenamefont {Cooper}, \citenamefont {Bloch},\ and\ \citenamefont
		{Goldman}}]{Aidelsburger-2015}%
	\BibitemOpen
	\bibfield  {author} {\bibinfo {author} {\bibfnamefont {M.}~\bibnamefont
			{Aidelsburger}}, \bibinfo {author} {\bibfnamefont {M.}~\bibnamefont {Lohse}},
		\bibinfo {author} {\bibfnamefont {C.}~\bibnamefont {Schweizer}}, \bibinfo
		{author} {\bibfnamefont {M.}~\bibnamefont {Atala}}, \bibinfo {author}
		{\bibfnamefont {J.~T.}\ \bibnamefont {Barreiro}}, \bibinfo {author}
		{\bibfnamefont {S.}~\bibnamefont {Nascimbene}}, \bibinfo {author}
		{\bibfnamefont {N.}~\bibnamefont {Cooper}}, \bibinfo {author} {\bibfnamefont
			{I.}~\bibnamefont {Bloch}}, \ and\ \bibinfo {author} {\bibfnamefont
			{N.}~\bibnamefont {Goldman}},\ }\href@noop {} {\bibfield  {journal} {\bibinfo
			{journal} {Nature Physics}\ }\textbf {\bibinfo {volume} {11}},\ \bibinfo
		{pages} {162} (\bibinfo {year} {2015})}\BibitemShut {NoStop}%
	\bibitem [{\citenamefont {Fl{\"a}schner}\ \emph {et~al.}(2016)\citenamefont
		{Fl{\"a}schner}, \citenamefont {Rem}, \citenamefont {Tarnowski},
		\citenamefont {Vogel}, \citenamefont {L{\"u}hmann}, \citenamefont
		{Sengstock},\ and\ \citenamefont {Weitenberg}}]{Flaschner-2016}%
	\BibitemOpen
	\bibfield  {author} {\bibinfo {author} {\bibfnamefont {N.}~\bibnamefont
			{Fl{\"a}schner}}, \bibinfo {author} {\bibfnamefont {B.}~\bibnamefont {Rem}},
		\bibinfo {author} {\bibfnamefont {M.}~\bibnamefont {Tarnowski}}, \bibinfo
		{author} {\bibfnamefont {D.}~\bibnamefont {Vogel}}, \bibinfo {author}
		{\bibfnamefont {D.-S.}\ \bibnamefont {L{\"u}hmann}}, \bibinfo {author}
		{\bibfnamefont {K.}~\bibnamefont {Sengstock}}, \ and\ \bibinfo {author}
		{\bibfnamefont {C.}~\bibnamefont {Weitenberg}},\ }\href@noop {} {\bibfield
		{journal} {\bibinfo  {journal} {Science}\ }\textbf {\bibinfo {volume}
			{352}},\ \bibinfo {pages} {1091} (\bibinfo {year} {2016})}\BibitemShut
	{NoStop}%
	\bibitem [{\citenamefont {Weinberg}\ \emph {et~al.}(2016)\citenamefont
		{Weinberg}, \citenamefont {Staarmann}, \citenamefont {{\"O}lschl{\"a}ger},
		\citenamefont {Simonet},\ and\ \citenamefont {Sengstock}}]{Weinberg-2016}%
	\BibitemOpen
	\bibfield  {author} {\bibinfo {author} {\bibfnamefont {M.}~\bibnamefont
			{Weinberg}}, \bibinfo {author} {\bibfnamefont {C.}~\bibnamefont {Staarmann}},
		\bibinfo {author} {\bibfnamefont {C.}~\bibnamefont {{\"O}lschl{\"a}ger}},
		\bibinfo {author} {\bibfnamefont {J.}~\bibnamefont {Simonet}}, \ and\
		\bibinfo {author} {\bibfnamefont {K.}~\bibnamefont {Sengstock}},\ }\href@noop
	{} {\bibfield  {journal} {\bibinfo  {journal} {2D Materials}\ }\textbf
		{\bibinfo {volume} {3}},\ \bibinfo {pages} {024005} (\bibinfo {year}
		{2016})}\BibitemShut {NoStop}%
	\bibitem [{\citenamefont {Gonz\'alez-Tudela}\ and\ \citenamefont
		{Cirac}(2019)}]{Gonzalez2019}%
	\BibitemOpen
	\bibfield  {author} {\bibinfo {author} {\bibfnamefont {A.}~\bibnamefont
			{Gonz\'alez-Tudela}}\ and\ \bibinfo {author} {\bibfnamefont {J.~I.}\
			\bibnamefont {Cirac}},\ }\href {\doibase 10.1103/PhysRevA.100.053604}
	{\bibfield  {journal} {\bibinfo  {journal} {Phys. Rev. A}\ }\textbf {\bibinfo
			{volume} {100}},\ \bibinfo {pages} {053604} (\bibinfo {year}
		{2019})}\BibitemShut {NoStop}%
	\bibitem [{\citenamefont {Wei\ss{}e}\ \emph {et~al.}(2006)\citenamefont
		{Wei\ss{}e}, \citenamefont {Wellein}, \citenamefont {Alvermann},\ and\
		\citenamefont {Fehske}}]{Weisse-2006}%
	\BibitemOpen
	\bibfield  {author} {\bibinfo {author} {\bibfnamefont {A.}~\bibnamefont
			{Wei\ss{}e}}, \bibinfo {author} {\bibfnamefont {G.}~\bibnamefont {Wellein}},
		\bibinfo {author} {\bibfnamefont {A.}~\bibnamefont {Alvermann}}, \ and\
		\bibinfo {author} {\bibfnamefont {H.}~\bibnamefont {Fehske}},\ }\href
	{\doibase 10.1103/RevModPhys.78.275} {\bibfield  {journal} {\bibinfo
			{journal} {Rev. Mod. Phys.}\ }\textbf {\bibinfo {volume} {78}},\ \bibinfo
		{pages} {275} (\bibinfo {year} {2006})}\BibitemShut {NoStop}%
	\bibitem [{\citenamefont {Altland}\ and\ \citenamefont
		{Zirnbauer}(1997)}]{AltlandZirnbauer1997}%
	\BibitemOpen
	\bibfield  {author} {\bibinfo {author} {\bibfnamefont {A.}~\bibnamefont
			{Altland}}\ and\ \bibinfo {author} {\bibfnamefont {M.~R.}\ \bibnamefont
			{Zirnbauer}},\ }\href@noop {} {\bibfield  {journal} {\bibinfo  {journal}
			{Physical Review B}\ }\textbf {\bibinfo {volume} {55}},\ \bibinfo {pages}
		{1142} (\bibinfo {year} {1997})}\BibitemShut {NoStop}%
	\bibitem [{\citenamefont {Devakul}\ and\ \citenamefont
		{Huse}(2017)}]{Devakul2017}%
	\BibitemOpen
	\bibfield  {author} {\bibinfo {author} {\bibfnamefont {T.}~\bibnamefont
			{Devakul}}\ and\ \bibinfo {author} {\bibfnamefont {D.~A.}\ \bibnamefont
			{Huse}},\ }\href {\doibase 10.1103/PhysRevB.96.214201} {\bibfield  {journal}
		{\bibinfo  {journal} {Phys. Rev. B}\ }\textbf {\bibinfo {volume} {96}},\
		\bibinfo {pages} {214201} (\bibinfo {year} {2017})}\BibitemShut {NoStop}%
	\bibitem [{\citenamefont {Tarnopolsky}\ \emph {et~al.}(2019)\citenamefont
		{Tarnopolsky}, \citenamefont {Kruchkov},\ and\ \citenamefont
		{Vishwanath}}]{Tarnopolsky-2019}%
	\BibitemOpen
	\bibfield  {author} {\bibinfo {author} {\bibfnamefont {G.}~\bibnamefont
			{Tarnopolsky}}, \bibinfo {author} {\bibfnamefont {A.~J.}\ \bibnamefont
			{Kruchkov}}, \ and\ \bibinfo {author} {\bibfnamefont {A.}~\bibnamefont
			{Vishwanath}},\ }\href {\doibase 10.1103/PhysRevLett.122.106405} {\bibfield
		{journal} {\bibinfo  {journal} {Phys. Rev. Lett.}\ }\textbf {\bibinfo
			{volume} {122}},\ \bibinfo {pages} {106405} (\bibinfo {year}
		{2019})}\BibitemShut {NoStop}%
	\bibitem [{\citenamefont {Chalker}\ and\ \citenamefont
		{Daniell}(1988)}]{Chalker1988}%
	\BibitemOpen
	\bibfield  {author} {\bibinfo {author} {\bibfnamefont {J.~T.}\ \bibnamefont
			{Chalker}}\ and\ \bibinfo {author} {\bibfnamefont {G.~J.}\ \bibnamefont
			{Daniell}},\ }\href {\doibase 10.1103/PhysRevLett.61.593} {\bibfield
		{journal} {\bibinfo  {journal} {Phys. Rev. Lett.}\ }\textbf {\bibinfo
			{volume} {61}},\ \bibinfo {pages} {593} (\bibinfo {year} {1988})}\BibitemShut
	{NoStop}%
	\bibitem [{\citenamefont {Chalker}(1990)}]{Chalker1990}%
	\BibitemOpen
	\bibfield  {author} {\bibinfo {author} {\bibfnamefont {J.}~\bibnamefont
			{Chalker}},\ }\href {\doibase http://dx.doi.org/10.1016/0378-4371(90)90056-X}
	{\bibfield  {journal} {\bibinfo  {journal} {Physica A: Statistical Mechanics
				and its Applications}\ }\textbf {\bibinfo {volume} {167}},\ \bibinfo {pages}
		{253 } (\bibinfo {year} {1990})}\BibitemShut {NoStop}%
	\bibitem [{\citenamefont {Feigel'man}\ \emph {et~al.}(2007)\citenamefont
		{Feigel'man}, \citenamefont {Ioffe}, \citenamefont {Kravtsov},\ and\
		\citenamefont {Yuzbashyan}}]{Feigelman2007}%
	\BibitemOpen
	\bibfield  {author} {\bibinfo {author} {\bibfnamefont {M.~V.}\ \bibnamefont
			{Feigel'man}}, \bibinfo {author} {\bibfnamefont {L.~B.}\ \bibnamefont
			{Ioffe}}, \bibinfo {author} {\bibfnamefont {V.~E.}\ \bibnamefont {Kravtsov}},
		\ and\ \bibinfo {author} {\bibfnamefont {E.~A.}\ \bibnamefont {Yuzbashyan}},\
	}\href {\doibase 10.1103/PhysRevLett.98.027001} {\bibfield  {journal}
		{\bibinfo  {journal} {Phys. Rev. Lett.}\ }\textbf {\bibinfo {volume} {98}},\
		\bibinfo {pages} {027001} (\bibinfo {year} {2007})}\BibitemShut {NoStop}%
	\bibitem [{\citenamefont {Foster}\ and\ \citenamefont
		{Yuzbashyan}(2012)}]{Foster2012}%
	\BibitemOpen
	\bibfield  {author} {\bibinfo {author} {\bibfnamefont {M.~S.}\ \bibnamefont
			{Foster}}\ and\ \bibinfo {author} {\bibfnamefont {E.~A.}\ \bibnamefont
			{Yuzbashyan}},\ }\href {\doibase 10.1103/PhysRevLett.109.246801} {\bibfield
		{journal} {\bibinfo  {journal} {Phys. Rev. Lett.}\ }\textbf {\bibinfo
			{volume} {109}},\ \bibinfo {pages} {246801} (\bibinfo {year}
		{2012})}\BibitemShut {NoStop}%
	\bibitem [{\citenamefont {Foster}\ \emph {et~al.}(2014)\citenamefont {Foster},
		\citenamefont {Xie},\ and\ \citenamefont {Chou}}]{Foster2014}%
	\BibitemOpen
	\bibfield  {author} {\bibinfo {author} {\bibfnamefont {M.~S.}\ \bibnamefont
			{Foster}}, \bibinfo {author} {\bibfnamefont {H.-Y.}\ \bibnamefont {Xie}}, \
		and\ \bibinfo {author} {\bibfnamefont {Y.-Z.}\ \bibnamefont {Chou}},\ }\href
	{\doibase 10.1103/PhysRevB.89.155140} {\bibfield  {journal} {\bibinfo
			{journal} {Phys. Rev. B}\ }\textbf {\bibinfo {volume} {89}},\ \bibinfo
		{pages} {155140} (\bibinfo {year} {2014})}\BibitemShut {NoStop}%
	\bibitem [{\citenamefont {Burmistrov}\ \emph {et~al.}(2012)\citenamefont
		{Burmistrov}, \citenamefont {Gornyi},\ and\ \citenamefont
		{Mirlin}}]{BurmistrovMirlin2012}%
	\BibitemOpen
	\bibfield  {author} {\bibinfo {author} {\bibfnamefont {I.~S.}\ \bibnamefont
			{Burmistrov}}, \bibinfo {author} {\bibfnamefont {I.~V.}\ \bibnamefont
			{Gornyi}}, \ and\ \bibinfo {author} {\bibfnamefont {A.~D.}\ \bibnamefont
			{Mirlin}},\ }\href {\doibase 10.1103/PhysRevLett.108.017002} {\bibfield
		{journal} {\bibinfo  {journal} {Phys. Rev. Lett.}\ }\textbf {\bibinfo
			{volume} {108}},\ \bibinfo {pages} {017002} (\bibinfo {year}
		{2012})}\BibitemShut {NoStop}%
	\bibitem [{\citenamefont {Benalcazar}\ \emph {et~al.}(2017)\citenamefont
		{Benalcazar}, \citenamefont {Bernevig},\ and\ \citenamefont
		{Hughes}}]{benalcazar2017HOTI}%
	\BibitemOpen
	\bibfield  {author} {\bibinfo {author} {\bibfnamefont {W.~A.}\ \bibnamefont
			{Benalcazar}}, \bibinfo {author} {\bibfnamefont {B.~A.}\ \bibnamefont
			{Bernevig}}, \ and\ \bibinfo {author} {\bibfnamefont {T.~L.}\ \bibnamefont
			{Hughes}},\ }\href@noop {} {\bibfield  {journal} {\bibinfo  {journal}
			{Science}\ }\textbf {\bibinfo {volume} {357}},\ \bibinfo {pages} {61}
		(\bibinfo {year} {2017})}\BibitemShut {NoStop}%
	\bibitem [{\citenamefont {Park}\ \emph {et~al.}(2019)\citenamefont {Park},
		\citenamefont {Kim}, \citenamefont {Cho},\ and\ \citenamefont
		{Lee}}]{Park2019}%
	\BibitemOpen
	\bibfield  {author} {\bibinfo {author} {\bibfnamefont {M.~J.}\ \bibnamefont
			{Park}}, \bibinfo {author} {\bibfnamefont {Y.}~\bibnamefont {Kim}}, \bibinfo
		{author} {\bibfnamefont {G.~Y.}\ \bibnamefont {Cho}}, \ and\ \bibinfo
		{author} {\bibfnamefont {S.}~\bibnamefont {Lee}},\ }\href {\doibase
		10.1103/PhysRevLett.123.216803} {\bibfield  {journal} {\bibinfo  {journal}
			{Phys. Rev. Lett.}\ }\textbf {\bibinfo {volume} {123}},\ \bibinfo {pages}
		{216803} (\bibinfo {year} {2019})}\BibitemShut {NoStop}%
	\bibitem [{\citenamefont {Fehske}\ \emph {et~al.}(2007)\citenamefont {Fehske},
		\citenamefont {Schneider},\ and\ \citenamefont {Wei{\ss}e}}]{Fehske-2007}%
	\BibitemOpen
	\bibfield  {author} {\bibinfo {author} {\bibfnamefont {H.}~\bibnamefont
			{Fehske}}, \bibinfo {author} {\bibfnamefont {R.}~\bibnamefont {Schneider}}, \
		and\ \bibinfo {author} {\bibfnamefont {A.}~\bibnamefont {Wei{\ss}e}},\
	}\href@noop {} {\emph {\bibinfo {title} {Computational many-particle
				physics}}},\ Vol.\ \bibinfo {volume} {739}\ (\bibinfo  {publisher}
	{Springer},\ \bibinfo {year} {2007})\BibitemShut {NoStop}%
	\bibitem [{\citenamefont {Huckestein}(1995)}]{Huckestein1995}%
	\BibitemOpen
	\bibfield  {author} {\bibinfo {author} {\bibfnamefont {B.}~\bibnamefont
			{Huckestein}},\ }\href {\doibase 10.1103/RevModPhys.67.357} {\bibfield
		{journal} {\bibinfo  {journal} {Rev. Mod. Phys.}\ }\textbf {\bibinfo {volume}
			{67}},\ \bibinfo {pages} {357} (\bibinfo {year} {1995})}\BibitemShut
	{NoStop}%
	\bibitem [{\citenamefont {Chamon}\ \emph {et~al.}(1996)\citenamefont {Chamon},
		\citenamefont {Mudry},\ and\ \citenamefont {Wen}}]{Chamon1996}%
	\BibitemOpen
	\bibfield  {author} {\bibinfo {author} {\bibfnamefont {C.~d.~C.}\
			\bibnamefont {Chamon}}, \bibinfo {author} {\bibfnamefont {C.}~\bibnamefont
			{Mudry}}, \ and\ \bibinfo {author} {\bibfnamefont {X.-G.}\ \bibnamefont
			{Wen}},\ }\href {\doibase 10.1103/PhysRevLett.77.4194} {\bibfield  {journal}
		{\bibinfo  {journal} {Phys. Rev. Lett.}\ }\textbf {\bibinfo {volume} {77}},\
		\bibinfo {pages} {4194} (\bibinfo {year} {1996})}\BibitemShut {NoStop}%
	\bibitem [{\citenamefont {Cuevas}\ and\ \citenamefont
		{Kravtsov}(2007)}]{Cuevas2007}%
	\BibitemOpen
	\bibfield  {author} {\bibinfo {author} {\bibfnamefont {E.}~\bibnamefont
			{Cuevas}}\ and\ \bibinfo {author} {\bibfnamefont {V.~E.}\ \bibnamefont
			{Kravtsov}},\ }\href {\doibase 10.1103/PhysRevB.76.235119} {\bibfield
		{journal} {\bibinfo  {journal} {Phys. Rev. B}\ }\textbf {\bibinfo {volume}
			{76}},\ \bibinfo {pages} {235119} (\bibinfo {year} {2007})}\BibitemShut
	{NoStop}%
	\bibitem [{\citenamefont {Chou}\ and\ \citenamefont {Foster}(2014)}]{Chou2014}%
	\BibitemOpen
	\bibfield  {author} {\bibinfo {author} {\bibfnamefont {Y.-Z.}\ \bibnamefont
			{Chou}}\ and\ \bibinfo {author} {\bibfnamefont {M.~S.}\ \bibnamefont
			{Foster}},\ }\href {\doibase 10.1103/PhysRevB.89.165136} {\bibfield
		{journal} {\bibinfo  {journal} {Phys. Rev. B}\ }\textbf {\bibinfo {volume}
			{89}},\ \bibinfo {pages} {165136} (\bibinfo {year} {2014})}\BibitemShut
	{NoStop}%
	\bibitem [{\citenamefont {Fyodorov}\ and\ \citenamefont
		{Mirlin}(1997)}]{Fyodorov1997}%
	\BibitemOpen
	\bibfield  {author} {\bibinfo {author} {\bibfnamefont {Y.~V.}\ \bibnamefont
			{Fyodorov}}\ and\ \bibinfo {author} {\bibfnamefont {A.~D.}\ \bibnamefont
			{Mirlin}},\ }\href {\doibase 10.1103/PhysRevB.55.R16001} {\bibfield
		{journal} {\bibinfo  {journal} {Phys. Rev. B}\ }\textbf {\bibinfo {volume}
			{55}},\ \bibinfo {pages} {R16001} (\bibinfo {year} {1997})}\BibitemShut
	{NoStop}%
	\bibitem [{\citenamefont {Ludwig}\ \emph {et~al.}(1994)\citenamefont {Ludwig},
		\citenamefont {Fisher}, \citenamefont {Shankar},\ and\ \citenamefont
		{Grinstein}}]{Ludwig1994}%
	\BibitemOpen
	\bibfield  {author} {\bibinfo {author} {\bibfnamefont {A.~W.~W.}\
			\bibnamefont {Ludwig}}, \bibinfo {author} {\bibfnamefont {M.~P.~A.}\
			\bibnamefont {Fisher}}, \bibinfo {author} {\bibfnamefont {R.}~\bibnamefont
			{Shankar}}, \ and\ \bibinfo {author} {\bibfnamefont {G.}~\bibnamefont
			{Grinstein}},\ }\href {\doibase 10.1103/PhysRevB.50.7526} {\bibfield
		{journal} {\bibinfo  {journal} {Phys. Rev. B}\ }\textbf {\bibinfo {volume}
			{50}},\ \bibinfo {pages} {7526} (\bibinfo {year} {1994})}\BibitemShut
	{NoStop}%
	\bibitem [{\citenamefont {Castillo}\ \emph {et~al.}(1997)\citenamefont
		{Castillo}, \citenamefont {de~C.~Chamon}, \citenamefont {Fradkin},
		\citenamefont {Goldbart},\ and\ \citenamefont {Mudry}}]{Castillo1997}%
	\BibitemOpen
	\bibfield  {author} {\bibinfo {author} {\bibfnamefont {H.~E.}\ \bibnamefont
			{Castillo}}, \bibinfo {author} {\bibfnamefont {C.}~\bibnamefont
			{de~C.~Chamon}}, \bibinfo {author} {\bibfnamefont {E.}~\bibnamefont
			{Fradkin}}, \bibinfo {author} {\bibfnamefont {P.~M.}\ \bibnamefont
			{Goldbart}}, \ and\ \bibinfo {author} {\bibfnamefont {C.}~\bibnamefont
			{Mudry}},\ }\href {\doibase 10.1103/PhysRevB.56.10668} {\bibfield  {journal}
		{\bibinfo  {journal} {Phys. Rev. B}\ }\textbf {\bibinfo {volume} {56}},\
		\bibinfo {pages} {10668} (\bibinfo {year} {1997})}\BibitemShut {NoStop}%
	\bibitem [{\citenamefont {Carpentier}\ and\ \citenamefont
		{Le~Doussal}(2001)}]{Carpentier2001}%
	\BibitemOpen
	\bibfield  {author} {\bibinfo {author} {\bibfnamefont {D.}~\bibnamefont
			{Carpentier}}\ and\ \bibinfo {author} {\bibfnamefont {P.}~\bibnamefont
			{Le~Doussal}},\ }\href {\doibase 10.1103/PhysRevE.63.026110} {\bibfield
		{journal} {\bibinfo  {journal} {Phys. Rev. E}\ }\textbf {\bibinfo {volume}
			{63}},\ \bibinfo {pages} {026110} (\bibinfo {year} {2001})}\BibitemShut
	{NoStop}%
	\bibitem [{\citenamefont {Horovitz}\ and\ \citenamefont
		{Doussal}(2002)}]{Horovitz2002}%
	\BibitemOpen
	\bibfield  {author} {\bibinfo {author} {\bibfnamefont {B.}~\bibnamefont
			{Horovitz}}\ and\ \bibinfo {author} {\bibfnamefont {P.~L.}\ \bibnamefont
			{Doussal}},\ }\href {\doibase 10.1103/PhysRevB.65.125323} {\bibfield
		{journal} {\bibinfo  {journal} {Phys. Rev. B}\ }\textbf {\bibinfo {volume}
			{65}},\ \bibinfo {pages} {125323} (\bibinfo {year} {2002})}\BibitemShut
	{NoStop}%
	\bibitem [{\citenamefont {Jackiw}\ and\ \citenamefont
		{Rebbi}(1976)}]{Jackiw1976}%
	\BibitemOpen
	\bibfield  {author} {\bibinfo {author} {\bibfnamefont {R.}~\bibnamefont
			{Jackiw}}\ and\ \bibinfo {author} {\bibfnamefont {C.}~\bibnamefont {Rebbi}},\
	}\href {\doibase 10.1103/PhysRevD.13.3398} {\bibfield  {journal} {\bibinfo
			{journal} {Phys. Rev. D}\ }\textbf {\bibinfo {volume} {13}},\ \bibinfo
		{pages} {3398} (\bibinfo {year} {1976})}\BibitemShut {NoStop}%
	\bibitem [{\citenamefont {Jackiw}\ and\ \citenamefont
		{Rossi}(1981)}]{Jackiw1981}%
	\BibitemOpen
	\bibfield  {author} {\bibinfo {author} {\bibfnamefont {R.}~\bibnamefont
			{Jackiw}}\ and\ \bibinfo {author} {\bibfnamefont {P.}~\bibnamefont {Rossi}},\
	}\href@noop {} {\bibfield  {journal} {\bibinfo  {journal} {Nuclear Physics
				B}\ }\textbf {\bibinfo {volume} {190}},\ \bibinfo {pages} {681} (\bibinfo
		{year} {1981})}\BibitemShut {NoStop}%
	\bibitem [{\citenamefont {Zhang}(2019)}]{Longzhang2018}%
	\BibitemOpen
	\bibfield  {author} {\bibinfo {author} {\bibfnamefont {L.}~\bibnamefont
			{Zhang}},\ }\href@noop {} {\bibfield  {journal} {\bibinfo  {journal} {Science
				Bulletin}\ }\textbf {\bibinfo {volume} {64}},\ \bibinfo {pages} {495}
		(\bibinfo {year} {2019})}\BibitemShut {NoStop}%
	\bibitem [{\citenamefont {K\"onig}\ \emph {et~al.}(2012)\citenamefont
		{K\"onig}, \citenamefont {Ostrovsky}, \citenamefont {Protopopov},\ and\
		\citenamefont {Mirlin}}]{Konig-2012}%
	\BibitemOpen
	\bibfield  {author} {\bibinfo {author} {\bibfnamefont {E.~J.}\ \bibnamefont
			{K\"onig}}, \bibinfo {author} {\bibfnamefont {P.~M.}\ \bibnamefont
			{Ostrovsky}}, \bibinfo {author} {\bibfnamefont {I.~V.}\ \bibnamefont
			{Protopopov}}, \ and\ \bibinfo {author} {\bibfnamefont {A.~D.}\ \bibnamefont
			{Mirlin}},\ }\href {\doibase 10.1103/PhysRevB.85.195130} {\bibfield
		{journal} {\bibinfo  {journal} {Phys. Rev. B}\ }\textbf {\bibinfo {volume}
			{85}},\ \bibinfo {pages} {195130} (\bibinfo {year} {2012})}\BibitemShut
	{NoStop}%
	\bibitem [{fn1()}]{fn1}%
	\BibitemOpen
	\href@noop {} {}\bibinfo {note} {The real hopping model in this work can be
		decomposed into two decoupled $\pi$-flux hopping models. It belongs to the
		chiral orthogonal class.}\BibitemShut {Stop}%
	\bibitem [{fn2()}]{fn2}%
	\BibitemOpen
	\href@noop {} {}\bibinfo {note} {$L=610$ is a large enough $L$ to suppress
		this rounding at the expansion orders we consider here.}\BibitemShut {Stop}%
	\bibitem [{\citenamefont {Chhabra}\ and\ \citenamefont
		{Jensen}(1989)}]{Chhabra1989}%
	\BibitemOpen
	\bibfield  {author} {\bibinfo {author} {\bibfnamefont {A.}~\bibnamefont
			{Chhabra}}\ and\ \bibinfo {author} {\bibfnamefont {R.~V.}\ \bibnamefont
			{Jensen}},\ }\href {\doibase 10.1103/PhysRevLett.62.1327} {\bibfield
		{journal} {\bibinfo  {journal} {Phys. Rev. Lett.}\ }\textbf {\bibinfo
			{volume} {62}},\ \bibinfo {pages} {1327} (\bibinfo {year}
		{1989})}\BibitemShut {NoStop}%
	\bibitem [{\citenamefont {Wu}\ \emph {et~al.}(2016)\citenamefont {Wu},
		\citenamefont {Zhang}, \citenamefont {Sun}, \citenamefont {Xu}, \citenamefont
		{Wang}, \citenamefont {Ji}, \citenamefont {Deng}, \citenamefont {Chen},
		\citenamefont {Liu},\ and\ \citenamefont {Pan}}]{WuPan2016}%
	\BibitemOpen
	\bibfield  {author} {\bibinfo {author} {\bibfnamefont {Z.}~\bibnamefont
			{Wu}}, \bibinfo {author} {\bibfnamefont {L.}~\bibnamefont {Zhang}}, \bibinfo
		{author} {\bibfnamefont {W.}~\bibnamefont {Sun}}, \bibinfo {author}
		{\bibfnamefont {X.-T.}\ \bibnamefont {Xu}}, \bibinfo {author} {\bibfnamefont
			{B.-Z.}\ \bibnamefont {Wang}}, \bibinfo {author} {\bibfnamefont {S.-C.}\
			\bibnamefont {Ji}}, \bibinfo {author} {\bibfnamefont {Y.}~\bibnamefont
			{Deng}}, \bibinfo {author} {\bibfnamefont {S.}~\bibnamefont {Chen}}, \bibinfo
		{author} {\bibfnamefont {X.-J.}\ \bibnamefont {Liu}}, \ and\ \bibinfo
		{author} {\bibfnamefont {J.-W.}\ \bibnamefont {Pan}},\ }\href@noop {}
	{\bibfield  {journal} {\bibinfo  {journal} {Science}\ }\textbf {\bibinfo
			{volume} {354}},\ \bibinfo {pages} {83} (\bibinfo {year} {2016})}\BibitemShut
	{NoStop}%
	\bibitem [{\citenamefont {Chen}\ \emph {et~al.}(2009)\citenamefont {Chen},
		\citenamefont {He}, \citenamefont {Chien},\ and\ \citenamefont
		{Levin}}]{ChenLevin2009}%
	\BibitemOpen
	\bibfield  {author} {\bibinfo {author} {\bibfnamefont {Q.}~\bibnamefont
			{Chen}}, \bibinfo {author} {\bibfnamefont {Y.}~\bibnamefont {He}}, \bibinfo
		{author} {\bibfnamefont {C.-C.}\ \bibnamefont {Chien}}, \ and\ \bibinfo
		{author} {\bibfnamefont {K.}~\bibnamefont {Levin}},\ }\href@noop {}
	{\bibfield  {journal} {\bibinfo  {journal} {Reports on Progress in Physics}\
		}\textbf {\bibinfo {volume} {72}},\ \bibinfo {pages} {122501} (\bibinfo
		{year} {2009})}\BibitemShut {NoStop}%
	\bibitem [{\citenamefont {Greiner}\ \emph {et~al.}(2001)\citenamefont
		{Greiner}, \citenamefont {Bloch}, \citenamefont {Mandel}, \citenamefont
		{H{\"a}nsch},\ and\ \citenamefont {Esslinger}}]{GreinerEsslinger2001}%
	\BibitemOpen
	\bibfield  {author} {\bibinfo {author} {\bibfnamefont {M.}~\bibnamefont
			{Greiner}}, \bibinfo {author} {\bibfnamefont {I.}~\bibnamefont {Bloch}},
		\bibinfo {author} {\bibfnamefont {O.}~\bibnamefont {Mandel}}, \bibinfo
		{author} {\bibfnamefont {T.~W.}\ \bibnamefont {H{\"a}nsch}}, \ and\ \bibinfo
		{author} {\bibfnamefont {T.}~\bibnamefont {Esslinger}},\ }\href@noop {}
	{\bibfield  {journal} {\bibinfo  {journal} {Physical Review Letters}\
		}\textbf {\bibinfo {volume} {87}},\ \bibinfo {pages} {160405} (\bibinfo
		{year} {2001})}\BibitemShut {NoStop}%
	\bibitem [{\citenamefont {K\"ohl}\ \emph {et~al.}(2005)\citenamefont {K\"ohl},
		\citenamefont {Moritz}, \citenamefont {St\"oferle}, \citenamefont
		{G\"unter},\ and\ \citenamefont {Esslinger}}]{KoehlEsslinger2005}%
	\BibitemOpen
	\bibfield  {author} {\bibinfo {author} {\bibfnamefont {M.}~\bibnamefont
			{K\"ohl}}, \bibinfo {author} {\bibfnamefont {H.}~\bibnamefont {Moritz}},
		\bibinfo {author} {\bibfnamefont {T.}~\bibnamefont {St\"oferle}}, \bibinfo
		{author} {\bibfnamefont {K.}~\bibnamefont {G\"unter}}, \ and\ \bibinfo
		{author} {\bibfnamefont {T.}~\bibnamefont {Esslinger}},\ }\href {\doibase
		10.1103/PhysRevLett.94.080403} {\bibfield  {journal} {\bibinfo  {journal}
			{Phys. Rev. Lett.}\ }\textbf {\bibinfo {volume} {94}},\ \bibinfo {pages}
		{080403} (\bibinfo {year} {2005})}\BibitemShut {NoStop}%
	\bibitem [{\citenamefont {Peterson}\ \emph {et~al.}(2018)\citenamefont
		{Peterson}, \citenamefont {Benalcazar}, \citenamefont {Hughes},\ and\
		\citenamefont {Bahl}}]{peterson2018resonator}%
	\BibitemOpen
	\bibfield  {author} {\bibinfo {author} {\bibfnamefont {C.~W.}\ \bibnamefont
			{Peterson}}, \bibinfo {author} {\bibfnamefont {W.~A.}\ \bibnamefont
			{Benalcazar}}, \bibinfo {author} {\bibfnamefont {T.~L.}\ \bibnamefont
			{Hughes}}, \ and\ \bibinfo {author} {\bibfnamefont {G.}~\bibnamefont
			{Bahl}},\ }\href@noop {} {\bibfield  {journal} {\bibinfo  {journal} {Nature}\
		}\textbf {\bibinfo {volume} {555}},\ \bibinfo {pages} {346} (\bibinfo {year}
		{2018})}\BibitemShut {NoStop}%
	\bibitem [{\citenamefont {Koll{\'a}r}\ \emph {et~al.}(2019)\citenamefont
		{Koll{\'a}r}, \citenamefont {Fitzpatrick},\ and\ \citenamefont
		{Houck}}]{Kollar2019}%
	\BibitemOpen
	\bibfield  {author} {\bibinfo {author} {\bibfnamefont {A.~J.}\ \bibnamefont
			{Koll{\'a}r}}, \bibinfo {author} {\bibfnamefont {M.}~\bibnamefont
			{Fitzpatrick}}, \ and\ \bibinfo {author} {\bibfnamefont {A.~A.}\ \bibnamefont
			{Houck}},\ }\href@noop {} {\bibfield  {journal} {\bibinfo  {journal}
			{Nature}\ }\textbf {\bibinfo {volume} {571}},\ \bibinfo {pages} {45}
		(\bibinfo {year} {2019})}\BibitemShut {NoStop}%
	\bibitem [{\citenamefont {Khanikaev}\ and\ \citenamefont
		{Shvets}(2017)}]{khanikaev2017photonic}%
	\BibitemOpen
	\bibfield  {author} {\bibinfo {author} {\bibfnamefont {A.~B.}\ \bibnamefont
			{Khanikaev}}\ and\ \bibinfo {author} {\bibfnamefont {G.}~\bibnamefont
			{Shvets}},\ }\href@noop {} {\bibfield  {journal} {\bibinfo  {journal} {Nature
				Photonics}\ }\textbf {\bibinfo {volume} {11}},\ \bibinfo {pages} {763}
		(\bibinfo {year} {2017})}\BibitemShut {NoStop}%
	\bibitem [{\citenamefont {Nash}\ \emph {et~al.}(2015)\citenamefont {Nash},
		\citenamefont {Kleckner}, \citenamefont {Read}, \citenamefont {Vitelli},
		\citenamefont {Turner},\ and\ \citenamefont {Irvine}}]{nash2015phononic}%
	\BibitemOpen
	\bibfield  {author} {\bibinfo {author} {\bibfnamefont {L.~M.}\ \bibnamefont
			{Nash}}, \bibinfo {author} {\bibfnamefont {D.}~\bibnamefont {Kleckner}},
		\bibinfo {author} {\bibfnamefont {A.}~\bibnamefont {Read}}, \bibinfo {author}
		{\bibfnamefont {V.}~\bibnamefont {Vitelli}}, \bibinfo {author} {\bibfnamefont
			{A.~M.}\ \bibnamefont {Turner}}, \ and\ \bibinfo {author} {\bibfnamefont
			{W.~T.}\ \bibnamefont {Irvine}},\ }\href@noop {} {\bibfield  {journal}
		{\bibinfo  {journal} {Proceedings of the National Academy of Sciences}\
		}\textbf {\bibinfo {volume} {112}},\ \bibinfo {pages} {14495} (\bibinfo
		{year} {2015})}\BibitemShut {NoStop}%
	\bibitem [{\citenamefont {Cao}\ \emph {et~al.}(2018{\natexlab{a}})\citenamefont
		{Cao}, \citenamefont {Fatemi}, \citenamefont {Demir}, \citenamefont {Fang},
		\citenamefont {Tomarken}, \citenamefont {Luo}, \citenamefont
		{Sanchez-Yamagishi}, \citenamefont {Watanabe}, \citenamefont {Taniguchi},
		\citenamefont {Kaxiras}, \citenamefont {Ashoori},\ and\ \citenamefont
		{Jarillo-Herrero}}]{tbg1}%
	\BibitemOpen
	\bibfield  {author} {\bibinfo {author} {\bibfnamefont {Y.}~\bibnamefont
			{Cao}}, \bibinfo {author} {\bibfnamefont {V.}~\bibnamefont {Fatemi}},
		\bibinfo {author} {\bibfnamefont {A.}~\bibnamefont {Demir}}, \bibinfo
		{author} {\bibfnamefont {S.}~\bibnamefont {Fang}}, \bibinfo {author}
		{\bibfnamefont {S.~L.}\ \bibnamefont {Tomarken}}, \bibinfo {author}
		{\bibfnamefont {J.~Y.}\ \bibnamefont {Luo}}, \bibinfo {author} {\bibfnamefont
			{J.~D.}\ \bibnamefont {Sanchez-Yamagishi}}, \bibinfo {author} {\bibfnamefont
			{K.}~\bibnamefont {Watanabe}}, \bibinfo {author} {\bibfnamefont
			{T.}~\bibnamefont {Taniguchi}}, \bibinfo {author} {\bibfnamefont
			{E.}~\bibnamefont {Kaxiras}}, \bibinfo {author} {\bibfnamefont {R.~C.}\
			\bibnamefont {Ashoori}}, \ and\ \bibinfo {author} {\bibfnamefont
			{P.}~\bibnamefont {Jarillo-Herrero}},\ }\href {\doibase 10.1038/nature26154}
	{\bibfield  {journal} {\bibinfo  {journal} {Nature}\ }\textbf {\bibinfo
			{volume} {556}},\ \bibinfo {pages} {80} (\bibinfo {year}
		{2018}{\natexlab{a}})}\BibitemShut {NoStop}%
	\bibitem [{\citenamefont {Cao}\ \emph {et~al.}(2018{\natexlab{b}})\citenamefont
		{Cao}, \citenamefont {Fatemi}, \citenamefont {Fang}, \citenamefont
		{Watanabe}, \citenamefont {Taniguchi}, \citenamefont {Kaxiras},\ and\
		\citenamefont {Jarillo-Herrero}}]{tbg2}%
	\BibitemOpen
	\bibfield  {author} {\bibinfo {author} {\bibfnamefont {Y.}~\bibnamefont
			{Cao}}, \bibinfo {author} {\bibfnamefont {V.}~\bibnamefont {Fatemi}},
		\bibinfo {author} {\bibfnamefont {S.}~\bibnamefont {Fang}}, \bibinfo {author}
		{\bibfnamefont {K.}~\bibnamefont {Watanabe}}, \bibinfo {author}
		{\bibfnamefont {T.}~\bibnamefont {Taniguchi}}, \bibinfo {author}
		{\bibfnamefont {E.}~\bibnamefont {Kaxiras}}, \ and\ \bibinfo {author}
		{\bibfnamefont {P.}~\bibnamefont {Jarillo-Herrero}},\ }\href {\doibase
		10.1038/nature26160} {\bibfield  {journal} {\bibinfo  {journal} {Nature}\
		}\textbf {\bibinfo {volume} {556}},\ \bibinfo {pages} {43} (\bibinfo {year}
		{2018}{\natexlab{b}})}\BibitemShut {NoStop}%
	\bibitem [{\citenamefont {Inui}\ \emph {et~al.}(1994)\citenamefont {Inui},
		\citenamefont {Trugman},\ and\ \citenamefont {Abrahams}}]{InuiAbrahams1994}%
	\BibitemOpen
	\bibfield  {author} {\bibinfo {author} {\bibfnamefont {M.}~\bibnamefont
			{Inui}}, \bibinfo {author} {\bibfnamefont {S.}~\bibnamefont {Trugman}}, \
		and\ \bibinfo {author} {\bibfnamefont {E.}~\bibnamefont {Abrahams}},\
	}\href@noop {} {\bibfield  {journal} {\bibinfo  {journal} {Physical Review
				B}\ }\textbf {\bibinfo {volume} {49}},\ \bibinfo {pages} {3190} (\bibinfo
		{year} {1994})}\BibitemShut {NoStop}%
\end{thebibliography}

%


\end{document}